\documentclass{article}

\usepackage{arxiv}

\usepackage[utf8]{inputenc} 
\usepackage[T1]{fontenc}    
\usepackage{hyperref}       
\usepackage{url}            
\usepackage{booktabs}       
\usepackage{multirow}
\usepackage{amsfonts}       
\usepackage{amsmath}       
\usepackage{nicefrac}       
\usepackage{microtype}      
\usepackage{lipsum}		
\usepackage{natbib}
\usepackage{doi}
\usepackage{graphicx}
\usepackage{subcaption}

\usepackage{color}
\usepackage{soul} 

\title{Neptune: An AI model for Global Ocean Subseasonal Prediction}

\author{ 
    \href{https://orcid.org/0009-0001-0440-7962}{\includegraphics[scale=0.06]{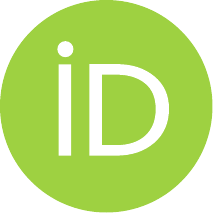}\hspace{1mm}Davide Donno\textsuperscript{1,2}} \\
    \And
    \href{https://orcid.org/0000-0002-6408-1335}{\includegraphics[scale=0.06]{orcid.pdf}\hspace{1mm}Italo Epicoco\textsuperscript{1,2}} \\
    \And
    \href{https://orcid.org/0000-0003-1118-7109}{\includegraphics[scale=0.06]{orcid.pdf}\hspace{1mm}Massimo Cafaro\textsuperscript{1}} \\
    \And
    \href{https://orcid.org/0000-0002-2969-1517}{\includegraphics[scale=0.06]{orcid.pdf}\hspace{1mm}Gabriele Accarino\textsuperscript{3,4}} \\
    \And
    \href{https://orcid.org/0000-0002-8402-5779}{\includegraphics[scale=0.06]{orcid.pdf}\hspace{1mm}Mohammad M. Amirian\textsuperscript{5}} \\
    \And
    \href{https://orcid.org/0000-0002-6788-6315}{\includegraphics[scale=0.06]{orcid.pdf}\hspace{1mm}Viviana Acquaviva\textsuperscript{5}} \\
    \And
    \href{https://orcid.org/0000-0003-2857-4391}{\includegraphics[scale=0.06]{orcid.pdf}\hspace{1mm}Paola Nassisi\textsuperscript{2}} \\
    \And
    \href{https://orcid.org/0000-0001-5132-7255}{\includegraphics[scale=0.06]{orcid.pdf}\hspace{1mm}Doroteaciro Iovino\textsuperscript{2}} \\
    \And
    \href{https://orcid.org/0000-0002-1619-6103}{\includegraphics[scale=0.06]{orcid.pdf}\hspace{1mm}Annalisa Bracco\textsuperscript{2}} \\
    \And
    \href{https://orcid.org/0000-0001-6273-7065}{\includegraphics[scale=0.06]{orcid.pdf}\hspace{1mm}Simona Masina\textsuperscript{2}} \\
    \And
    \href{https://orcid.org/0000-0002-0845-8345}{\includegraphics[scale=0.06]{orcid.pdf}\hspace{1mm}Pierre Gentine\textsuperscript{3,4}} \\
}

\renewcommand{\shorttitle}{Donno et al.}

\begin{document}
\maketitle

1 Department of Engineering for Innovation, University of Salento, Via per Monteroni, Lecce, Italy \\
2 CMCC Foundation - Euro-Mediterranean Center on Climate Change, Italy \\
3 Department of Earth and Environmental Engineering, Columbia University, New York, NY, USA \\
4 Learning the Earth with Artificial Intelligence \& Physics (LEAP) Center, Columbia University, New York, NY, USA \\
5 CUNY New York City College of Technology, 300 Jay Street, Brooklyn NY 11201 \\

\begin{abstract}
Subseasonal-to-seasonal (S2S) forecasting is societally critical, supporting decision-making in sectors ranging from water and agricultural management to disaster risk reduction, energy planning, and insurance. Achieving reliable predictions at these timescales requires representing the ocean and its dynamics, but traditional physics-based ocean numerical models, also known as Ocean General Circulation Models (OGCMs),  are computationally expensive and difficult to develop and improve because of the code complexity. 
In this work, we propose Neptune, an end-to-end data-driven framework for global ocean and sea-ice components emulation tailored for S2S timescales, up to 60 days. Neptune combines Convolutional Neural Networks (CNNs) and Spherical Fourier Neural Operators (SFNOs) to effectively capture local features and global cross-scale interactions, thereby obtaining a coherent representation of the ocean state. Forced by prescribed daily atmospheric fields, Neptune emulates ocean state variables, from temperature and salinity, to zonal and meridional currents, from sea surface height to sea ice thickness and concentration, with daily outputs at the ocean surface and through the water column. Specifically, we propose two variants of Neptune, Neptune-1 and Neptune-025, capable of emulating the ocean state at $1^\circ$ and $0.25^\circ$ horizontal resolution, respectively.
Evaluated against a suite of metrics, including statistics (RMSE, CRPS and ACC), physical coherency (Ocean Heat Content, Eddy Kinetic Energy and Ice Brier Score) and climate indices (ENSO and Z20 metric, IOD), Neptune successfully reproduces the spatio-temporal evolution of the oceanic fields up to 60 days, and is stable over long timescales.
Neptune provides compelling evidence that end-to-end data-driven ocean emulators can become a powerful component of next-generation S2S forecasting systems, emulating ocean state at high spatio-temporal resolution.
\end{abstract}


\section{Introduction and State of the art}
\label{sect:intro}

Subseasonal-to-seasonal (S2S) forecasting is critical for protecting communities and economies because it provides the actionable weeks-to-months lead times needed to prepare for high-impact large-scale events such as heatwaves, droughts, floods, and marine extremes \citep{Vitart2018}. It supports decision-making in sectors ranging from agriculture \citep{LIANG2026} and fishery management \citep{pikitch2004} to disaster risk reduction, energy planning, and insurance \citep{troccoli2010,meehl2021}. 
Achieving reliable predictions at these timescales, though, is more complicated than weather forecasting, because it requires accounting for, and therefore adequately representing, the ocean and its dynamics \citep{BALMASEDA2026271}. Oceanic processes constitute a key source of predictability because the ocean stores and redistributes heat, freshwater, and other climate-relevant properties over much longer timescales than the atmosphere, thereby sustaining the forecast skill. In other words, after two weeks the ocean memory and ocean-atmosphere interactions, including large-scale modes of variability and regional circulation changes, strongly influence the development and persistence of climate anomalies, making their explicit incorporation in longer forecasting essential. For this reason, Ocean General Circulation Models (OGCMs), physics-based numerical models that simulate the complex interactions underlying ocean dynamics by solving a set of partial differential equations (PDE) are a key component of S2S forecast systems. 
OGCMs, however,  are computationally intensive, difficult to maintain and improve, and are mostly impractical for both quick evaluations usually needed by end-users or extensive forecast ensembles performed by climate centers. 

During the last few years, with the rise of Artificial Intelligence (AI), Deep Learning (DL) techniques have progressively been adopted for weather applications, leading to a revolution in atmospheric weather forecasting \citep{Bracco2025}.
Being hundreds of times faster than their PDE-based counterparts, data-driven emulators provide a powerful framework for generating large ensembles, rapidly exploring scenarios, establishing benchmarks, and enabling downstream applications for stakeholders, policymakers, and researchers. There are downsides as well. DL models' higher inference speed often comes at the cost of reduced physical interpretability \citep{Bauer2015}, as they are usually considered as a black box that learns hidden non-linear patterns within the data in high-dimensional spaces. For systems with inherently short memory, such as the atmosphere, this limitation is often of secondary importance because forecast skill rapidly decays and predictive performance may outweigh the need for detailed physical interpretability. In this landscape, DL was first employed for skillful weather forecasting, with many remarkable works such as FourCastNet \citep{pathak2022}, GraphCast \citep{lam2023}, Pangu Weather \citep{bi2023}, NeuralGCM \citep{kochkov2024}, ExtremeCast \citep{xu2024}, FuXi-ENS \citep{zhong2024}, Aurora \citep{bodnar2025}, FGN \citep{alet2025} and GenCast \citep{price2025}. 
Emulators or generative data-driven approaches for probabilistic forecasting (e.g. DLWP \citep{weyn2021}, ACE \citep{watt-meyer2023}, ClimaX \citep{nguyen2023}, Fuxi-S2S \citep{chen2024}, ACE2 \citep{watt-meyer2025,kent2025}, CodensNet \citep{wang2025}) have opened new avenues not just in weather science, but also in weather services, sometimes even outperforming well-established numerical models.

Conversely, the development of data-driven ocean models has been slower, mostly because of the different and significant challenges that the intrinsic complexity of ocean dynamics poses. Continents and islands define the boundaries of ocean basins, shaping complex fluid-dynamical interactions along coastlines; unlike the atmosphere, which is continuously monitored by ground stations, weather balloons, and satellites, the ocean remains largely invisible to remote sensing, with extensive areas in the deep ocean and polar regions where observations have historically been sparse \citep{amirian2026compilation}; ocean dynamics emerge from the coupled effects of winds, tides, heat exchange, and freshwater fluxes, making it difficult for AI models to represent all relevant processes with physical realism; lastly and most importantly, ocean dynamics evolve on much longer timescales than the atmosphere, posing a challenge for AI models that are typically optimized to learn from short-term feedback.

Recent work has addressed these challenges by focusing on global medium-range forecasting at relatively coarse resolution. The DL emulation task has been framed as a global forecasting engine in XiHe \citep{wang2024}, OceanNet \citep{chattopadhyay2024,lowe2025}, GLONET \citep{aouni2025}, FuXi-Ocean \citep{huang2025} and TianHai \citep{niu2025}. Oppositely, in MedFormer \citep{epicoco2025} and SeaCast \citep{holmberg2025}, the authors develop very high-resolution ocean emulators tailored to the Mediterranean region for medium-term operational forecasting, achieving comparable or even better skills than the underlying numerical model, MedFS \citep{pinardi2003,pinardi2010,coppini2023}. Aforementioned emulators are not meant to predict ocean state at S2S timescale as they are not explicitly trained to guarantee stability over long rollouts, thereby accumulating large errors on such timescales. 

Data-driven oceanic emulation on decadal timescales has been investigated in different impactful works. In ORCA-DL \citep{guo2025}, the authors leveraged ORAS5 \citep{oras52021}, GODAS \citep{behringer1998} and other datasets at coarse temporal resolution (monthly scale) to achieve global ocean predictions from seasonal to decadal timescales, obtaining good skills. They propose a Transformer architecture \citep{vaswani2023}  for their fusion module. 
ORCA-DL on some metrics can outperform state-of-the-art numerical models, as it is capable of accurately simulating the structure of events of climatic relevance, including the El Ni\~no Southern Oscillation (ENSO) \citep{wang2017} and the development of upper-ocean heat-waves. 
In \cite{subel2024}, on the other hand, the authors proposed a simple and powerful U-Net \citep{ronneberger2015} network to investigate the role of the atmosphere in DL-based ocean emulation,  achieving stable roll-outs over 8 years. They trained the U-Net model leveraging GFDL CM2.6 \citep{stouffer2006} coupled climate model, focusing on a set of regions where the ocean dynamics play key but different roles in climate prediction, namely the Tropics, the Gulf Stream, the African Cape and the South Pacific. 
Building on the aforementioned work, the Samudra emulator, based on ConvNeXt \citep{liu2022} UNet model, was able to deliver stable ocean simulations for several centuries \citep{dheeshjith2025}. The model was trained to emulate key ocean state variables, i.e. potential temperature, salinity, zonal (u) and meridional (v) currents and sea surface height (SSH), leveraging the OM4 \citep{adcroft2019} dataset. The ocean variables, originally at $0.25^{\circ}$, were remapped to a coarser horizontal resolution of $1.0^{\circ}$, aggregating the data in depth levels and in 5-day averages. 
Despite the relevant results they have achieved, aforementioned works share some limitations. ORCA-DL predicts global ocean at monthly timescale, whereas in Samudra the authors preprocess the data with Gaussian filter and average the ocean data over 5 days, thereby smoothing out high-frequency details. In addition, the two global ocean emulators work at a coarse spatial resolution of $1^\circ \times 1^\circ$.

To bridge the gap between weather and climate scales, here we propose Neptune, a high spatio-temporal resolution AI ocean model that combines Convolutional Neural Networks (CNNs) and Spherical Fourier Neural Operators (SFNO) \citep{bonev2023} for simulating both the sea-ice component and the global ocean on S2S at high spatio-temporal resolution. Leveraging prescribed atmospheric forcing, Neptune can stably forecast the daily global ocean state up to 60 days, thus providing a powerful AI solution for S2S applications. 
We pre-train the Neptune model using the ORAS5 ocean reanalysis and ERA5 \citep{hersbach2020} as atmospheric forcing, both at $1^\circ \times 1^\circ$ horizontal resolution, hereafter named Neptune-1. In addition to the key variables describing the ocean dynamics (i.e., temperature, salinity, zonal (u) and meridional (v) currents, SSH, mixed layer depth (MLD)), Neptune also includes ice-related state variables (i.e., sea-ice concentration, sea-ice thickness). After the pre-training phase of Neptune-1, we perform a fine-tuning of the DL emulator at higher spatial resolution of $0.25^\circ \times 0.25^\circ$ to further enhance Neptune's resolution, also improving the range of possible downstream applications. We refer to the fine-tuned version as Neptune-025.
Our model provides physically consistent realizations of the ocean dynamics and of sea ice coverage and thickness at daily frequency up to 60 days at a maximum horizontal resolution of $0.25^\circ \times 0.25^\circ$, and can accurately portray ocean indices and modes of variability such as ENSO (and Isotherm depth at $20^\circ C$, Z20 metric) or the Indian Ocean Dipole (IOD). 
The spatial and temporal scales resolved by Neptune make it an ideal global ocean component of a S2S forecasting system.

The remainder of the paper is organized as follows: Section \ref{sect:materials_and_methods} describes the Neptune workflow starting from data preparation, to the model architecture and training details. Section \ref{sect:results} provides a detailed assessment of Neptune skills, while Section \ref{sect:discussion} contextualizes our findings, highlighting key differences with respect to the current literature. Finally, Section \ref{sect:conclusion} draws the most relevant conclusions regarding the work, pointing out potential future activities. 

\section{Materials and Methods}
\label{sect:materials_and_methods}
In this section, we describe in detail the datasets used, their preparation, and the materials used for replicating the experiments.

    \subsection{Data Sources: ORAS5 and ERA5}
    \label{sect:data_sources}

    Here we provide the dataset specifications used for training and evaluating Neptune model. Table \ref{tab:data} summarizes the key oceanic and atmospheric variables described in the following.

\begin{table}[hb]
        \centering
        \begin{tabular}{@{}lllccc@{}}
        \toprule
        \textbf{System}   & \textbf{Variable} & \textbf{Description}    & \textbf{Post-processed Shape}         & \textbf{Depth Levels}    & \textbf{Category}             \\ \midrule
        \multirow{8}{*}{Ocean}      & $\theta_o$        & Temperature             & \multirow{4}{*}{$ 181 \times 360 $} & \multirow{4}{*}{14 Levels} & \multirow{4}{*}{Input/Output} \\
                                    & $S_o$             & Salinity                &                                     &                                  & \\
                                    & $u_o$             & Eastward Velocity       &                                     &                                  & \\
                                    & $v_o$             & Northward Velocity      &                                     &                                  & \\
        \cmidrule(lr){2-6}
                                    & SSH               & Sea Surface Height      & \multirow{4}{*}{$ 181 \times 360 $} & \multirow{4}{*}{Surface} & \multirow{4}{*}{Input/Output}\\
                                    & SIT               & Sea Ice Thickness       &                                     &                          & \\
                                    & SIC              & Sea Ice Concentration   &                                     &                          & \\
                                    & MLD               & Mixed Layer depth   &                                     &                          & \\
        \hline
        \multirow{5}{*}{Atmosphere} & T2M               & 2m Temperature          & \multirow{5}{*}{$ 181 \times 360 $} & \multirow{5}{*}{Surface} & \multirow{5}{*}{Input} \\
                                    & D2M               & 2m Dewpoint             &                                     &                          &  \\
                                    & U10               & 10m u component of wind &                                     &                          &  \\
                                    & V10               & 10m v component of wind &                                     &                          &  \\
                                    & MSLP              & Mean Sea Level Pressure &                                     &                          &  \\
        \hline
        \end{tabular}
        \caption{
        Description of input and output variables used for Neptune Ocean emulator. Variables are categorized as input/output (i.e., prognostic) and input-only (i.e., forcing).
        }
        \label{tab:data}
        \end{table}

        \subsubsection{ORAS5 Oceanic Dataset}
        \label{sect:data_oras5}
        The ORAS5 Ocean Dataset \citep{zuo2019} is an ensemble of reanalysis and real-time analysis of both global ocean and sea ice. Among the key innovations of the dataset, we account for the perturbation of initial conditions and the generic perturbation scheme for both observations and forcing fields. The quality of the data is ensured by the verification against reference climate datasets from the European Space Agency Climate Change Initiative (ESA CCI) project. 
        We retrieve ORAS5 data from the Global Ocean Ensemble Physics Reanalysis \citep{gounou2024} (GLOBAL\_MULTIYEAR\_PHY\_ENS\_001\_031), comprising ORAS5, GLORYS \citep{glorys2018} and C-GLORS \citep{cipollone2021} experiments. The dataset is produced by the Copernicus Marine Service and contains several key ocean state variables, including sea ice. The ocean datasets are forced by ERA-Interim \citep{dee2011} atmospheric forcing, and contain daily data from 1 January 1993 to 31 December 2022, at a native resolution of $0.25^{\circ} \times 0.25^{\circ}$. 
        
        Covering the period 1993-2021, for our experiments we select \textit{temperature} ($\theta_o$), \textit{salinity} ($S_o$), \textit{eastward} and \textit{northward velocities} (\textit{U} and \textit{V}, respectively) as 3D variables, followed by \textit{ocean mixed layer thickness}, \textit{sea ice thickness}, \textit{sea ice concentration} as surface variables.
        Additionally, among the 75 depth levels prioritizing resolution near the ocean surface (from surface to 5,000 meters of depth) comprising ORAS5 product, we select a subset of 14 depth levels ($0.5$ m, $1.5$ m, $2.7$ m, $8.1$ m, $26.5$ m, $53.8$ m, $108.0$ m, $199.8$ m, $300.9$ m, $508.6$ m, $1045.8$ m, $1945.3$ m, $2955.6$ m, $5089.5$ m).

        \subsubsection{ERA5 Atmospheric Dataset}
        \label{sect:data_era5}
        The ERA5 reanalysis dataset \citep{hersbach2023a,hersbach2023b} includes key variables essential for driving and constraining ocean dynamics. ERA5 reanalysis combines a global numerical model with an innovative ensemble-based data assimilation technique to produce a consistent estimate of the atmospheric state \citep{hersbach2020}.
        ERA5 climate variables are provided on a regular grid at a spatial resolution of $0.25^{\circ} \times 0.25^{\circ}$, representing the global atmospheric state and covering a period that starts from 1st January 1940 to present on an hourly basis. 
        
        Among the ERA5 variables, we selected \textit{2m temperature}, \textit{2m dewpoint}, \textit{10m u and v components of wind} and \textit{mean sea level pressure} as surface forcing variables. This selection aims to learn a meaningful indirect (latent) representation of the wind stress, variables driving the heat, moisture and momentum exchange between the atmosphere and the ocean. 
        We gather the atmospheric forcing at a daily temporal resolution from the WeatherBench2 dataset \citep{rasp2023}, and select the same temporal extent (i.e., 1993 - 2021) to ensure temporal consistency with ORAS5 dataset.

    \subsection{Experimental Setup}
    \label{sect:exp_setup}

    Let $O_t$ and $A_t$ respectively be the real ocean and atmospheric state at time $t$. We can represent the dynamical evolution of the ocean state with a discrete-time underlying function $\Phi$ such that $O_{t+\Delta{t}} = \Phi (O_t, A_t)$. We can thereby obtain a trajectory of the future ocean states by auto-regressively applying the $\Phi$ function:

    \begin{align}
        O_{t+\Delta{t}} &= \Phi (O_t, A_t) \\
        \nonumber O_{t+2\Delta{t}} &= \Phi(\Phi(O_t, A_t), A_{t+\Delta{t}}) \\
        \nonumber O_{t+3\Delta{t}} &= \Phi(\Phi(\Phi(O_t, A_t), A_{t+\Delta{t}}), A_{t+2\Delta{t}}) \\
        \nonumber \cdots
    \end{align}
    
    Our goal is to find a set of learnable parameters $\theta$ and a DL model $\mathcal{F}_{\theta}$ that accurately approximates the true dynamical function $\Phi$ over a S2S temporal horizon, $T\Delta{t}$. Since we only have a partial view on the real ocean state $O_t$ caused by dimensionality reduction and truncation \citep{brolly2026}, we account for the unresolved dynamics by adding temporal information to the overall system, thereby including the previous state $O_{t-\Delta{t}}$.

    Therefore, we can write the learning problem formulation as:

    \begin{equation}        
    \left( \Delta O_{t+\Delta{t}}, \Delta O_{t+2\Delta{t}} \right) \simeq \left( \Delta\hat{O}_{t+\Delta{t}}, \Delta\hat{O}_{t+2\Delta{t}} \right) = \mathcal{F}_{\theta}(A_{t-\Delta{t}}, A_{t}, O_{t-\Delta{t}}, O_{t})
    \end{equation}

    Where $\Delta O_{t+\Delta{t}} = O_{t+\Delta{t}} - O_{t}$ is the residual of the ocean state, computed as the difference between two consecutive ocean states, while $\Delta \hat{O}_{t+\Delta{t}}$ is the prediction of the DL model. 
    Following \cite{brolly2026} and taking advantage of results obtained in \cite{epicoco2025}, where we used 4 timesteps in input to the emulator, we require additional ocean states by predicting two consecutive residuals to effectively reconstruct the steady ocean dynamics.
    In contrast with other works, the innovation of our experimental setup lies in the prediction of two consecutive timesteps, as in \cite{dheeshjith2025}, with the exception that we predict residuals instead of full ocean fields, leading to a better forecasting skills.
    
    \subsubsection{Data processing}
    \label{sect:data_process}

    Before feeding the data into the Deep Learning (DL) model, we pre-processed them according to the following procedure. Both atmosphere and ocean data are provided on the same regular grid at a daily $0.25^{\circ} \times 0.25^{\circ}$ resolution, thus corresponding to a $1440 \times 721$ ($H \times W$) matrix. Land pixels in ocean data are represented with NaN values. Moreover, since ocean data are not defined on the Antarctic, their corresponding grid is $1440 \times 681$. Therefore, as a preliminary step, we fill the missing ocean latitudes with NaNs to match the atmospheric grid size. 
    We split the 1993-2021 dataset into training (1993-2016), validation (2017-2018) and test (2019-2021).
    Then, due to the large amount of data, we computed the global mean and standard deviation on the training set leveraging the Welford algorithm \citep{efanov2021}. The Welford algorithm enables the computation of both the running mean and standard deviation, iterating over each dataset sample, with high numerical stability. Before feeding the Neptune emulator with ocean and atmospheric predictors, we scale them and replace NaN values (i.e., in the land and Antarctic) with zeros.
    Furthermore, we interpolated ocean and atmosphere data at a coarser resolution of $1^{\circ} \times 1^{\circ}$ grid using the bilinear interpolation algorithm, resulting in a $181 \times 360$ ($H \times W$) data matrix. 
    The resulting dataset at $1^\circ$ horizontal resolution was used for pre-training (Neptune-1) while the dataset at native $0.25^\circ$ resolution was used for fine-tuning (Neptune-025).

    \subsubsection{Deep Learning Architecture}
    \label{sect:architecture}

    We designed the Neptune Ocean architecture to process global geophysical data, adopting a hierarchical structure that integrates neural operators on the $S^2$ sphere and convolutions. The spherical formulation of the network eliminates geometrical distortions introduced by grid projections, enabling rotational invariance. As introduced in Section \ref{sect:intro}, (S)FNOs have proven their effectiveness in Climate Science both in atmosphere and ocean emulation due to their ability to learn operators that map between function spaces, thereby being able to learn complex PDE families.

    In our work, mixing SFNO to convolutions enables capturing the strong multi-scale behavior of the ocean dynamics, thus modeling long-range spatial correlations while maintaining efficiency across the scales. Moreover, Neptune is able to reproduce response time to atmospheric forcing, which varies a lot.

    \noindent \textbf{Ocean Neural Operator}
    \label{sect:arch_overview}

\begin{figure}
        \centering
        \includegraphics[width=\textwidth]{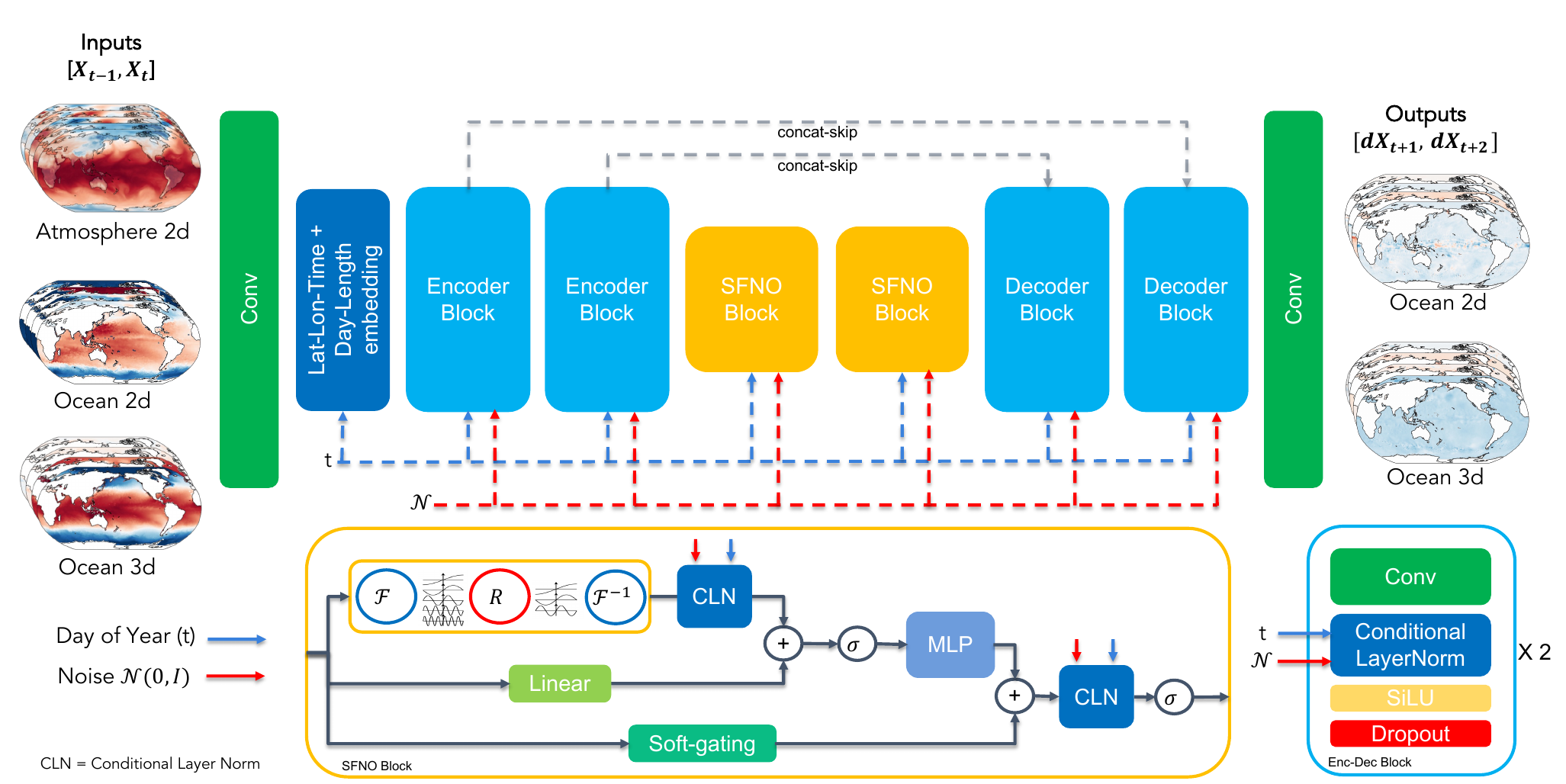}
        \caption{
        Neptune architecture is a Unet-shaped SFNO operating on dimensions  $H \times W$. Ocean data along with atmospheric forcing at time-steps $t-1$ and $t$ are projected into a 256-dimensional latent space using convolutions. Encoder and decoder blocks capture local non-linear relationships within the latent data, whereas the SFNO blocks capture global relationships. Noise and day-of-year are injected into the conditional layer norms of the network to drive during the residual state forecasting. A convolution-based recovery restores the data into the original physical space, thus producing ocean forecasts at times $t+1$ and $t+2$.
        }
        \label{fig:arch}
    \end{figure}

    We designed the Neptune model as an encoder-decoder network, Figure \ref{fig:arch}. The input data consists of both 3D and 2D ocean data together with the 2D surface atmosphere forcing. We project each variable into a common latent space of dimension 256 through convolutions with kernel size 3, padding 1 and stride 1.
    We end up with a single tensor of shape $B \times D \times H \times W = B \times 256 \times 181 \times 360$. 
    Moreover, inspired by the Fourier-based encoding used in the Aurora Foundation model \citep{bodnar2025}, we used a positional encoding that encodes the data in latitude, longitude and day-of-year. This module provides the network with absolute spatio-temporal context, thus learning specific geographical biases. Using the day-of-year $t$, we further enriched the positional encoding by computing and encode the day-length map (i.e., containing the number of hours of sunlight) in the latent space. As a result, this operation increased the conservation of the Ocean Heat Content (OHC) across the forecasting horizon. 

    The latent tensor is first processed by two encoder blocks that hierarchically capture multi-scale local patterns within the data. Then, two SFNO blocks compute global spectral convolution, truncating the modes at $H \times W = 2 \times 2$ to keep only the largest scale frequency modes. The SFNO is linked to the decoder which reconstructs the spatial fields leveraging both the global SFNOs context and the local encoder information provided through the skip connection.
    
    Finally, the latent data is projected back into the original space, reconstructing the residuals $dX_{t+1}$ and $dX_{t+2}$. 
    We inject day-of-year $t$ and noise $\mathcal{N}(0,I)$ into the Conditional Layer Norms (CLN) \citep{chen2021} to drive the model during the residuals' reconstruction. The day-of-year enhances the DL model representation of the seasonality of the ocean state, driving its response depending on the seasonal cycle. The noise conditioning enables the DL model to produce ensemble forecasts, improving its long-term accuracy, given the chaotic nature of the ocean \citep{li2024}. In Section \ref{sect:inference} we explain the procedure to construct the ensemble.

    We dimensioned the network configuration to work with $1^\circ$ data, resulting in a 13M parameters model. To adapt the architecture for the fine-tuning at higher resolution data (i.e., $0.25^\circ$), after the first and before the last convolutions (green blocks in Figure \ref{fig:arch}), we added two $S^2$ bilinear resample operations that map $0.25^\circ$ (i.e., $721 \times 1440$ grid size) to $1^\circ$ (i.e., $181 \times 360$ grid size). This operation helped Neptune to keep the same latent space representation size at the working resolution of the pre-training phase, thus limiting the number of epochs needed for fine-tuning.
    
    \noindent \textbf{Encoder-Decoder Block}
    \label{sect:encdec_block}
    
    Both the Encoder and Decoder blocks share the same architecture, inspired by the UNet layers. Each block contains two sequential layers composed of convolution, CLN, Sigmoid Linear Unit (SiLU) and Dropout. The convolution layer extracts local features within the latent tensor, whereas the CLN applies a stochastic layer normalization to the features based on both the day-of-year and the noise conditioning. 
    
    \noindent \textbf{SFNO Block}
    \label{sect:sfno_block}

    The architecture of the SFNO block is based on the NeuralOperator \citep{kovachki2024} implementation. As shown in Figure \ref{fig:arch} (yellow panel), the latent tensor is fed into the spectral convolution layer that computes global spectral convolution leveraging SHT transform. After the spherical convolution, the input is normalized through CLN and added to densely connected skip connection. 
    Then, Gaussian Linear Unit (GeLU) non-linearity is applied, followed by a Multi-Layer Perceptron (MLP) and added to a Soft-Gating skip connection. Finally, CLN and GeLU non-linearity are sequentially applied. 
    We condition the SFNO outputs to drive the forecasting leveraging the stochastic layer normalization.
    
    \noindent \textbf{Pre-training and Fine-tuning}
    \label{sect:train}
    
    We pre-train Neptune-1 ocean emulator for 150 epochs using a global batch size of 8 samples, using $1^\circ$ data. We select AdamW as weight optimizer, with $0.01$ of weight decay, $\beta_1 = 0.9$ and $\beta_2 = 0.95$. We first warm up the learning rate to $5 * 10^{-4}$ for the first 10 epochs, and then apply a cosine annealing schedule that decreases it up to $5 * 10^{-8}$. We leverage the \textit{Mean Absolute Error} (MAE) loss function, to optimize the model for simply forecasting the global ocean residuals $dX_{t+1}, \space dX_{t+2}$.
    After the pre-training phase, we fine-tune the model (Neptune-025) on higher spatial resolution data at $0.25^\circ$ for 50 epochs, using the same configuration of the pre-train.
    We conduct all our experiments on the Juno HPC Cluster, a supercomputer belonging to the CMCC (Euro-Mediterranean Center on Climate Change) Foundation's High Performance Computing (HPC) facilities \citep{cmcc_hpcc}. We used a total of 8 NVidia Volta A100 GPUs, 40GB, for a total of 48 hours for pre-training and $\sim$100 hours for fine-tuning.

    \noindent \textbf{Inference on Test set}
    \label{sect:inference}
    
    For our evaluations and results reported in Section \ref{sect:results}, we perform inference on 2019-2021 test set. 
    We use each day of the test set as the initial condition of a 60-day long forecast. Using the Neptune-1 emulator, we build an ensemble of forecasts composed of 8 members. Starting from the same initial condition, we obtain each member by sampling a different noise tensor drawn from a Gaussian normal distribution $\mathcal{N}(0, I)$ and use it for latent space conditioning.
    Due to computational limitations during both inference and evaluation across the 3 years of data, we could not perform the ensemble for higher spatial resolution emulator, Neptune-025.
    
    \subsubsection{Evaluation Metrics}
    \label{sect:eval_metrics}

    Here we provide a brief definition and explanation of the metrics that will be used to evaluate Neptune-1 and Neptune-025 performance in the following Section \ref{sect:results}.

    \noindent \textbf{Statistical Evaluation}
    \begin{itemize}
        \item Root Mean Squared Error (RMSE): measures the magnitude of the mean quadratic deterministic error, severely penalizing huge discrepancies in the point-wise values. After after some time, due to the chaoticity of the flow, the RMSE is less accurate in determining the skills of a DL emulator.
        \item Continuous Ranked Probability Score (CRPS): extends the analysis to the probabilistic domain, evaluating the accuracy of the predicted ocean-state trajectories.
        \item Anomaly Correlation Coefficient (ACC): quantifies the phase correspondence and the morphological faithfulness of spatial patterns between the predicted and true anomalies, with values near to 1 indicating perfect structural synchrony.
    \end{itemize}

    \noindent \textbf{Physical Coherency}
    \begin{itemize}
        \item Ocean Heat Content (OHC): To evaluate Neptune's ability to emulate the fundamental dynamics of the Tropical Pacific Ocean, we examine the evolution of the Ocean Heat Content (OHC). 
        We compute OHC using the following formula: 
        
        $$ H = \rho C \int^{z}_{0} T(z) dz $$

        where T is the temperature, $\rho = 1026$ $kg/m^3$ is the density of sea water, $C = 3990$ $J / kg K$ is the specific heat of sea water and $ z = 300$ $m$.
        Unlike surface temperature, OHC integrated over the first $300$ $m$ provides a more robust representation of the upper-ocean thermal structure and additionally highlights charge-discharge processes associated with ENSO variability. 
        
        \item Eddy Kinetic Energy (EKE): We compute the Eddy Kinetic Energy (EKE) to evaluate Neptune's dynamical variability. Analyzing the EKE is crucial for assessing model's ability to resolve large-scale transient flow and the variability of the major current systems. In Figure \ref{fig:eke}, we can identify systematic errors in Neptune's representation of the ocean energy distribution. We compute EKE as:
        
        $$EKE = \frac{1}{2} (u'^{2} + v'^{2})$$
        
        where $u' = u - \bar u$ and $v' = v - \bar v$ are the anomalies in zonal and meridional velocities, respectively, while $\bar u$ and $\bar v$ are their daily mean values. 
        
        \item Ice Brier Score (IBS): To assess Neptune's skills in reproducing ice fields, we compute the Brier Score (BS) \citep{brier1950} on the Sea Ice Concentration (SIC) field. The score measures the accuracy of probabilistic prediction in binary outcomes. In our work, we define a $15\%$ threshold to retrieve the binarized reanalysis observation ($o_t$) and then compute the score as:
        
        $$BS_k = \frac{1}{N} \sum_{t=0}^{N} (f_t^k - o_t)^{2}$$
        
        where N is the number of days within the 2019-2021 test set, $f_t^k$ is the forecast SIC at the $k$-th lead time, treated as a probability (i.e., not binarized).
    \end{itemize}

    \noindent \textbf{Oceanic Indices}
    \begin{itemize}
        \item El Ni\~no Southern Oscillation (ENSO): is a large-scale phenomenon occurring in the tropical Pacific Ocean region and oscillating every few years \citep{wang2017}. In the ocean, ENSO is characterized by positive (El Ni\~no) and negative (La Ni\~na) Sea Surface Temperature (SST) anomalies. Although with irregular occurrence, it shows an oscillatory behavior that generally lasts 3-5 years. Moreover, the phenomenon is irregular in the occurrence rate and is asymmetric, showing larger and often longer El Ni\~no warm events with respect to La Ni\~na cold events. 

        Included in our ENSO analysis we also include the isotherm depth at $20^{\circ}C$ (Z20), since it gives insights regarding the movement of water masses. Z20 is the depth at which the temperature reaches $20^{\circ}C$. It is widely considered the primary physical proxy to identify the thermal variability in the Equatorial Pacific \citep{kessler1990}. It acts as a frontier, separating warm surface waters from the deeper, colder waters of the deep ocean.
        We computed Z20 over a Eastern ($5^{\circ}S - 5^{\circ}N$, $100^{\circ}W - 90^{\circ}W$) and Western ($5^{\circ}S - 5^{\circ}N$, $165^{\circ}E - 175^{\circ}E$) Boxes surrounding El Ni\~no 3.4 region. During El Ni\~no phases, since hot waters moved eastward, the Western box depth decreases accompanied by sharp increases in the Eastern. During La Ni\~na phases, the opposite happens. The Western box depth increases while Eastern ones will show decreasing trends.
        
        \item Indian Ocean Dipole (IOD): is the main inter-annual variability mode of the Tropical Indian Ocean (TIO), tightly linked with the atmosphere-ocean coupling \citep{saji1999,webster1999,yamagata2013}. It is characterized by a strong zonal thermal gradient that can be numerically quantified by the Dipole Mode Index (DMI) that expresses the difference between SST anomalies of the Western ($50^{\circ}E–70^{\circ}E$, $10^{\circ}S–10^{\circ}N$) and Eastern ($90^{\circ}E–110^{\circ}E$, $10^{\circ}S–0^{\circ}S$) basins. 
        During a positive IOD phase (pIOD), equatorial trade winds intensify towards West, causing an anomalous SST cooling near the Sumatra and Java regions. At the same time, thermocline depth increases in the Arabian Sea, causing a warming in the Western pole \citep{liu2024}.
        
    \end{itemize}

\section{Results}
\label{sect:results}

In this section, we evaluate Neptune-1's skills based on various metrics. 
As mentioned in Section \ref{sect:inference}, we use each day of the 2019-2021 test set as the initial condition of a 60-day long forecast. Subsequently, we perform the benchmarks using the i-th forecast of each day to provide the skills for such lead time. 

We divide our results into: Statistical Evaluation, Physical Coherency and Oceanic Indices. Within Statistical Evaluation \ref{sect:stat_eval}, we discuss RMSE, CRPS and ACC. In Physical Coherency \ref{sect:phys_coh}, we provide skills about OHC, EKE and IBS.

In Oceanic Indices Section \ref{sect:oce_idxs}, we analyze how Neptune reproduces the occurrence and intensity of ENSO and Z20 metric related to it and IOD across the 60 days of forecast.

Finally, in Section \ref{sect:eval_on_025} we analyze Neptune-025's skills in terms of RMSE, OHC, EKE and IBS to highlight benefits and limitations of the fine-tuning on downstream tasks for S2S applications.

All evaluation metrics involving the climatology are computed relative to a daily climatology derived from the training set (1993-2016), to prevent any data leakage from the test period and providing a consistent baseline for evaluating the emulator skills.

    \subsection{Statistical Evaluation}
    \label{sect:stat_eval}

        \subsubsection{Root Mean Squared Error}
        \label{sect:rmse}

\begin{figure}[t!]
            \centering
            \begin{subfigure}{\textwidth}
                \centering
                \caption{RMSE Difference between Neptune and Climatology}
                \includegraphics[width=\textwidth]{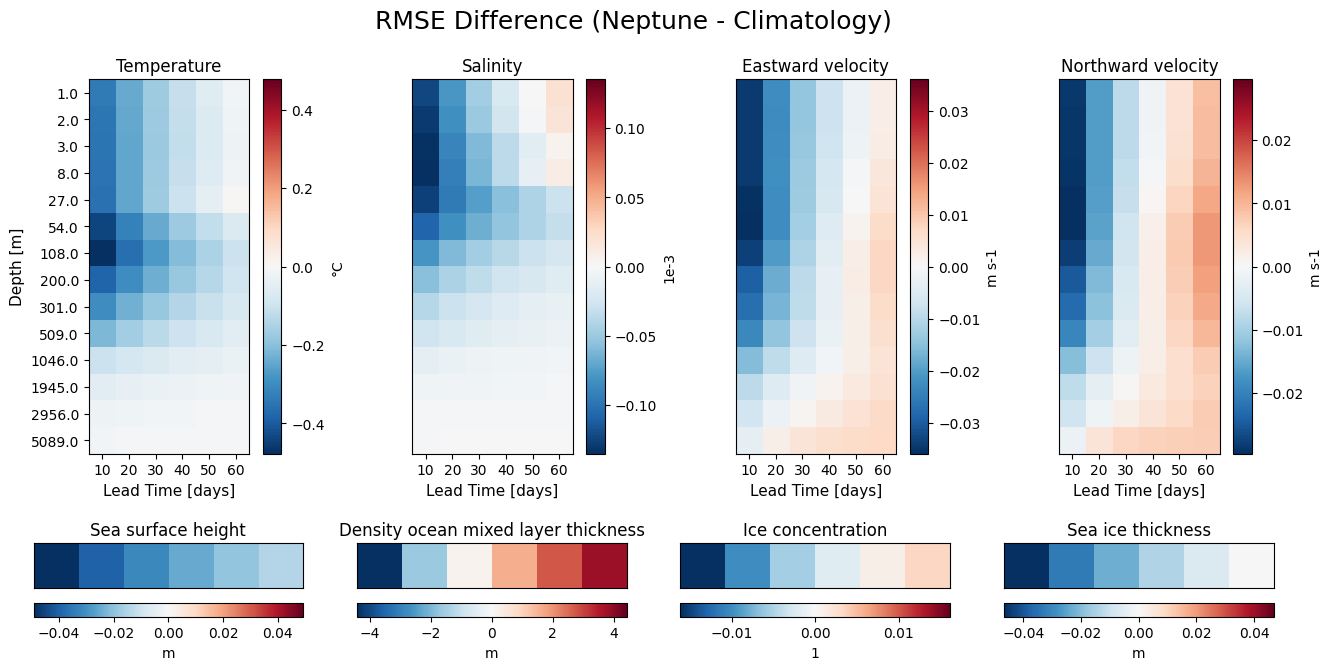}
                \label{fig:rmse_diff}
            \end{subfigure}
            \hfill
            \begin{subfigure}{\textwidth}
                \centering
                \caption{Neptune RMSE at t+60 lead time}
                \includegraphics[width=\textwidth]{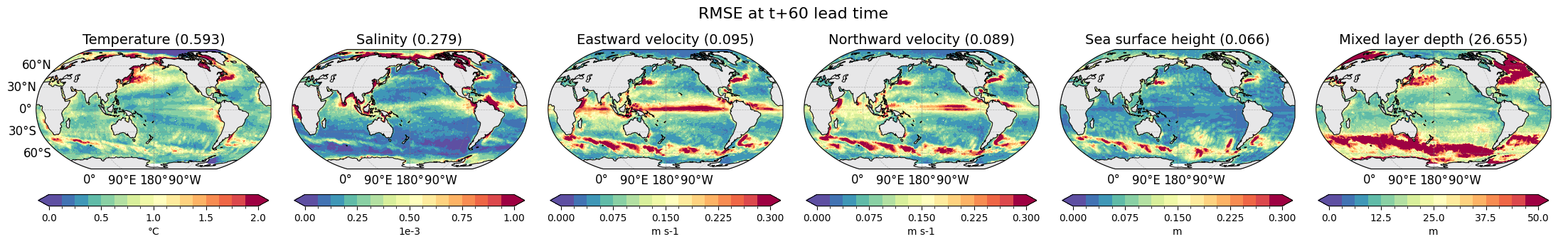}
                \label{fig:rmse_lt}
            \end{subfigure}
            \caption{
            Neptune's RMSE for some selected lead times from t+10 to t+60 days.
            Panel a) reports RMSE difference between Neptune and climatology, comprising both surface and depth variables. 
            Blue colors, associated with negative values, mean that Neptune has a lower RMSE than climatology. Red colors, associated with positive values, mean the opposite.
            Panel b) contains spatial Neptune's RMSE at t+60 days for a subset of the predicted variables.
            }
            \label{fig:rmse}
        \end{figure}

        Neptune's RMSE difference with respect to the climatology (Figure \ref{fig:rmse_diff}) is characterized by a regular asymptotic decay of the quadratic error, documenting the robustness of the emulator with respect to the climatology in representing the seasonal cycle.
        This behavior is a direct consequence of the structural topology of the DL model (i.e., Encoder-SFNO-Decoder) that combines multiple convolutional blocks with an SFNO bottleneck to capture complex global and local structure, successfully combining them to obtain a robust predicted ocean state. Moreover, predicting the residual regularizes the error accumulation during iterative long rollouts, limiting the RMSE at t+60 days. 
        Except for $u_o$ and $v_o$ velocities, MLD and SIT, Neptune has much lower RMSE with respect to the climatology 
        Figure \ref{fig:rmse_diff}. While approaching the 60 days of forecast, the scorecards transition from blue to more neutral colors confirming that the emulator slowly converges towards the climatology without drifting with respect to it. 
        Deterministic RMSE validation with respect to the climatology confirms that Neptune shows improved skills in emulating the ocean dynamics. Finally, temperature scores reveal that Neptune has a reduced RMSE with respect to the climatology, in correspondence of the $50$-$300 \space m$ depth, which is a well-known bias of OGCMs.
        From a spatial analysis, Neptune RMSE shows exceptional consistency with the predicted variables, revealing that Neptune has learned a highly faithful representation of the physics underlying global ocean dynamics.
        Due to their intrinsically chaotic nature \citep{lorenz1969}, regions of highest mesoscale variability across the globe like the Gulf Stream, Kuroshio region, Agulhas, Brazil-Malvinas Confluence, and Antarctic Circumpolar Currents, present great modeling challenges for data-driven emulators. 
        Neptune coherently models all the ocean state variables across the lead times, Figure \ref{fig:rmse_lt}, showing low RMSE for each of the selected variables.
        
        A degradation in Neptune's predictive skill occurs for MLD as well as horizontal current components (u,v) at longer lead times approaching 60 days. We largely attribute the MLD skill drop to the coarse upper-ocean vertical resolution, which restricts the emulator's ability to resolve steep density gradients near the surface. As a consequence, in regions of large S2S MLD variability (e.g., Labrador Sea or the Antarctic Circumpolar region), Neptune's error inflates, thereby increasing the overall RMSE (Figure \ref{fig:rmse_lt}). We hypothesize that further enhancing the vertical grid resolution in the upper ocean could improve the representation of MLD dynamics.
        Moreover, regarding the velocity fields, we observe a higher prediction error with respect to the climatology beyond 40-50 days. Due to the high variability of these fields, Neptune struggles more to follow the exact velocity patterns, causing greater RMSE values. As emerges in Figure \ref{fig:rmse_lt}, the emulator's surface RMSE is higher in regions of high kinetic variability (e.g., the Equator, the Kuroshio currents, the Gulf Stream, etc.). In Figure \ref{fig:rmse_diff}, we observe higher RMSE earlier in time. While upper ocean layers are constrained by the prescribed atmospheric wind stress, deeper ocean dynamics relies more on the internal physics. As a consequence, deeper levels lack direct forcing to limit the model drifts.
        
        \subsubsection{Continuous Ranked Probability Score}
        \label{sect:crps}

        While RMSE penalizes deterministic quadratic discrepancies between ground truth and forecasts, CRPS evaluates the global probabilistic accuracy of the emulator, which is critical for chaotic systems after some time, as the trajectories naturally diverge so that RMSE remains irrelevant.

\begin{figure}[t!]
            \centering
            \begin{subfigure}{\textwidth}
                \centering
                \caption{CRPS Difference between Neptune and Climatology}
                \includegraphics[width=\textwidth]{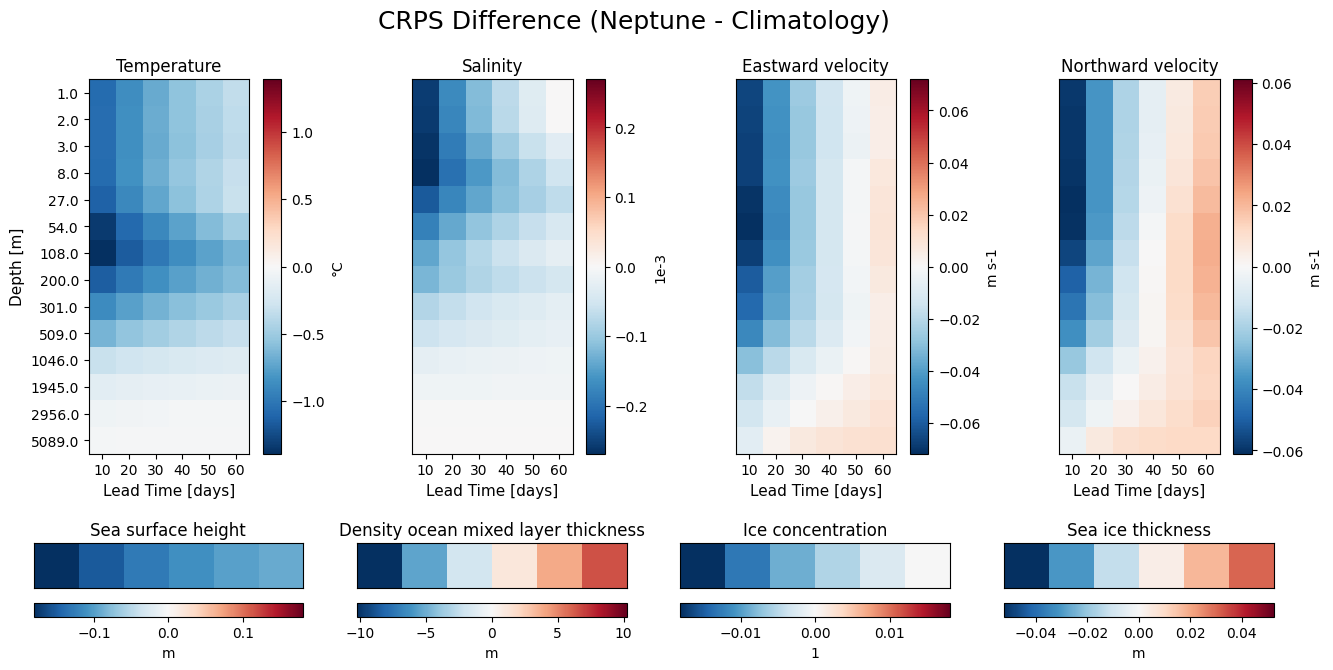}
                \label{fig:crps_diff}
            \end{subfigure}
            \hfill
            \begin{subfigure}{\textwidth}
                \centering
                \caption{Neptune CRPS at t+60 lead time}
                \includegraphics[width=\textwidth]{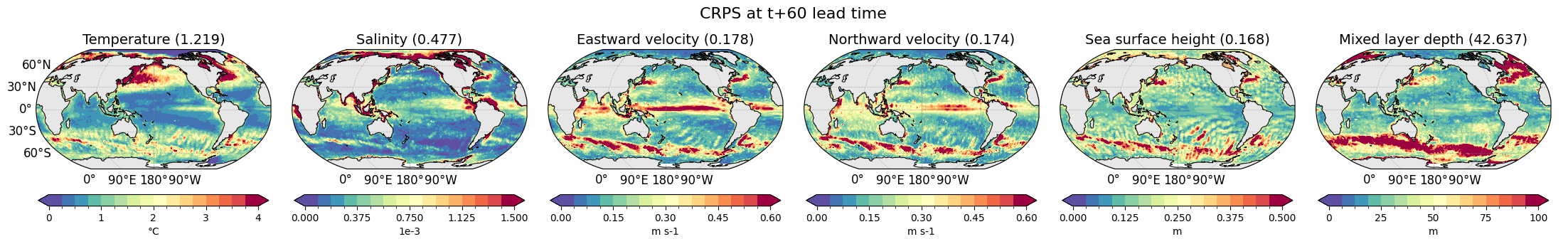}
                \label{fig:crps_lt}
            \end{subfigure}
            \caption{
            Neptune's CRPS for selected lead times from t+10 to t+60 days.
            Panel a) reports the CRPS difference between Neptune and climatology for each of the predicted variables. Blue colors, associated with negative values, mean that Neptune has a lower CRPS than climatology. Red colors, associated with positive values, mean the opposite.
            Panel b) contains spatial Neptune's CRPS at t+60 days for a subset of the predicted variables.
            }
            \label{fig:crps}
        \end{figure}

        The CRPS scorecard (Figure \ref{fig:crps_diff}) shows Neptune's advantage with respect to the climatology for most of the predicted variables up to 30 days, followed by a slight decrease when approaching the 60 days limit.
        We attribute this temporal behavior to the structural stability of the network, that preserves the geometrical coherence of latent spaces, reducing the spread of the error. In the meantime, noise injection gives more forecasting robustness, improving its skills at t+60 days.
        Figure \ref{fig:crps_diff} confirms our findings about RMSE, showing similar scorecard patterns with respect to the climatology. Except for some variables at lead times above 40 days, most Neptune's CRPS skills are better than the climatology.
        Further analyzing spatial the CRPS distribution after 60 days of forecast, it emerges that Neptune's skill is not distributed homogeneously across the globe.
        Neptune's forecasts have higher CRPS in highly chaotic areas (e.g., Western Boundary Currents, Antarctic Circumpolar Current, North Atlantic Subpolar Gyre, etc.), reflecting the inherent behavior of these ocean regions \citep{germe2022,larson2024,sohail2025}.
        In Figure \ref{fig:crps_lt}, we report the spatial CRPS score after 60 days of forecast for a subset of thermodynamic ocean variables. In agreement with the analysis shown in Section \ref{sect:rmse}, Neptune skill drops in such regions, with both $\theta_o$ and $S_o$ showing higher error in the Northern Hemisphere. In contrast, other variables reveal similar spatial patterns to the ones observed in Figure \ref{fig:rmse_lt}. 
        We attribute the spatial error to the continent distribution, as the Northern Hemisphere is dominated by land, creating small basins with jagged coastlines. Moreover, Neptune-1 spatial resolution makes it structurally incapable of resolving eddies that typically characterize high energetic regions like Kuroshio currents (North Pacific) and Gulf Stream (North Atlantic).
        
        \subsubsection{Anomaly Correlation Coefficient}
        \label{sect:acc}

        Anomaly Correlation Coefficient (ACC) is a dimensionless metric defined between $-1$ and $1$, quantifying the spatial correlation between the forecasted anomalies and the ground truth ones. A score of $1$ indicates perfect morphological and phase correspondence between spatial patterns. 
        At the S2S timescale, this metric helps us determine whether Neptune is preserving consistent correlation skills after 60 days.
        
        Neptune preserves a skillful ACC for each ocean field across lead times, highlighting a performance that faithfully reflects the intrinsic memory of each oceanic variable. 
        
\begin{figure}[t!]
            \centering
            \begin{subfigure}{\textwidth}
                \centering
                \caption{Neptune Anomaly Correlation Coefficient Scorecard}
                \includegraphics[width=\textwidth]{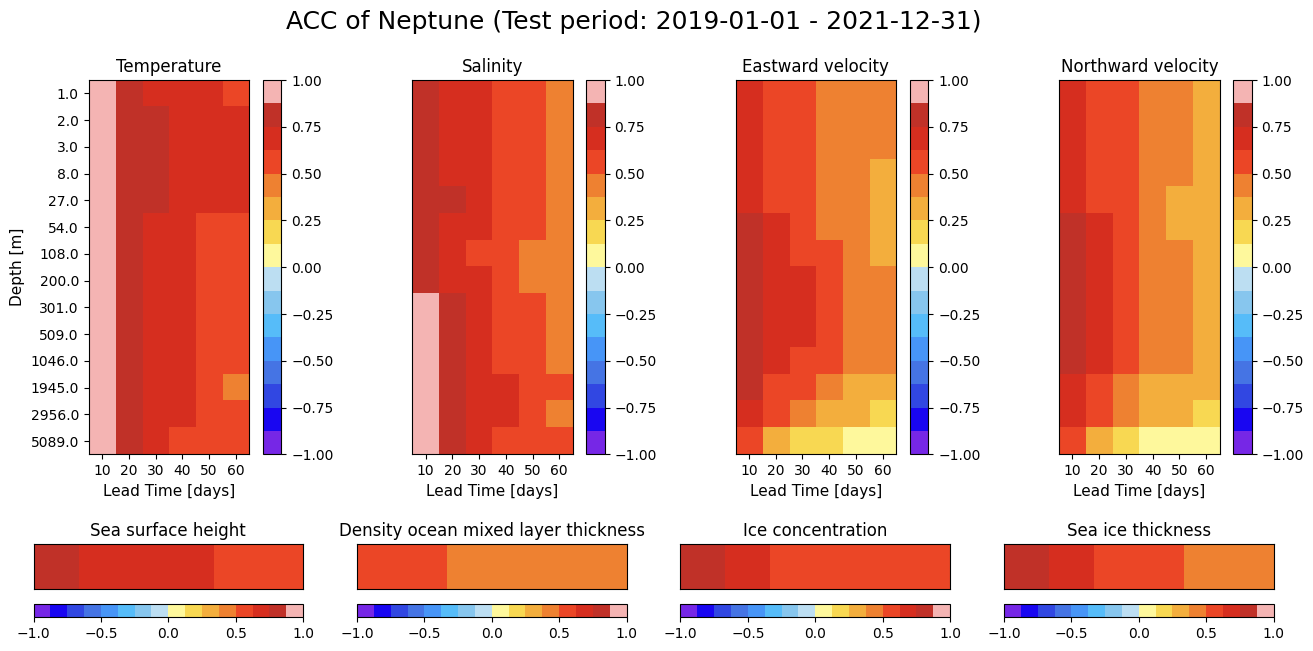}
                \label{fig:acc_score}
            \end{subfigure}
            \hfill
            \begin{subfigure}{\textwidth}
                \centering
                \caption{Neptune ACC at t+60 lead time}
                \includegraphics[width=\textwidth]{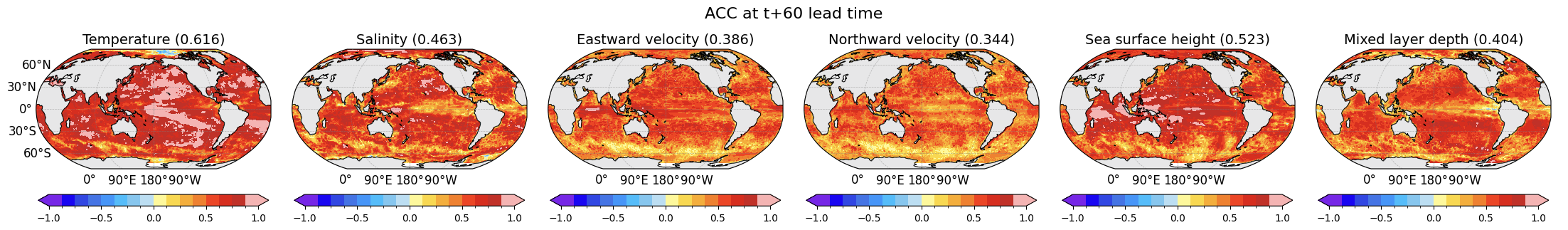}
                \label{fig:acc_lt}
            \end{subfigure}
            \caption{
            Neptune's ACC for selected lead times from t+10 to t+60 days.
            Panel a) reports, for each ocean variable, the ACC scorecard of Neptune computed using climatology.
            Panel b) contains spatial Neptune's ACC at t+60 days for a subset of the predicted variables.
            }
            \label{fig:acc}
        \end{figure}

        As shown in Figure \ref{fig:acc}, $\theta_o$ is the variable that provides the highest ACC skills across every lead time and depth level. Up to 30 days, the ACC is very high with values above $0.75$ across all depths. After 60 days, the skills are lower, but preserving ACC above $0.5$ and diminishing with depth. 
        Spatial ACC patterns (Figure \ref{fig:acc_lt}) reveal that Neptune's skills surrounding the Pacific ENSO region are very high after 60 days, highlighting its excellent performance in capturing large-scale patterns typically associated with atmosphere-ocean coupling. 
        $S_o$, instead, shows a decrease in the skills after 30 days concentrated around the $100$ $m$ layer where the gradients are more difficult to emulate. After 60 days (Figure \ref{fig:acc_lt}), Neptune shows a noisy pattern of lower ACC along the coastline and in the higher latitudes. 
        Neptune's currents provide a rapid decrease in the ACC skills. 
        Regarding surface variables, both SSH and SIC provide the highest correlation, while delivering lower ACC skills in determining the phase of MLD and SIT. More in detail, linked to the ENSO and IOD dynamics, spatial ACC reveals excellent SSH skills in the Indian Ocean and the Equatorial Pacific, with values above $0.9$ after 60 days. However, our model struggles in the Southern Ocean at higher latitudes, where the ACC sensibly decreases, likely due to the presence of the Antarctic Circumpolar Current.
        
    \subsection{Physical Coherency}
    \label{sect:phys_coh}

        \subsubsection{Ocean Heat Content}
        \label{sect:ohc}

\begin{figure}[t!]
            \centering
            \begin{subfigure}{\textwidth}
                \centering
                \caption{Spatial Ocean Heat Content}
                \includegraphics[width=\textwidth]{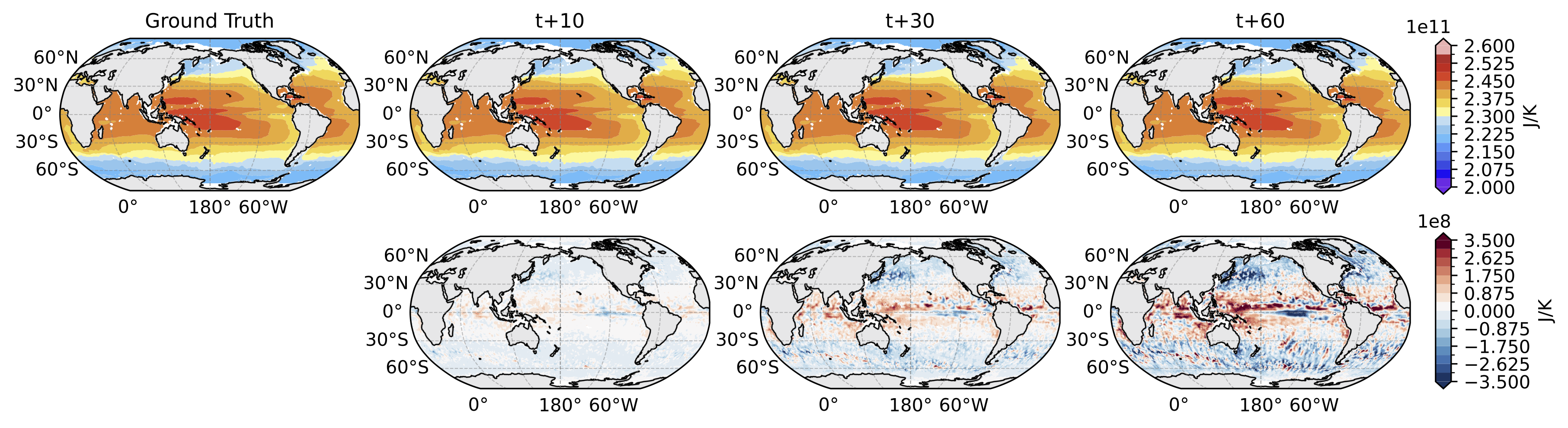}
                \label{fig:avg_ohc}
            \end{subfigure}
            \hfill
            \begin{subfigure}{\textwidth}
                \centering
                \caption{Temporal Ocean Heat Content}
                \includegraphics[width=\textwidth]{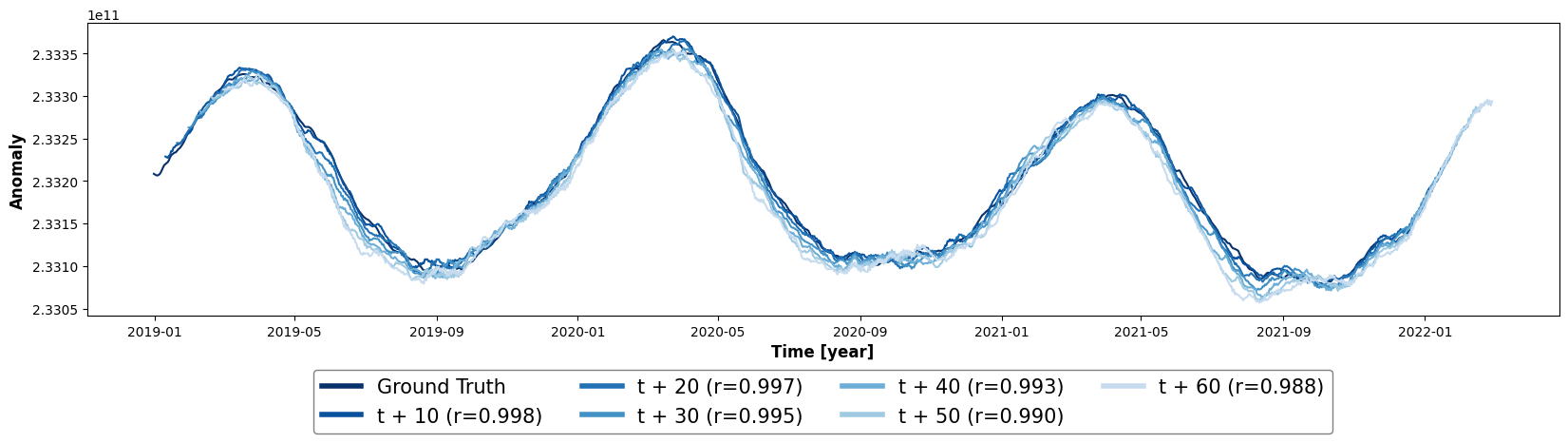}
                \label{fig:time_ohc}
            \end{subfigure}
            \hfill
            \begin{subfigure}{\textwidth}
                \centering
                \caption{Ocean Heat Content anomaly Hovmoller on Pacific Equatorial region}
                \includegraphics[width=\textwidth]{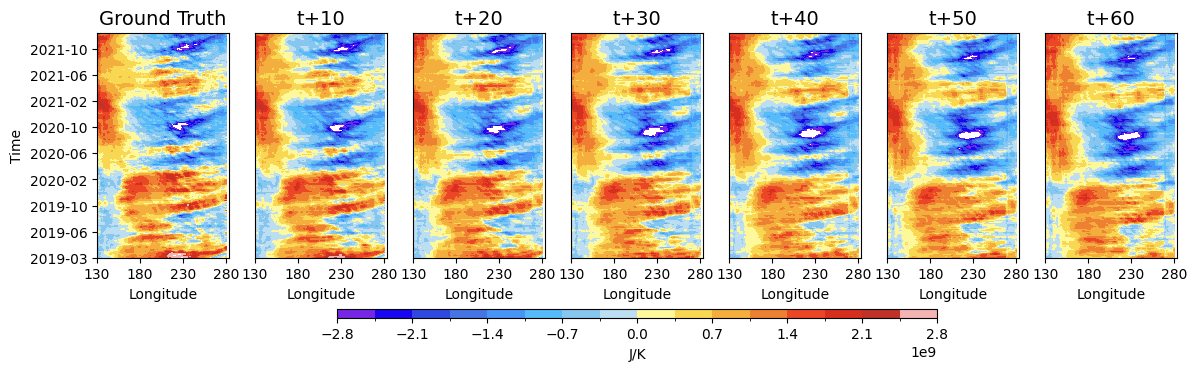}
                \label{fig:hov_ohc}
            \end{subfigure}
            \caption{
            Ocean Heat Content (OHC) computed over 2019-2021 Test Set. For both ground truth and forecast lead time (t+10, t+30 and t+60), panel a) top subplots show spatial OHC averaged over each day of the Test Set. Panel a) bottom subplots represent the bias between true and predicted OHC at a specific lead time. 
            Panel b) shows the temporal OHC timeseries. Reanalysis is represented in dark blue, while Neptune forecasts (from t+10 to t+60) are reported in progressively lighter blue colors. Each forecast reports the Pearson correlation between predicted and true timeseries.
            Panel c) shows the OHC anomaly Hovmoller diagram computed on the Pacific Equatorial Region (2S-2N, 130-280E). The OHC anomaly is obtained by subtracting the climatology from the OHC.
            }
            \label{fig:ohc}
        \end{figure}

        Analyzing the difference between the true and predicted OHC, Figure \ref{fig:avg_ohc}, the error reveals that Neptune struggles in regions of high variability. Slowly, Neptune progressively gains heat on the Equatorial region, mostly on the Pacific and Atlantic, while losing most heat in correspondence of Kuroshio, Gulf Stream, Agulhas and Brazil-Malvinas Confluence.
        
        The timeseries reveals that Neptune has a very high OHC Pearson correlation, Figure \ref{fig:time_ohc}, with the lowest value of $0.988$ at 60 days, meaning that the DL emulator is almost perfectly reproducing the OHC seasonality. Moreover, Neptune shows an exceptionally stable trend in the total OHC content. We attribute the stability in the introduction of day-length map into the latent space, that enhances the physical representation of the ocean temperature. 

        To further support our OHC analysis, in Figure \ref{fig:hov_ohc} we report the OHC anomaly Hovmoller diagram computed in the Pacific Equatorial Region (i.e., 2S-2N, 130-280E). The Hovmoller diagram reveals a clear pattern of eastward-propagating signals across the equatorial Pacific basin, suggesting that Neptune captures propagating OHC anomalies consistent with large-scale tropical Pacific dynamics even after 60 days, showing good autoregressive stability. As a result, the model successfully captures the coherent structure of these waves as they propagate from the western boundary ($130^{\circ} E$) towards the South American coast.
        Both the reanalysis and the DL model reveal that 2019 has been characterized by positive OHC anomalies, associated with El Ni\~no events. Conversely, 2020 and the last half of 2021 are characterized by negative OHC anomalies associated with La Ni\~na events. This is further confirmed in Section \ref{sect:enso}.
        In agreement with other panels of Figure \ref{fig:ohc}, we notice that the Hovmoller extremes shown in the ground truth panel (i.e., pink spots in early 2019, white spots at the end of 2020 and 2021) are progressively under-represented by Neptune predictions, meaning that the DL emulator is very slowly smoothing out the extreme values except for 2020 La Ni\~na event. 

        In summary, Neptune accurately reproduces the temperature fields with high consistency and stability with respect to the reanalysis up to $300$ $m$ of depth, showing interesting physical accuracy of the OHC representation across the 60 day forecasts. 
        
        \subsubsection{Eddy Kinetic Energy}
        \label{sect:eke}

\begin{figure}
        \centering
            \begin{subfigure}{\textwidth}
                \centering
                \caption{Eddy Kinetic Energy with bias}
                \includegraphics[width=\textwidth]{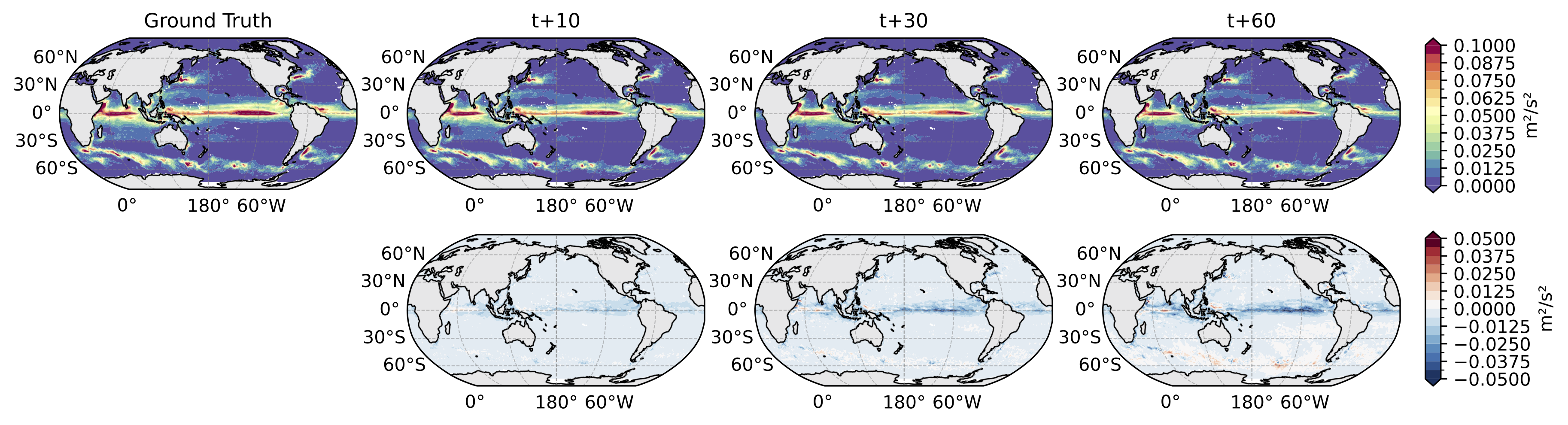}
                \label{fig:bias_eke}
            \end{subfigure}
            \hfill
            \centering
            \begin{subfigure}{0.53\textwidth}
                \centering
                \caption{Highly energetic regions at t+60 detail}
                \includegraphics[width=\textwidth]{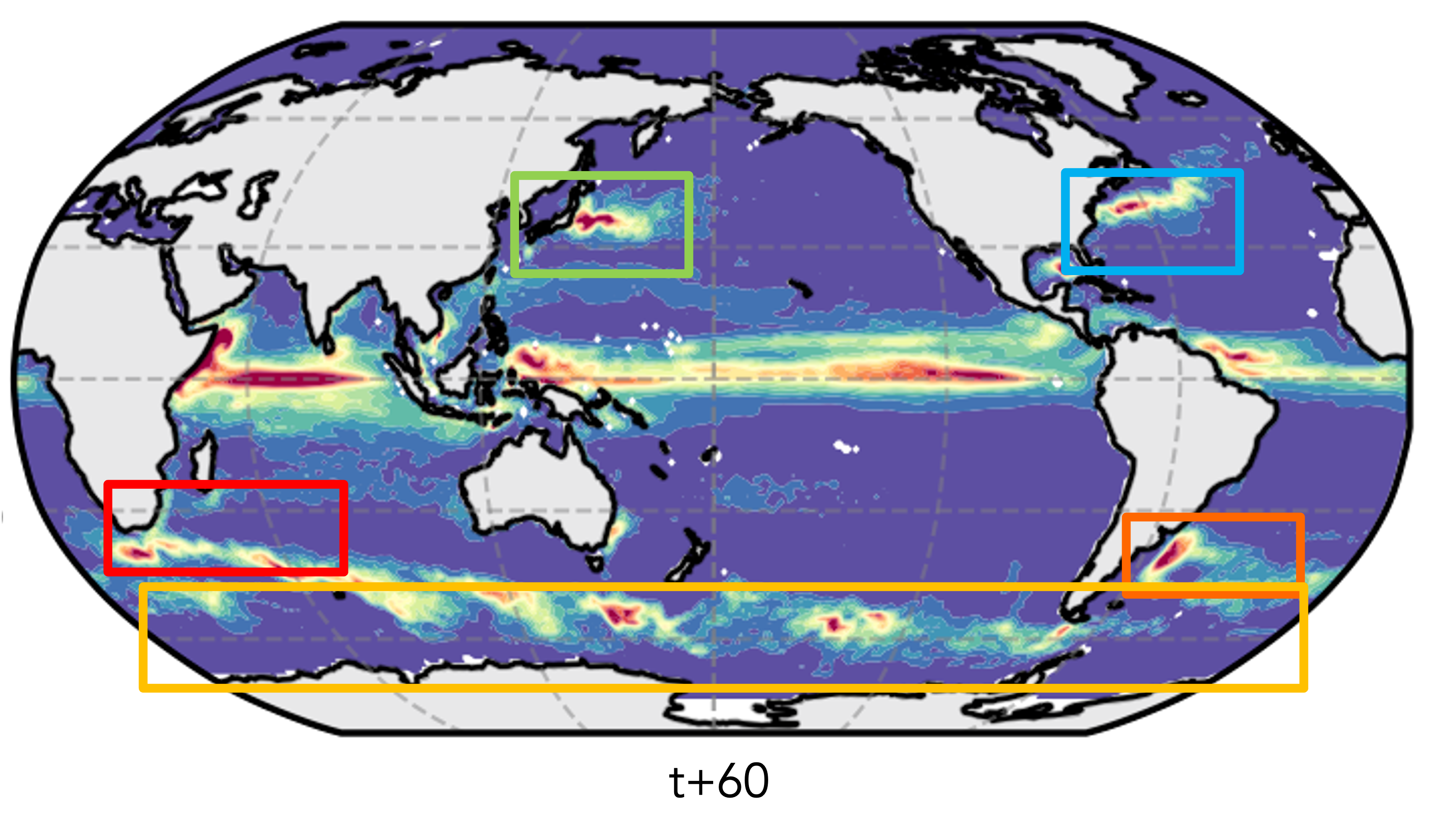}
                \label{fig:high_eke}
            \end{subfigure}
            \centering
            \begin{subfigure}{0.4\textwidth}
                \centering
                \caption{EKE Power Spectral Density by lead time}
                \includegraphics[width=\textwidth]{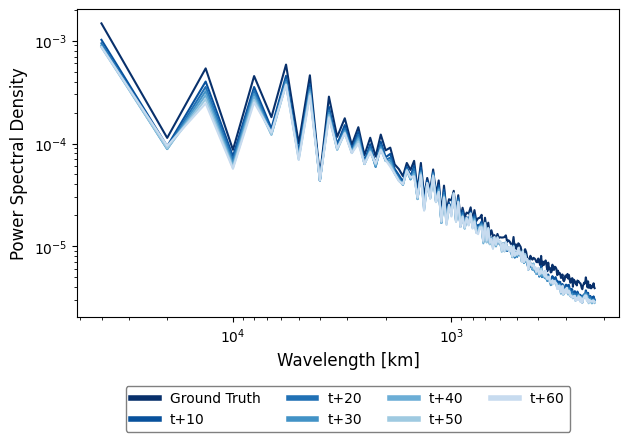}
                \label{fig:psd_eke}
            \end{subfigure}
            \caption{
            Eddy Kinetic Energy (EKE) averaged over the 2019-2021 Test Set. For ground truth and forecasts at t+10, t+30 and t+60, panel a) top row reports the EKE averaged over each day of the Test Set. Bottom row contains the bias between true and predicted EKE at the specified lead time (column). 
            Panel b) reports a detail of the forecasted EKE at t+60. High energetic regions of the globe are represented with colored boxes: Kuroshio region (green), Gulf Stream (blue), Agulhas (red), Brazil-Malvinas Confluence (orange), Antarctic Circumpolar Currents (yellow).
            Panel c) shows the Power Spectral Density (PSD) of the EKE at different wavelengths, expressed in kilometers. Ground truth is reported in dark blue, whereas forecasts are reported with progressively lighter blue colors.
            }
            \label{fig:eke}
        \end{figure}

        During the forecasting horizon, the spatial EKE distribution \ref{fig:bias_eke} remains remarkably stable. This consistency suggests that the DL model's energy is established early in the simulation and does not significantly drift over the 60-day forecast horizon. 
        
        As the bias row (second) reports in Figure \ref{fig:bias_eke}, we observe slightly negative values ($0.05$ $\frac{m^2}{s^2}$) on the Equatorial Pacific, where most EKE is localized in the ground truth. During the forecasting towards the S2S scale, Neptune tends to lose the ability to forecast the exact position of the energy details, smoothing the velocity anomaly fields $u'$ and $v'$. Since EKE is quadratic with respect to the velocities, small smoothing causes wide decreases in the EKE.
        To better analyze Neptune's ability in modeling the EKE spectra, in Figure \ref{fig:psd_eke} we report EKE's Power Spectral Density (PSD), obtained using Spherical Harmonic Transform (SHT) following \cite{lam2023}. The spectra reveal that Neptune consistently preserves the EKE spectral components at almost every wavelength. Wavelengths at $10^3$ $km$ are correctly modeled for each lead time (light blue colors), as they overlap with the ground truth (in dark blue). Lowest wavelengths (i.e., small scale details) are slightly under-represented, as we observe lower PSD in the right side of Figure \ref{fig:psd_eke}, whereas the highest wavelengths are reproduced with good fidelity, suggesting Neptune's ability to faithfully reproduce the spatial EKE patterns at the S2S timescale.
        
        Moreover, as Figure \ref{fig:high_eke} shows, after 60 days of forecast, Neptune exhibits good skills in modeling the energetic regions of the globe, like the Kuroshio region (green), the Gulf Stream (blue), the Agulhas (red), the Brazil-Malvinas Confluence (orange), and the Antarctic Circumpolar Currents (yellow).
        
        \subsubsection{Ice Brier Score}
        \label{sect:ibs}
        
\begin{figure}
        \centering
            \begin{subfigure}{0.48\textwidth}
                \centering
                \caption{North Pole Ice Brier Score}
                \includegraphics[width=\textwidth]{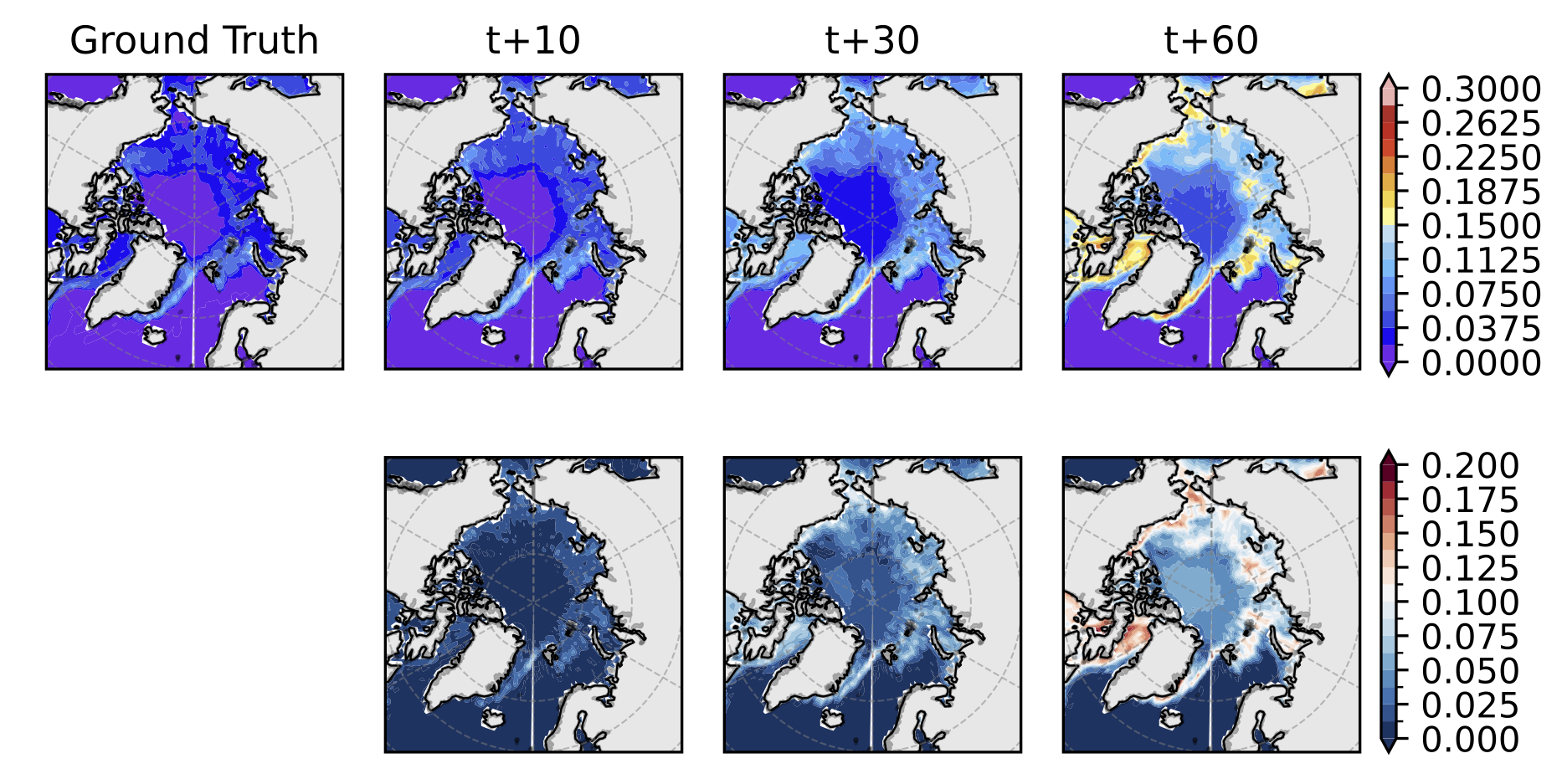}
                \label{fig:nor_ibs}
            \end{subfigure}
            \centering
            \begin{subfigure}{0.48\textwidth}
                \centering
                \caption{South Pole Ice Brier Score}
                \includegraphics[width=\textwidth]{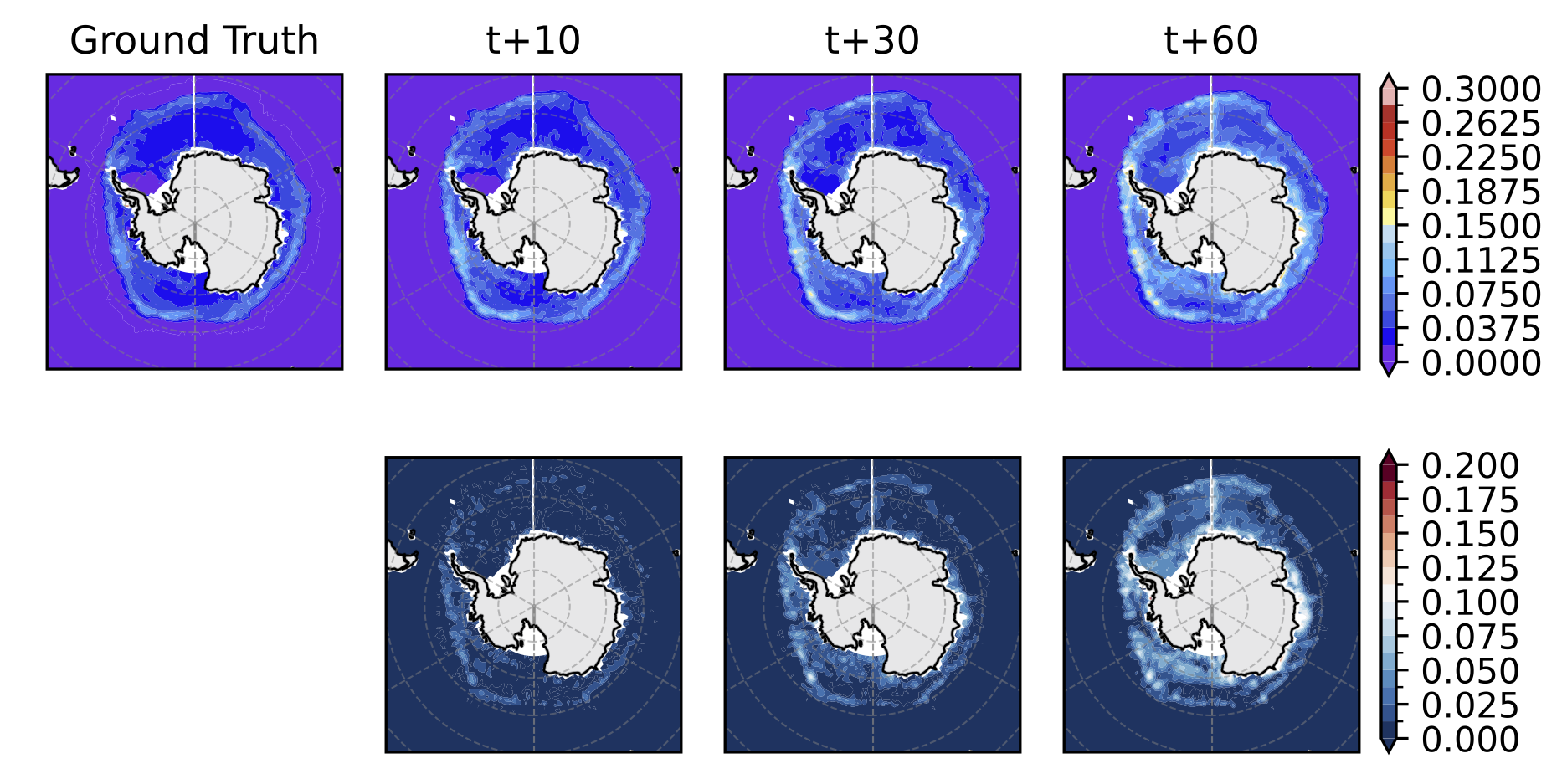}
                \label{fig:sou_ibs}
            \end{subfigure}
            \caption{
            Ice Brier Score (IBS) computed over North (panel a) and South (panel b) Poles using $15\%$ ice thresholding and averaged over the 2019-2021 Test Set. 
            Both panels a) and b) contain IBS ground truth and forecasts at t+10, t+30 and t+60 on the top row. Bottom row contains the difference between the IBS forecast specified by the column and the IBS ground truth.
            }
            \label{fig:ibs}
        \end{figure}

        The analysis of BS, Figure \ref{fig:ibs}, which defines the intrinsic error of the system, highlights that up to t+30 days, both poles are spatially faithful with respect to the analysis, providing smooth values similar to the ground truth ones. Moreover, Neptune correctly models the physical position of the ice, placing it only in physically plausible regions.
        At t+60, we observe a slight asymmetry in the hemispheric forecasting skills. In the Arctic (North), the model exhibits a slightly higher degradation after 60 days, with errors that spread from the marginal regions into the central pack. Such an error is likely caused by the jagged coastline, which does not contain information that causes sea-ice misrepresentation in Neptune's forecast. In contrast, the Antarctic (South) shows a higher resilience, with errors primarily localized to the ice-edge dynamics throughout the 60-day forecast. The persistence of low BS values in the Southern Ocean's central pack shows that the model effectively captures the large-scale seasonal stability of the Antarctic ice cover. 
        
    \subsection{Oceanic Indices}
    \label{sect:oce_idxs}
    
    In this Section we assess Neptune's skills in capturing and reproducing crucial S2S oceanic indices, namely ENSO and Z20 metric associated to it and IOD.
    
    \subsubsection{El Ni\~no Southern Oscillation}
    \label{sect:enso}

\begin{figure}
    \centering
        
        \begin{subfigure}{\textwidth}
            \centering
            \caption{El Ni\~no index timeseries in the 3.4 region}
            \includegraphics[width=\textwidth]{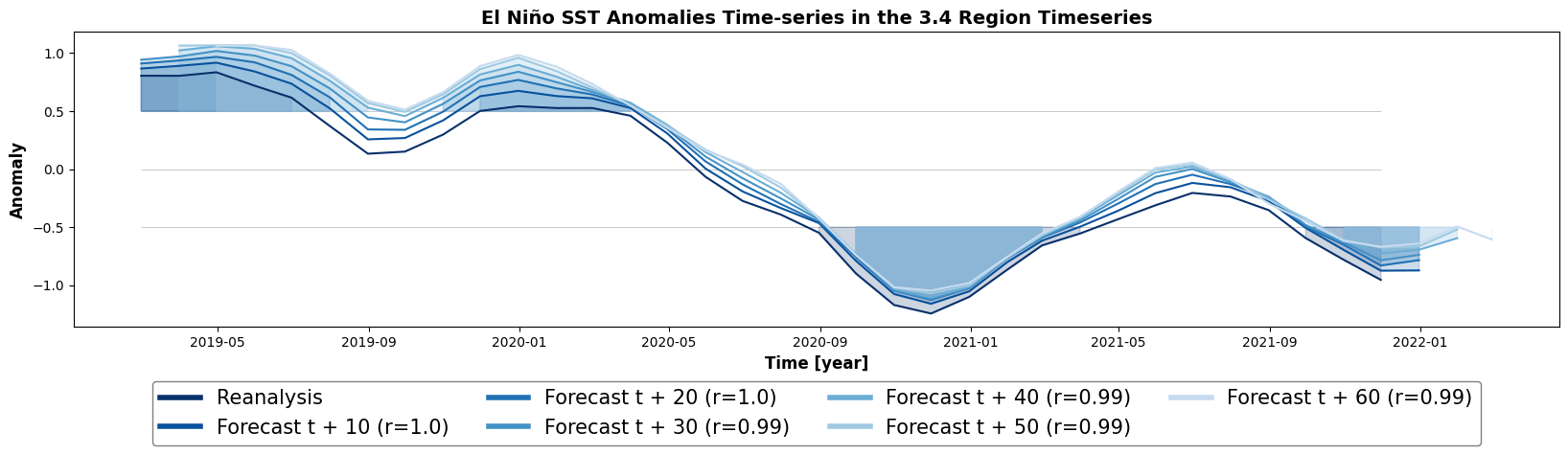}
            \label{fig:time_enso}
        \end{subfigure}

        \hfill

        \begin{subfigure}{\textwidth}
            \centering
            \caption{El Ni\~no index SST spatial composite}
            \includegraphics[width=\textwidth]{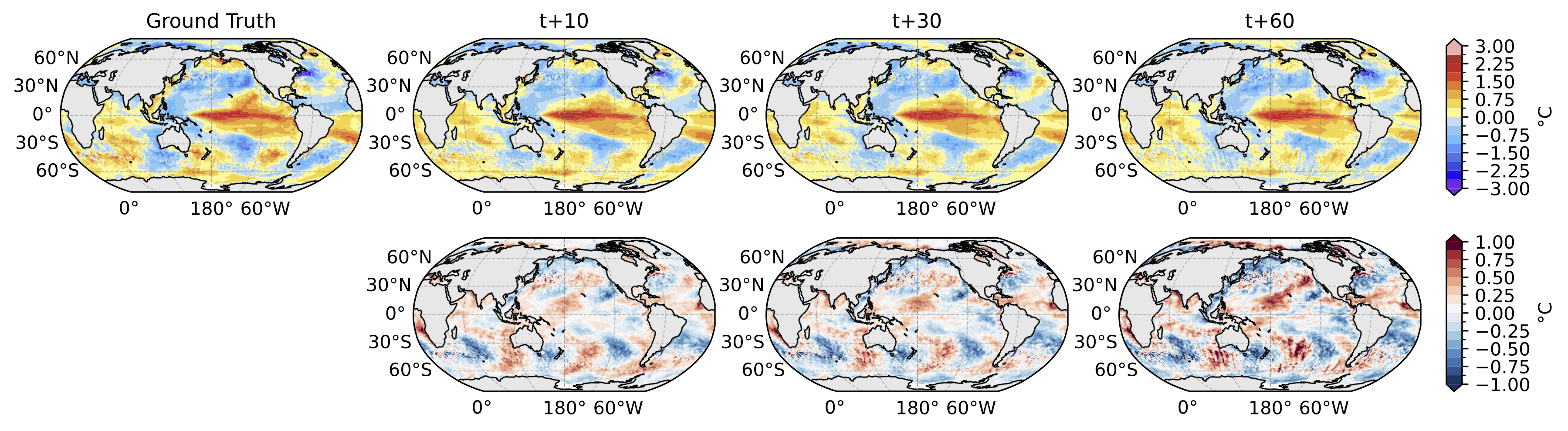}
            \label{fig:sst_enso}
        \end{subfigure}

        \caption{
        Panel a) shows the SST anomaly timeseries on the 3.4 region. Dark blue line represents the reanalysis, whereas gradually lighter lines represent Neptune's forecasts from $t+10$ to $t+60$. 
        Panel b) top row shows the ENSO spatial composite during 2019-2021 test set with respect to the lead time. The reanalysis of ENSO is shown in the first column. Panel b) bottom row depicts the SST bias to highlight potential cold or warm biases of the emulator. 
        }
        \label{fig:enso}
    \end{figure}

    ENSO is key to seasonal forecasting, and is therefore important to verify how Nepture captures its characteristics envisioning its future S2S applications. Neptune accurately reproduces the large-scale characteristics of ENSO phenomena as well as its global spatial symmetry and timing, showing an Equatorial thermic amplitude damping after 60 days of forecast.
    Coupled ENSO dynamics is based on slow ocean oscillations that our emulator effectively learned the Ni\~no-Ni\~na sequence within the test set, Figure \ref{fig:sst_enso}. The SST Spatial Composite (i.e., $SST_{pos} - SST_{neg}$) reveals that Neptune preserves the morphology of the "warm tongue" in the Equatorial Eastern Pacific across every lead time. 
    Bias maps of Figure \ref{fig:sst_enso} (bottom row) quantify the spatial regions where Neptune emulator misplaces the SST morphological structures.
    Neptune reveals a remarkable geometrical coherence of the SST spatial composite, showing that it learned the dynamics characterizing this phenomena. The test set period 2019-2021 is characterized by one El Ni\~no event (mid 2019) and two strong La Ni\~na events (during 2020 and the end of 2021). Neptune, indeed, successfully captures the phase of the event, showing slight warming of the extremes. Indeed, at t+60 days due to the compound autoregressive error accumulation, we observe a slight increase in the extreme temperature anomalies.
    
\begin{figure}
    \centering
        \begin{subfigure}{\textwidth}
            \centering
            \caption{Isotherm at $20^{\circ}C$ over Western Box ($5^{\circ}S - 5^{\circ}N$, $165^{\circ}E - 175^{\circ}E$)}
            \includegraphics[width=\textwidth]{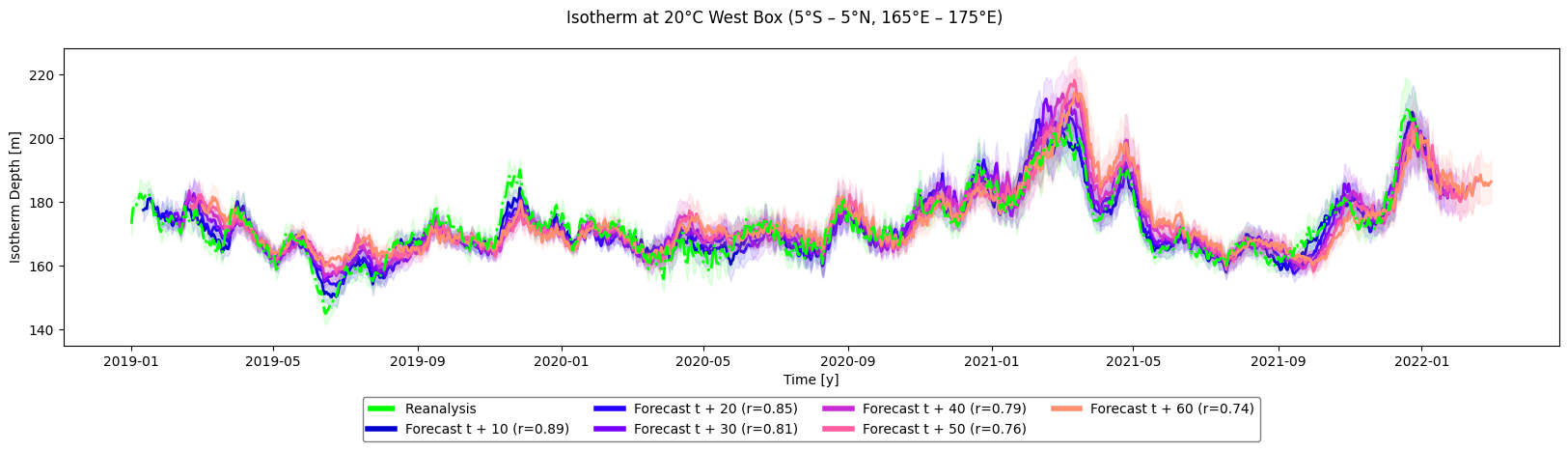}
            \label{fig:z20_west}
        \end{subfigure}
        \hfill
        \begin{subfigure}{\textwidth}
            \centering
            \caption{Isotherm at $20^{\circ}C$ over Eastern Box ($5^{\circ}S - 5^{\circ}N$, $100^{\circ}W - 90^{\circ}W$)}
            \includegraphics[width=\textwidth]{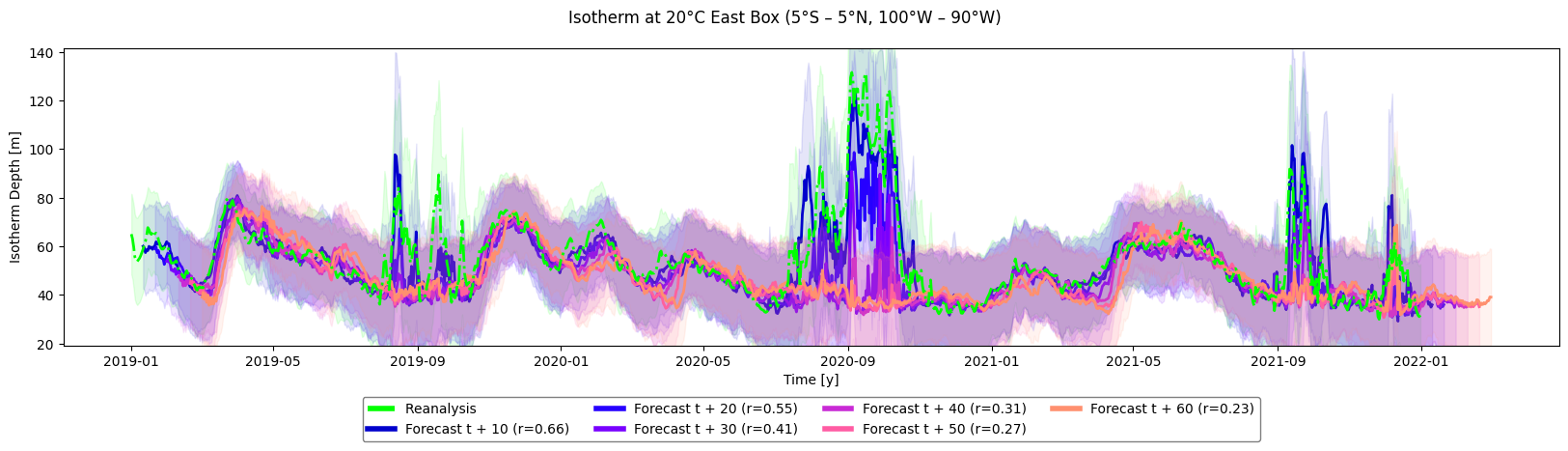}
            \label{fig:z20_east}
        \end{subfigure}
        \caption{
        Isotherm at $20^{\circ}C$ (Z20) is computed. Reanalysis is represented with green dashed line, whereas Forecasts from t+10 to t+60 are represented using colors spanning from blue to orange. Along with the Z20 timeseries, each line is surrounded by the standard deviation of the Z20 measure. The Z20 is computed over two boxes surrounding the El N\~no 3.4 region. Western box is shown in panel a), while the Eastern box is shown in panel b).
        }
        \label{fig:z20}
    \end{figure}

    \noindent \textbf{Isotherm at $20^{\circ}C$}
    \label{sect:iso} - 
    The Neptune emulator demonstrates noticeable accuracy in reproducing vertical fluctuations of thermocline depth, successfully capturing the different dynamic sensitivity that characterizes the East and West sides of the central Pacific Ocean.
    Neptune preserves the stability of the Z20 variability across the 60 days of forecast, without drifting with respect to the reanalysis' ground truth. The west box is usually deeper, with values across 150-200 meters, and varies slowly. Conversely, the Eastern box is shallower and wind-forced \citep{kessler1990}, thereby characterized by wider variability. Neptune's skills reflect the intrinsic difference between the two basins. Since the Eastern basin is less predictable due to its wind-driven nature, we observe a progressive amplitude damping when progressing towards the S2S horizon at t+60.
    Analyzing Figure \ref{fig:z20}, we have clear evidence of the difference between the two basins: the West box (Figure \ref{fig:z20_west}) is dominated by slow and deep thermal dynamics, Neptune's forecast establishes an excellent stability and coherence, with Pearson correlation coefficients varying from $0.89$ at t+10 to $0.77$ at t+60. As aforementioned above, the East box (Figure \ref{fig:z20_east}) shows a highly reactive wind forcing, and the emulator provides a much lower Pearson coefficient (from $0.63$ at t+10 to $0.34$ at t+60), despite correctly forecasting the timing of the event.
    The ability to correctly emulate the Z20 variability is a key criterion in the Pacific, and it demonstrates that Neptune has learned to accurately reproduce it on the considered test set. 
    
    \subsubsection{Indian Ocean Dipole}
    \label{sect:iod}

    During a positive IOD phase (pIOD), equatorial trade winds intensify towards West, causing an anomalous SST cooling near the Sumatra and Java regions. At the same time, thermocline depth increases in the Arabian Sea, causing a warming in the Western pole \citep{liu2024}.
    Our 2019-2021 Test Set shows an extraordinarily intense pIOD period during the second half of 2019 and a smaller pIOD in 2020.
    
\begin{figure}
    \centering
        
        \begin{subfigure}{\textwidth}
            \centering
            \caption{Indian Ocean Dipole index timeseries}
            \includegraphics[width=\textwidth]{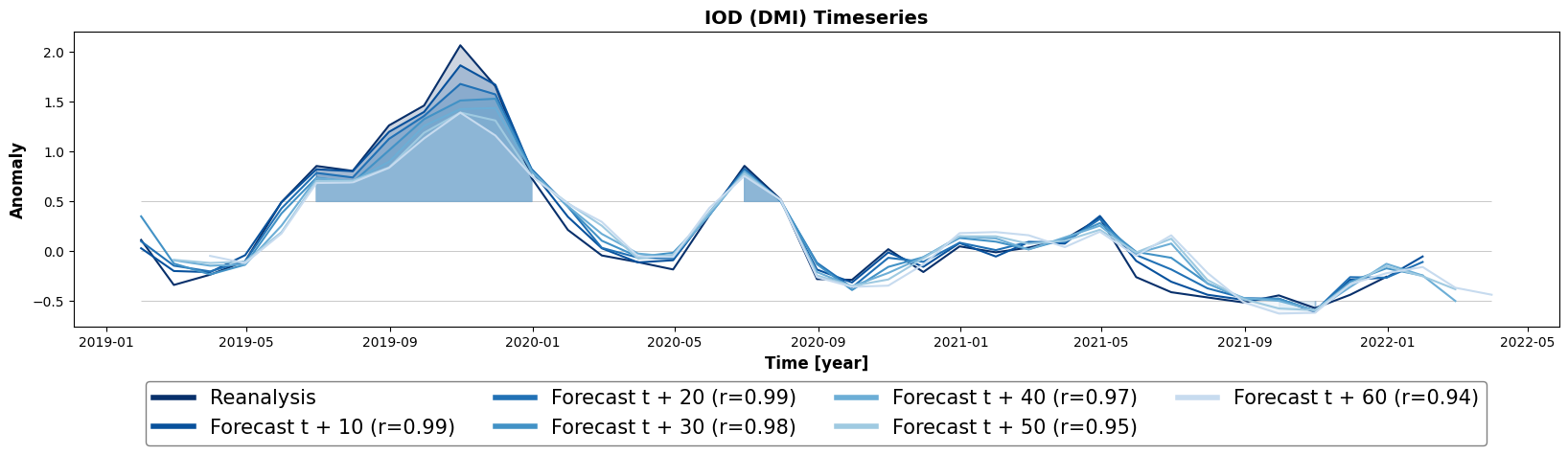}
            \label{fig:iod_time}
        \end{subfigure}

        \hfill

        \begin{subfigure}{\textwidth}
            \centering
            \caption{Indian Ocean Dipole index SST spatial composite}
            \includegraphics[width=\textwidth]{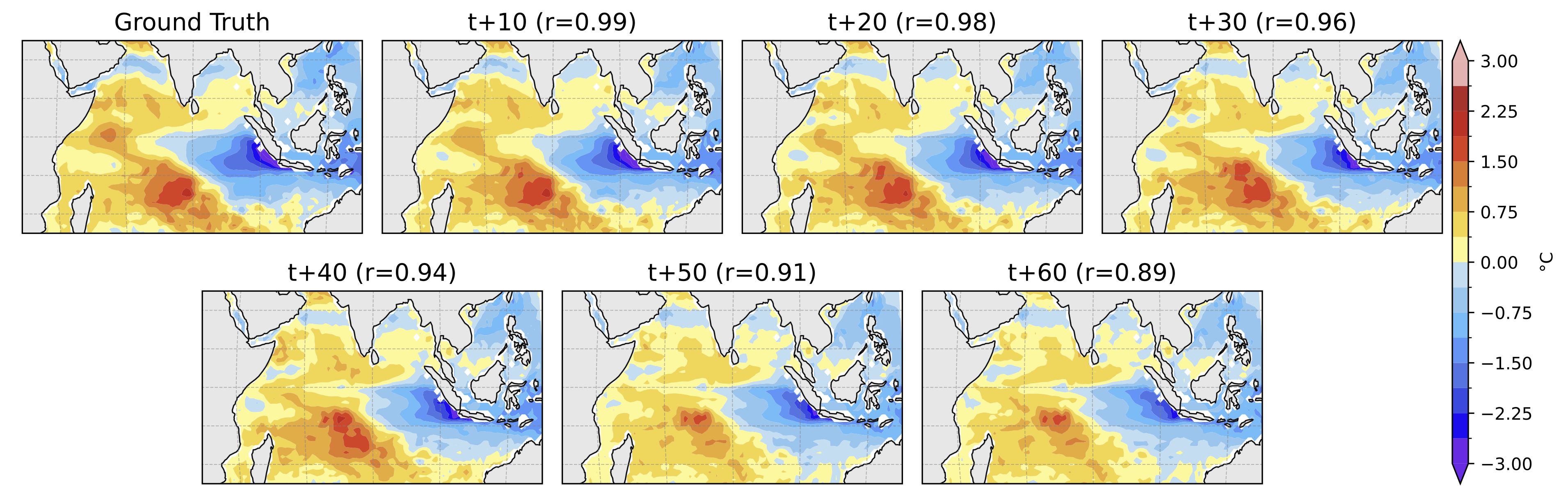}
            \label{fig:iod_sst}
        \end{subfigure}

        \caption{
        Panel a) shows the IOD index timeseries computed using the SST anomalies in the Indian Ocean region. Dark blue line represents the reanalysis, whereas gradually lighter lines represent Neptune's forecasts from $t+10$ to $t+60$. Each line is associated with the Pearson correlation computed between the forecast and ground truth timeseries.
        Panel b) reports the IOD index spatial SST composite for every lead time, spanning from t+10 to t+60. Each sub-panel is associated with the spatial Pearson correlation coefficient computed between forecast and ground truth.
        }
        \label{fig:iod}
    \end{figure}

    The Neptune DL model has a high predictive skill in the IOD representation within the considered evaluation period, preserving the morphological structure and polarity of the zonal gradient at S2S scale.
    We attribute the good conservation of the geometrical IOD patterns characterizing the dipole to the emulator's ability to learn the patterns underlying the ocean dynamics in the TIO region. Mixing local convolutions and global convolutions in Neptune's architecture (see Section \ref{sect:arch_overview}) enabled to effectively merge global ocean patterns with local features of this region, resulting in a good representation of the morphological IOD patterns.
    This robustness in reproducing the IOD is proved by the high Pearson correlation coefficient computed in both Figure \ref{fig:iod_time} and \ref{fig:iod_sst}. The two Figures highlight a stable and slow decay: the spatial correlation pattern degrades from an initial $0.99$ at t+10 and reaches a still high $0.89$ at the end of the rollout (t+60). Moreover, Figure \ref{fig:iod_time} shows a high stability in the phase of the IOD events, correctly reproducing the timeseries at a S2S scale. Except for the highest anomaly observed in the 2019 pIOD event, Neptune faithfully reproduces the amplitude of the index.
    Despite a small amplitude decay during the 2019 pIOD period that limits the direct use in operational contexts, the robustness of the spatial and temporal correlation coefficients suggests potential application scenarios of Neptune's predictions in this region.
    
    \subsection{Evaluation on Neptune-025}
    \label{sect:eval_on_025}
    
    To conclude our evaluations of the Neptune global ocean emulator, we report results regarding the fine-tuned version at the native data resolution of $0.25^\circ$, Neptune-025. 
    
\begin{figure}
    \centering
        \begin{subfigure}{\textwidth}
            \centering
            \caption{Surface RMSE at t+60}
            \includegraphics[width=\textwidth]{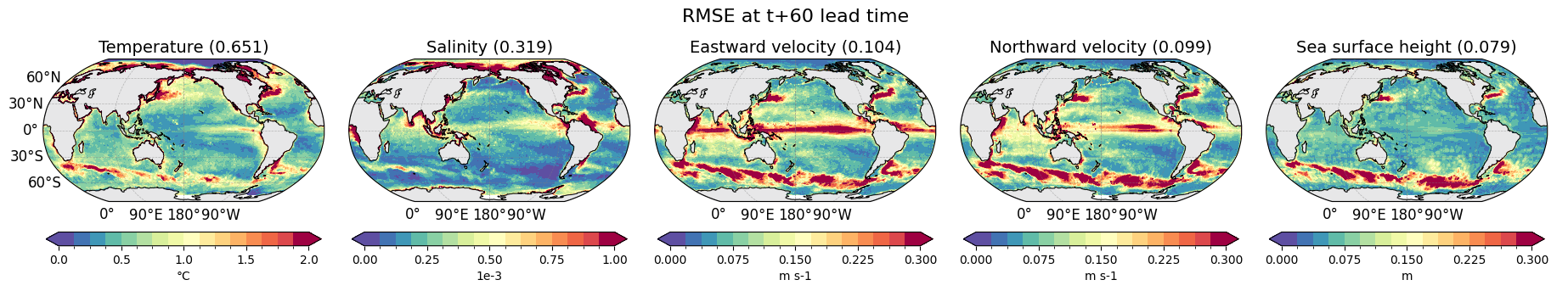}
            \label{fig:rmse_025}
        \end{subfigure}
        
        \begin{subfigure}{0.49\textwidth}
            \centering
            \caption{Spatial Ocean Heat Content}
            \includegraphics[width=\textwidth]{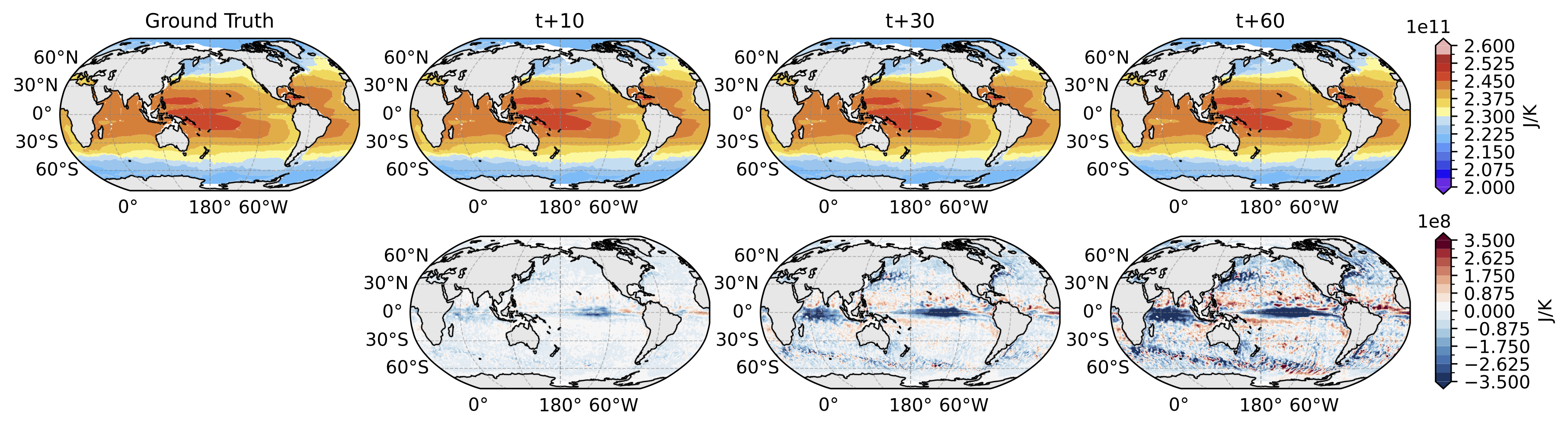}
            \label{fig:ohc_025}
        \end{subfigure}
        \begin{subfigure}{0.49\textwidth}
            \centering
            \caption{Temporal Ocean Heat Content}
            \includegraphics[width=\textwidth]{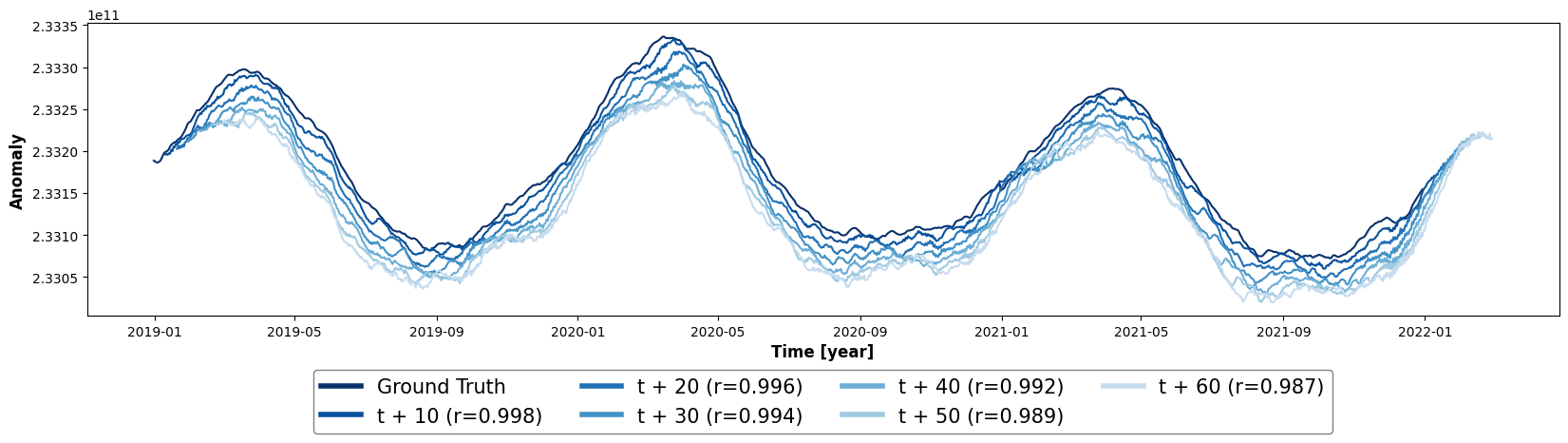}
            \label{fig:ohc_time_025}
        \end{subfigure}

        \hfill

        \begin{subfigure}{0.73\textwidth}
            \centering
            \caption{Eddy Kinetic Energy with bias}
            \includegraphics[width=\textwidth]{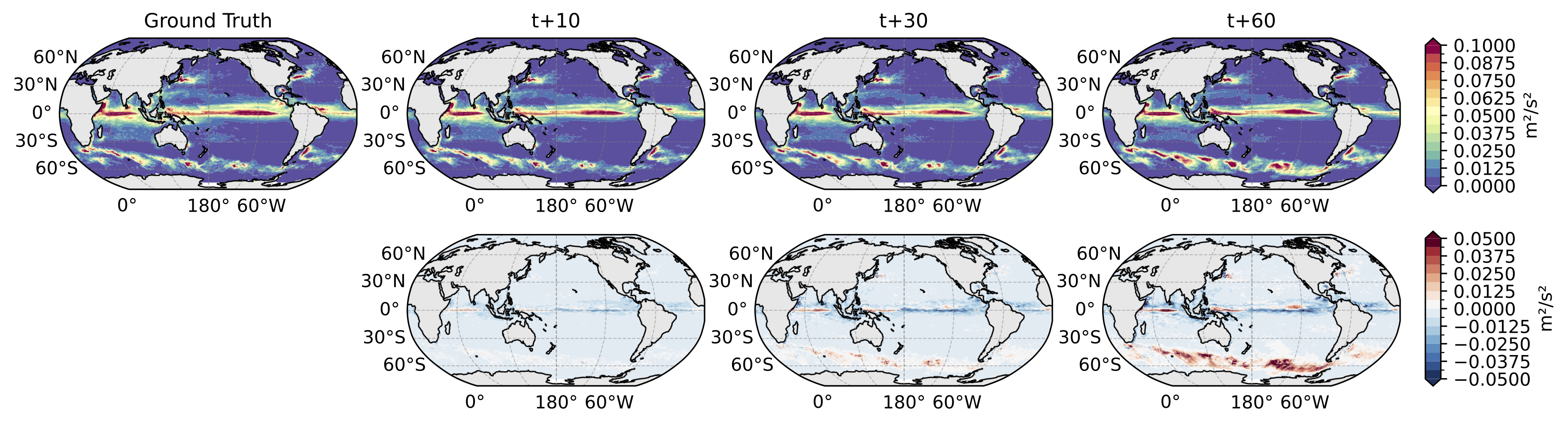}
            \label{fig:eke_025}
        \end{subfigure}
        \begin{subfigure}{0.26\textwidth}
            \centering
            \caption{EKE Power Spectral Density by lead time}
            \includegraphics[width=\textwidth]{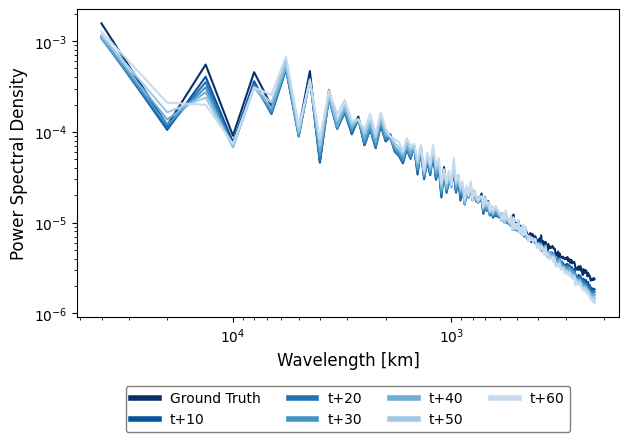}
            \label{fig:eke_psd_025}
        \end{subfigure}

        \hfill

        \begin{subfigure}{0.49\textwidth}
            \centering
            \caption{North Pole Ice Brier Score}
            \includegraphics[width=\textwidth]{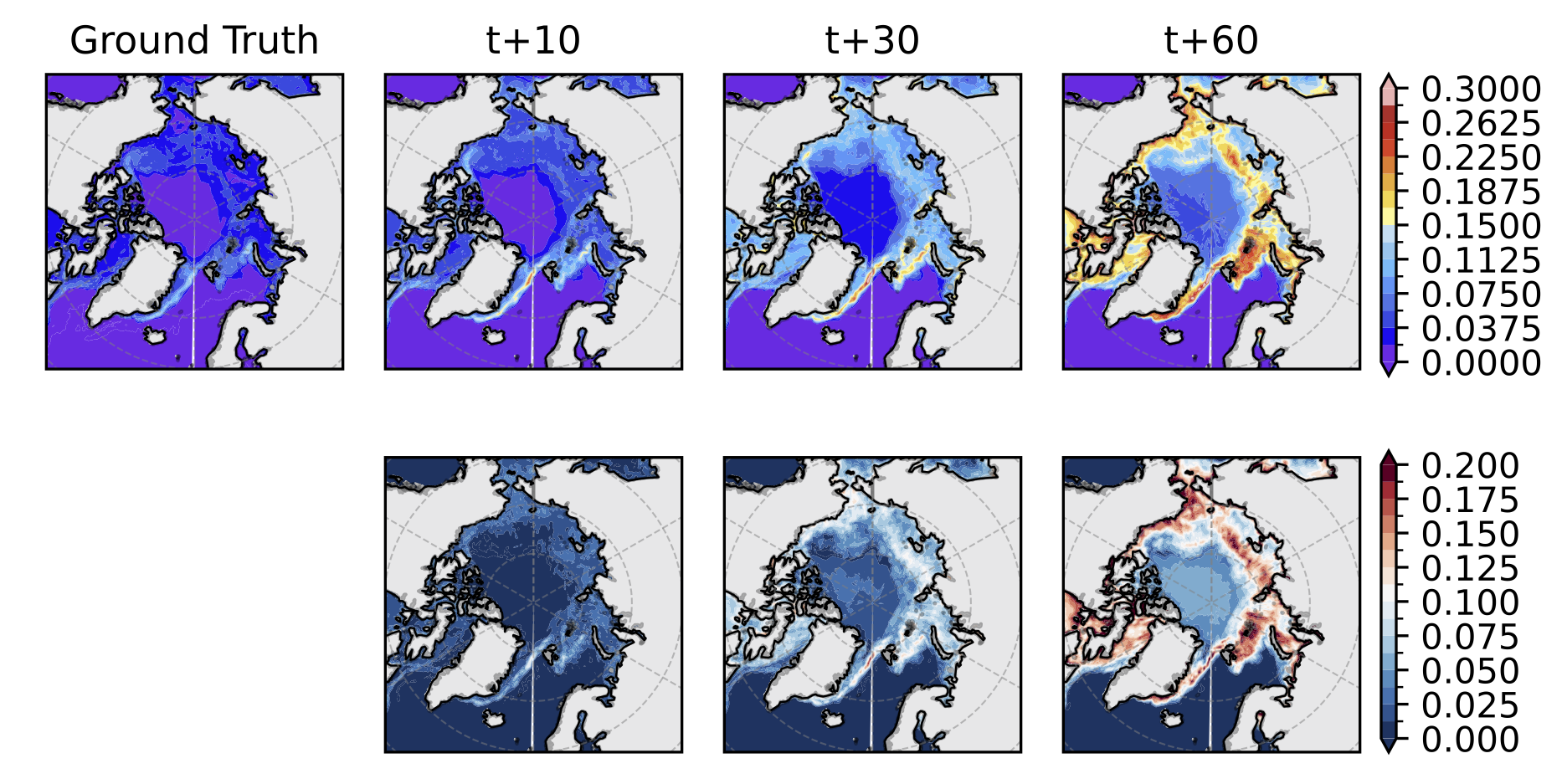}
            \label{fig:ibs_north_025}
        \end{subfigure}
        \begin{subfigure}{0.49\textwidth}
            \centering
            \caption{South Pole Ice Brier Score}
            \includegraphics[width=\textwidth]{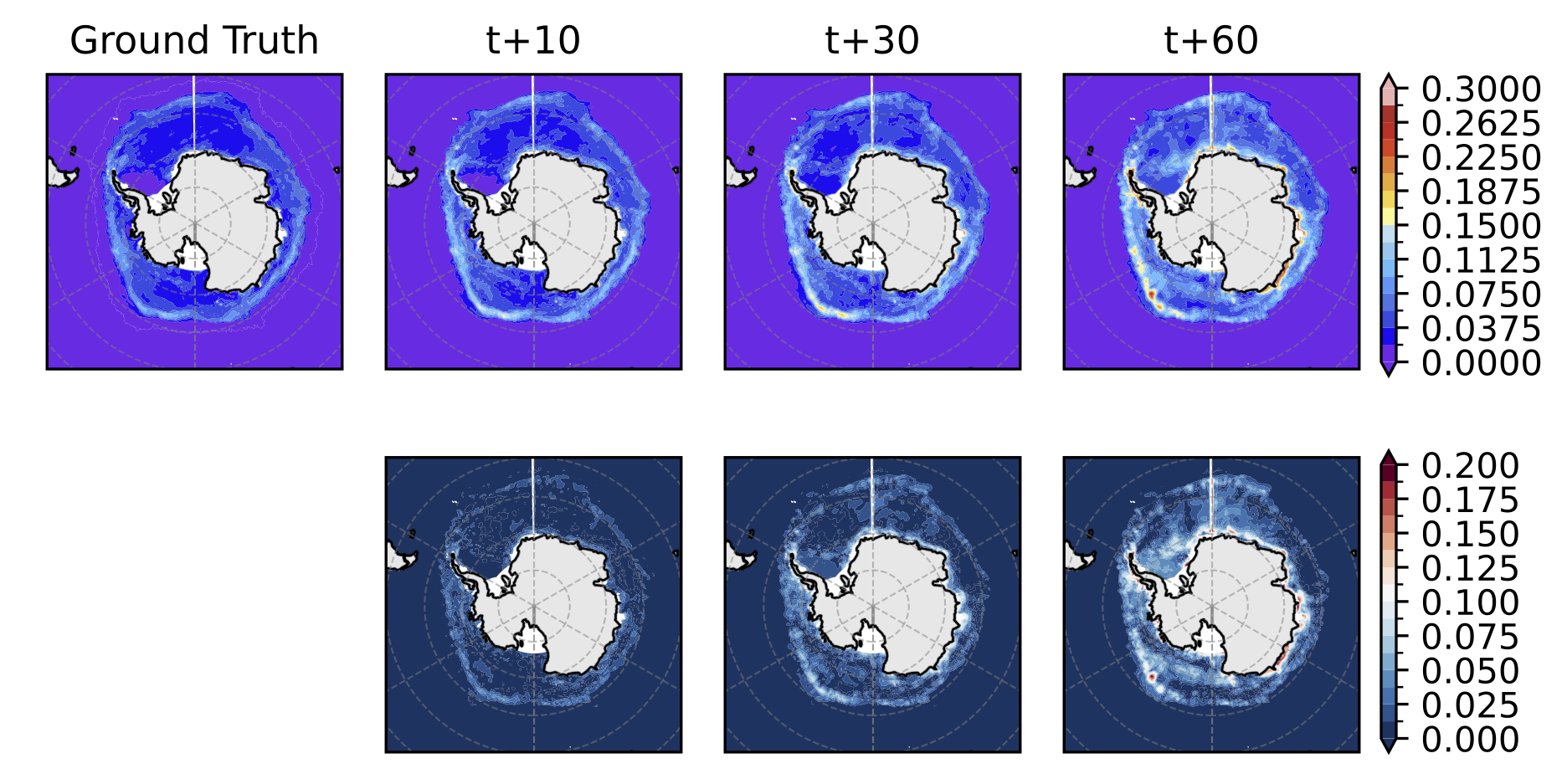}
            \label{fig:ibs_south_025}
        \end{subfigure}
        \caption{
        Neptune-025 skills are reported in the figure.
        Panel a) contains spatial Neptune’s RMSE at t+60 days for a subset of the predicted variables.
        Panels b-c) contain OHC computed over the test set. Panel b) top row contains OHC of ground truth and forecasts at t+10, t+30, t+60 while bottom row contains OHC bias between true and predicted OHC at a specific lead time. Panel c) shows the temporal OHC timeseries. Ground truth is in dark blue while forecasts from t+10 to t+60 are reported in progressively lighter blue colors. 
        Similarly to panel b), panel d) reports EKE of ground truth and forecasts at t+10, t+30, t+60 along with the bias between true and predicted EKE. Panel e) the EKE PSD is reported by lead time. The panel's color code is shared with panel c).
        Panels f) and g) report Neptune-025's IBS over North and South Pole, respectively. Top panels report reanalysis and forecasts at t+10, t+30 and t+60 while bottom panels report the bias between true and predicted IBS.
        }
        \label{fig:eval_025}
    \end{figure}

    In Figure \ref{fig:rmse_025} we report the surface RMSE of Neptune-025 at t+60 days, revealing the spatial distribution of the error for each of the most meaningful variables emulated by the DL model. After the fine-tuning, the RMSE is slightly higher than the pre-training and is distributed similarly to Neptune-1, as most of the error is localized in energetic areas of the globe. 
    
    Regarding OHC (Figures \ref{fig:ohc_025} \ref{fig:ohc_time_025}), we observe that Neptune-025 faithfully represents the OHC spatial patterns along the 60-day forecast. However, Figure \ref{fig:ohc_025} bottom row reveals a cool bias in the Pacific and Indian Equatorial regions. Additionally, Figure \ref{fig:ohc_time_025} shows that Neptune-025 has a slight decreasing trend in the average OHC timeseries, likely linked to the cooling we observe at the Equator. Despite the scale of the error is negligible with respect to the variable's scale, it is worth mentioning this unusual behavior of the DL emulator. 
    
    The EKE (Figure \ref{fig:eke_025} \ref{fig:eke_psd_025}) reveals that after the fine-tuning phase, it reproduces physically coherent current patterns, but after 60 days the DL model slightly over-estimates the Antarctic Circumpolar Current, while subtly under-estimating the energy at the equator. The PSD (Figure \ref{fig:eke_psd_025}) shows that the highest frequency details are slightly under-represented by the DL model, likely caused by the smoothing of the high frequencies of the MAE loss function. 
    
    Lastly, Neptune-025 preserves similar skills on the sea-ice representation, as we observe in Figure \ref{fig:ibs_north_025} and \ref{fig:ibs_south_025}. Similarly to the coarser resolution version, Neptune-025's IBS skills are asymmetric between North- and South- Pole. At t+60 days, IBS has a higher error over the North Pole, while the RMSE at the South Pole increases more slowly.
    
    We attribute the slight decrease in the emulator's skills to the need of a greater number of fine-tuning epochs or a better dimensioning (e.g., increasing latent dimension/model parameters) of the model to improve the learning of the complex interactions that drive the small-scale ocean dynamics. 
    Although Neptune-025 exhibits slightly lower performance than Neptune-1 in certain metrics, the fine-tuned model remains an effective eddy-resolving DL model suitable for a broader range of S2S applications. This demonstrates the feasibility of adapting the Neptune framework to downstream tasks in future works.
    
\section{Discussion}
\label{sect:discussion}

Neptune-1 and Neptune-025 combine Convolutional Neural Networks (CNNs) and Spherical Fourier Neural Operators (SFNO), establishing a skillful emulator for the global ocean and sea-ice state prediction at S2S timescales. We attribute its success to the inherent ability of SFNOs to learn the underlying Partial Differential Equations (PDEs) governing the ocean system. 
To simulate the complex interaction between the large- and small-scale processes intrinsic to the ocean dynamics, we add CNN layers to the SFNOs, thus learning complex local patterns and further enhancing Neptune's skills.

With small fine-tuning effort (Neptune-025), Neptune-1 framework can be adapted to high resolution data to effectively broaden the range of applications of the emulator on the S2S timescales. Neptune-025, indeed, results in an eddy resolving emulator capable of reproducing physically faithful predictions of the ocean and sea-ice state up to 60 days, using prescribed atmospheric forcing.

Statistical evaluations (RMSE, CRPS and ACC) confirm the model's performance. The low RMSE for temperature variable suggests Neptune's ability in reproducing the mean thermal state, lowering the known biases in traditional OGCMs in the mixed layer region. Neptune-025 version preserves similar RMSE skills, showing the adaptability of our framework.
Moreover, Neptune-1's high performance in preserving ACC across long lead times highlights its capacity to reproduce coherent morphological large-scale ocean patterns, successfully capturing the slow, deterministic evolution of the system. 
The spatial analysis of CRPS and ACC reveals fundamental information regarding its strengths and weaknesses. While Neptune-1 excels at reproducing large-scale patterns (i.e., ENSO, IOD), we observe a decline in performance in highly dynamic mesoscale regions (e.g., the Gulf Stream, Kuroshio currents, etc.) and at higher latitudes. 
This spatial degradation aligns with the theoretical challenge: the model is trained at $1^\circ \times 1^\circ$ horizontal resolution and the intrinsic noise in the observational data, it struggles to resolve high-frequency, small-scale eddies that dominate such regions, leading to the widely known double-penalty issue where smaller-scale details that are shifted in space cause two compound errors: spatial drift and under/over estimation \citep{subich2025}.
Additionally, the CRPS probabilistic metric shows asymmetric spatial irregularities, causing higher error in the Northern Hemisphere likely caused by the land displacement.

Physical coherence evaluations, specifically regarding the Ocean Heat Content (OHC) and Eddy Kinetic Energy (EKE) and Ice Brier Score (IBS), provide powerful insights about the surrogate's model ability in emulating the dynamic processes beyond the simple ocean state prediction. Both Neptune-1 and Neptune-025 successfully reproduce the spatio-temporal evolution of OHC, confirming its ability to reproduce large-scale thermal structures of the ocean. Despite the high temporal correlation, in Neptune-025 we observe a slight and progressive decrease in the total OHC budget over the forecast horizon, likely caused by few fine-tuning epochs or the small network dimension (i.e., 13M parameters).
Therefore, the model accurately predicts dominant, slowly evolving large-scale signals, like ENSO, but smooths out the fine-scale energetic variability.

The EKE analysis further assesses Neptune's dynamical skills. The emulator exhibits good skill in modeling the overall EKE distribution and preserves the EKE power spectra components at large wavelengths  (above $10^3$ $km$), but has lower skill in accurately reproducing the fine-scale, high-frequency variability associated with mesoscale patterns (both Neptune-1 and Neptune-025). The loss of fine-scale EKE information in the S2S context reflects the structural limitation of Neptune due to the input resolution and the use of a deterministic loss function. 

IBS score for the Sea Ice Concentration (SIC) provides us with a crucial evaluation of Neptune's skill in emulating ice fields at S2S scale. Results show that both the emulators maintain a notable spatiotemporal coherency of the morphological ice patterns at large scale, above all in the South Pole where we observe most of the emulator's stability. Along highly dynamic regions, like the coastline, BS degrades faster at 60 days.
This result highlights that, even using a strong physics-informed architecture like SFNO, ice forecasting at S2S timescale remains a significant challenge.

Finally, the successful emulation of El Ni\~no Southern Oscillation (ENSO) and Indian Ocean Dipole (IOD) indices confirms that Neptune-1 captures and reproduces the crucial atmosphere-ocean coupling mechanisms that drive such large-scale patterns. Indeed, since Neptune provides high temporal and spatial correlation coefficients for both ENSO and IOD, even after 60 days, it demonstrated that the model captures the inter-annual variability. This result, along with Isotherm at $20^{\circ}C$ (Z20) skills, further confirms our analysis.

\section{Conclusion}
\label{sect:conclusion}
 
In this work, we introduced and validated Neptune, a novel data-driven architecture mixing Convolutional Neural Networks (CNNs) and Spherical Fourier Neural Operators (SFNOs) for skillful global ocean forecasting at the subseasonal-to-seasonal (S2S) scale.
By integrating the SFNO on a spherical geometry, Neptune introduces a significant advancement over the current literature, demonstrating a remarkable ability to emulate the complex, multi-scale dynamics of the ocean system.
Unlike computationally expensive physics-based models, Neptune offers a faster alternative suitable for ensemble forecasting on S2S domain. 
Mixing CNN encoder and decoder with SFNO components reflects the inherent global ocean structure, characterized by complex interactions between large and small scales. Through encoder and decoder, we emulate the small-scale ocean thermo-dynamics, whereas through SFNO, we emulate the global teleconnections in an effective manner, thus improving Neptune's stability over long S2S rollouts. 
Unlike other data-driven ocean emulators tailored on the S2S scale, Neptune offers a huge advantage in terms of spatial and temporal resolution. Indeed, while other DL models such as Samudra and ORCA-DL are trained to forecast 5-day and monthly averages, respectively, Neptune-1 produces accurate daily forecasts and its fine-tuned version, Neptune-025, produces physically realistic high-resolution forecasts. Moreover, with respect to other works, Neptune is trained to emulate the full ocean state, including MLD, SIC and SIT, widening the range of possible evaluations, benchmarks and downstream tasks. Neptune-025 serves as a clear demonstration of the framework's adaptability to new tasks.
Finally, we evaluated Neptune against a wide variety of statistics (RMSE, CRPS, ACC), physical scores (OHC, EKE, IBS) and oceanic indices (ENSO and Z20 metric, IOD), contributing to the literature with advanced evaluation skills for S2S global ocean emulation.

However, limitations still remain, defining the trajectory for future research. While the SFNO architecture is structured on a sphere and learns global spherical convolutions, the encoder and decoder are still defined on rectangular grid. Future work will pave spherical local convolutions to provide more coherence to the network structure.
Despite high spatio-temporal resolution, Neptune-025 can be further enhanced by increasing the model's number of parameters or enhancing its training strategy (e.g., stochastic loss functions, generative approaches, etc.).
Moreover, model's performance decreases in highly dynamic regions (e.g., boundary currents and mesoscale). Observed smoothing of EKE over long lead times suggests exploring even more advanced probabilistic forecasting techniques. 

In summary, Neptune offers a powerful, adaptable, quick and accurate emulator for S2S ocean forecasting that successfully reproduces the dynamics of the ocean system. While the DL model demonstrates high forecasting skills for large-scale patterns, our work highlights the ongoing challenge of accurately resolving high-frequency, mesoscale variability. 
Future research should further enhance Neptune's S2S skills by exploiting advanced probabilistic forecasting techniques (e.g., representation learning) or generative AI methods (e.g., flow matching, diffusion, etc.). Despite the high-quality reanalysis dataset we used (ORAS5), including other reanalyses and simulation data could further improve Neptune's skills, posing the basis for global ocean foundation modeling.
At last, another direction of research should exploit a fully coupled S2S atmosphere-ocean emulator.

\bibliographystyle{unsrtnat}
\bibliography{main}

@article{amirian2026compilation,
  title={A Compilation of Marine Photosynthesis--Irradiance Data from $^{14}${C} Incubation Experiments},
  author={Amirian, Mohammad M and Devred, Emmanuel and Clay, Stephanie and Finkel, Zoe V and Irwin, Andrew J},
  journal={Earth System Science Data Discussions},
  volume={2026},
  year={2026},
  pages={1--34},
   URL = {	
https://essd.copernicus.org/preprints/essd-2026-651/},
    DOI = {10.5194/essd-2026-651},
  publisher={G{\"o}ttingen, Germany}
}

@article{zuo2019,
    AUTHOR = {Zuo, H. and Balmaseda, M. A. and Tietsche, S. and Mogensen, K. and Mayer, M.},
    TITLE = {The ECMWF operational ensemble reanalysis--analysis system for ocean and sea ice: a description of the system and assessment},
    JOURNAL = {Ocean Science},
    VOLUME = {15},
    YEAR = {2019},
    NUMBER = {3},
    PAGES = {779--808},
    URL = {https://os.copernicus.org/articles/15/779/2019/},
    DOI = {10.5194/os-15-779-2019}
}

@article{hersbach2023a,
    author = {Hersbach, H. and Bell, B. and Berrisford, P. and Biavati, G. and Hor\'anyi, A. and Mu\~noz Sabater, J. and Nicolas, J. and Peubey, C. and Radu, R. and Rozum, I. and Schepers, D. and Simmons, A. and Soci, C. and Dee, D. and Th\'epaut, J-N.},
    title = "{ERA5 hourly data on single levels from 1940 to present. Copernicus Climate Change Service (C3S) Climate Data Store (CDS)}",
    year = {2023},
    journal = "-",
    doi = {10.24381/cds.adbb2d47}
}

@article{hersbach2023b,
    author = {Hersbach, H. and Bell, B. and Berrisford, P. and Biavati, G. and Hor\'anyi, A. and Mu\~noz Sabater, J. and Nicolas, J. and Peubey, C. and Radu, R. and Rozum, I. and Schepers, D. and Simmons, A. and Soci, C. and Dee, D. and Th\'epaut, J-N.},
    title = "{ERA5 hourly data on pressure levels from 1940 to present. Copernicus Climate Change Service (C3S) Climate Data Store (CDS)}",
    year = {2023},
    journal = "-",
    doi = {10.24381/cds.bd0915c6}
}

@article{hersbach2020,
    author = {Hersbach, Hans and Bell, Bill and Berrisford, Paul and Hirahara, Shoji and Horányi, András and Muñoz-Sabater, Joaquín and Nicolas, Julien and Peubey, Carole and Radu, Raluca and Schepers, Dinand and Simmons, Adrian and Soci, Cornel and Abdalla, Saleh and Abellan, Xavier and Balsamo, Gianpaolo and Bechtold, Peter and Biavati, Gionata and Bidlot, Jean and Bonavita, Massimo and De Chiara, Giovanna and Dahlgren, Per and Dee, Dick and Diamantakis, Michail and Dragani, Rossana and Flemming, Johannes and Forbes, Richard and Fuentes, Manuel and Geer, Alan and Haimberger, Leo and Healy, Sean and Hogan, Robin J. and Hólm, Elías and Janisková, Marta and Keeley, Sarah and Laloyaux, Patrick and Lopez, Philippe and Lupu, Cristina and Radnoti, Gabor and de Rosnay, Patricia and Rozum, Iryna and Vamborg, Freja and Villaume, Sebastien and Thépaut, Jean-Noël},
    title = {The ERA5 global reanalysis},
    journal = {Quarterly Journal of the Royal Meteorological Society},
    volume = {146},
    number = {730},
    pages = {1999-2049},
    doi = {https://doi.org/10.1002/qj.3803},
    url = {https://rmets.onlinelibrary.wiley.com/doi/abs/10.1002/qj.3803},
    eprint = {https://rmets.onlinelibrary.wiley.com/doi/pdf/10.1002/qj.3803},
    year = {2020}
}

@misc{rasp2023,
    title={WeatherBench 2: A benchmark for the next generation of data-driven global weather models}, 
    author={Stephan Rasp and Stephan Hoyer and Alexander Merose and Ian Langmore and Peter Battaglia and Tyler Russel and Alvaro Sanchez-Gonzalez and Vivian Yang and Rob Carver and Shreya Agrawal and Matthew Chantry and Zied Ben Bouallegue and Peter Dueben and Carla Bromberg and Jared Sisk and Luke Barrington and Aaron Bell and Fei Sha},
    year={2023},
    eprint={2308.15560},
    archivePrefix={arXiv},
    primaryClass={physics.ao-ph}
}

@misc{gounou2024,
    title={Global Ocean Reanalysis Product - PRODUCT USER MANUAL}, 
    author={Gounou, A. and Drévillon, M. and Clavier, M.},
    year={2024},
    url={https://documentation.marine.copernicus.eu/PUM/CMEMS-GLO-PUM-001-031.pdf}, 
}

@article{troccoli2010,
	title = {Seasonal climate forecasting: {SEASONAL} {CLIMATE} {FORECASTING}: {A} {REVIEW}},
	volume = {17},
	copyright = {http://doi.wiley.com/10.1002/tdm\_license\_1.1},
	issn = {13504827},
	shorttitle = {Seasonal climate forecasting},
	url = {https://onlinelibrary.wiley.com/doi/10.1002/met.184},
	doi = {10.1002/met.184},
	language = {en},
	number = {3},
	urldate = {2026-01-15},
	journal = {Meteorological Applications},
	author = {Troccoli, Alberto},
	month = sep,
	year = {2010},
	pages = {251--268},
}

@article{meehl2021,
	title = {Initialized {Earth} {System} prediction from subseasonal to decadal timescales},
	volume = {2},
	issn = {2662-138X},
	url = {https://www.nature.com/articles/s43017-021-00155-x},
	doi = {10.1038/s43017-021-00155-x},
	language = {en},
	number = {5},
	urldate = {2026-01-15},
	journal = {Nature Reviews Earth \& Environment},
	author = {Meehl, Gerald A. and Richter, Jadwiga H. and Teng, Haiyan and Capotondi, Antonietta and Cobb, Kim and Doblas-Reyes, Francisco and Donat, Markus G. and England, Matthew H. and Fyfe, John C. and Han, Weiqing and Kim, Hyemi and Kirtman, Ben P. and Kushnir, Yochanan and Lovenduski, Nicole S. and Mann, Michael E. and Merryfield, William J. and Nieves, Veronica and Pegion, Kathy and Rosenbloom, Nan and Sanchez, Sara C. and Scaife, Adam A. and Smith, Doug and Subramanian, Aneesh C. and Sun, Lantao and Thompson, Diane and Ummenhofer, Caroline C. and Xie, Shang-Ping},
	month = apr,
	year = {2021},
	pages = {340--357},
}

@article{pikitch2004,
	title = {Ecosystem-{Based} {Fishery} {Management}},
	volume = {305},
	issn = {0036-8075, 1095-9203},
	url = {https://www.science.org/doi/10.1126/science.1098222},
	doi = {10.1126/science.1098222},
	language = {en},
	number = {5682},
	urldate = {2026-01-15},
	journal = {Science},
	author = {Pikitch, E. K. and Santora, C. and Babcock, E. A. and Bakun, A. and Bonfil, R. and Conover, D. O. and Dayton, P. and Doukakis, P. and Fluharty, D. and Heneman, B. and Houde, E. D. and Link, J. and Livingston, P. A. and Mangel, M. and McAllister, M. K. and Pope, J. and Sainsbury, K. J.},
	month = jul,
	year = {2004},
	pages = {346--347},
}

@article{bi2023,
	title = {Accurate medium-range global weather forecasting with {3D} neural networks},
	volume = {619},
	issn = {0028-0836, 1476-4687},
	url = {https://www.nature.com/articles/s41586-023-06185-3},
	doi = {10.1038/s41586-023-06185-3},
	language = {en},
	number = {7970},
	urldate = {2024-07-08},
	journal = {Nature},
	author = {Bi, Kaifeng and Xie, Lingxi and Zhang, Hengheng and Chen, Xin and Gu, Xiaotao and Tian, Qi},
	month = jul,
	year = {2023},
	pages = {533--538},
}

@article{bodnar2025,
	title = {A foundation model for the {Earth} system},
	volume = {641},
	issn = {0028-0836, 1476-4687},
	url = {https://www.nature.com/articles/s41586-025-09005-y},
	doi = {10.1038/s41586-025-09005-y},
	language = {en},
	number = {8065},
	urldate = {2026-01-15},
	journal = {Nature},
	author = {Bodnar, Cristian and Bruinsma, Wessel P. and Lucic, Ana and Stanley, Megan and Allen, Anna and Brandstetter, Johannes and Garvan, Patrick and Riechert, Maik and Weyn, Jonathan A. and Dong, Haiyu and Gupta, Jayesh K. and Thambiratnam, Kit and Archibald, Alexander T. and Wu, Chun-Chieh and Heider, Elizabeth and Welling, Max and Turner, Richard E. and Perdikaris, Paris},
	month = may,
	year = {2025},
	pages = {1180--1187},
}

@misc{chen2024,
	title = {{FuXi}-{S2S}: {A} machine learning model that outperforms conventional global subseasonal forecast models},
	shorttitle = {{FuXi}-{S2S}},
	url = {http://arxiv.org/abs/2312.09926},
	language = {en},
	urldate = {2024-07-08},
	publisher = {arXiv},
	author = {Chen, Lei and Zhong, Xiaohui and Li, Hao and Wu, Jie and Lu, Bo and Chen, Deliang and Xie, Shangping and Chao, Qingchen and Lin, Chensen and Hu, Zixin and Qi, Yuan},
	month = jul,
	year = {2024},
	note = {arXiv:2312.09926 [physics]},
}

@misc{xu2024,
	title = {{ExtremeCast}: {Boosting} {Extreme} {Value} {Prediction} for {Global} {Weather} {Forecast}},
	shorttitle = {{ExtremeCast}},
	url = {http://arxiv.org/abs/2402.01295},
	language = {en},
	urldate = {2024-07-08},
	publisher = {arXiv},
	author = {Xu, Wanghan and Chen, Kang and Han, Tao and Chen, Hao and Ouyang, Wanli and Bai, Lei},
	month = may,
	year = {2024},
	note = {arXiv:2402.01295 [cs]},
}

@misc{zhong2024,
	title = {{FuXi}-{ENS}: {A} machine learning model for medium-range ensemble weather forecasting},
	shorttitle = {{FuXi}-{ENS}},
	url = {http://arxiv.org/abs/2405.05925},
	language = {en},
	urldate = {2024-07-08},
	publisher = {arXiv},
	author = {Zhong, Xiaohui and Chen, Lei and Li, Hao and Liu, Jun and Fan, Xu and Feng, Jie and Dai, Kan and Luo, Jing-Jia and Wu, Jie and Qi, Yuan and Lu, Bo},
	month = jul,
	year = {2024},
	note = {arXiv:2405.05925 [physics]},
}

@misc{lam2023,
	title = {{GraphCast}: {Learning} skillful medium-range global weather forecasting},
	shorttitle = {{GraphCast}},
	url = {http://arxiv.org/abs/2212.12794},
	language = {en},
	urldate = {2024-07-08},
	publisher = {arXiv},
	author = {Lam, Remi and Sanchez-Gonzalez, Alvaro and Willson, Matthew and Wirnsberger, Peter and Fortunato, Meire and Alet, Ferran and Ravuri, Suman and Ewalds, Timo and Eaton-Rosen, Zach and Hu, Weihua and Merose, Alexander and Hoyer, Stephan and Holland, George and Vinyals, Oriol and Stott, Jacklynn and Pritzel, Alexander and Mohamed, Shakir and Battaglia, Peter},
	month = aug,
	year = {2023},
	note = {arXiv:2212.12794 [physics]},
}

@article{kochkov2024,
	title = {Neural {General} {Circulation} {Models} for {Weather} and {Climate}},
	volume = {632},
	issn = {0028-0836, 1476-4687},
	url = {http://arxiv.org/abs/2311.07222},
	doi = {10.1038/s41586-024-07744-y},
	language = {en},
	number = {8027},
	urldate = {2024-12-20},
	journal = {Nature},
	author = {Kochkov, Dmitrii and Yuval, Janni and Langmore, Ian and Norgaard, Peter and Smith, Jamie and Mooers, Griffin and Klöwer, Milan and Lottes, James and Rasp, Stephan and Düben, Peter and Hatfield, Sam and Battaglia, Peter and Sanchez-Gonzalez, Alvaro and Willson, Matthew and Brenner, Michael P. and Hoyer, Stephan},
	month = aug,
	year = {2024},
	note = {arXiv:2311.07222 [physics]},
	pages = {1060--1066},
}

@misc{pathak2022,
	title = {{FourCastNet}: {A} {Global} {Data}-driven {High}-resolution {Weather} {Model} using {Adaptive} {Fourier} {Neural} {Operators}},
	shorttitle = {{FourCastNet}},
	url = {http://arxiv.org/abs/2202.11214},
	doi = {10.48550/arXiv.2202.11214},
	language = {en},
	urldate = {2024-12-20},
	publisher = {arXiv},
	author = {Pathak, Jaideep and Subramanian, Shashank and Harrington, Peter and Raja, Sanjeev and Chattopadhyay, Ashesh and Mardani, Morteza and Kurth, Thorsten and Hall, David and Li, Zongyi and Azizzadenesheli, Kamyar and Hassanzadeh, Pedram and Kashinath, Karthik and Anandkumar, Animashree},
	month = feb,
	year = {2022},
	note = {arXiv:2202.11214 [physics]},
}

@article{price2025,
	title = {Probabilistic weather forecasting with machine learning},
	volume = {637},
	issn = {0028-0836, 1476-4687},
	url = {https://www.nature.com/articles/s41586-024-08252-9},
	doi = {10.1038/s41586-024-08252-9},
	language = {en},
	number = {8044},
	urldate = {2025-05-08},
	journal = {Nature},
	author = {Price, Ilan and Sanchez-Gonzalez, Alvaro and Alet, Ferran and Andersson, Tom R. and El-Kadi, Andrew and Masters, Dominic and Ewalds, Timo and Stott, Jacklynn and Mohamed, Shakir and Battaglia, Peter and Lam, Remi and Willson, Matthew},
	month = jan,
	year = {2025},
	pages = {84--90},
}

@misc{alet2025,
	title = {Skillful joint probabilistic weather forecasting from marginals},
	url = {http://arxiv.org/abs/2506.10772},
	doi = {10.48550/arXiv.2506.10772},
	language = {en},
	urldate = {2025-07-03},
	publisher = {arXiv},
	author = {Alet, Ferran and Price, Ilan and El-Kadi, Andrew and Masters, Dominic and Markou, Stratis and Andersson, Tom R. and Stott, Jacklynn and Lam, Remi and Willson, Matthew and Sanchez-Gonzalez, Alvaro and Battaglia, Peter},
	month = jun,
	year = {2025},
	note = {arXiv:2506.10772 [cs]},
}

@article{watt-meyer2025,
	title = {{ACE2}: accurately learning subseasonal to decadal atmospheric variability and forced responses},
	volume = {8},
	issn = {2397-3722},
	shorttitle = {{ACE2}},
	url = {https://www.nature.com/articles/s41612-025-01090-0},
	doi = {10.1038/s41612-025-01090-0},
	language = {en},
	number = {1},
	urldate = {2026-01-15},
	journal = {npj Climate and Atmospheric Science},
	author = {Watt-Meyer, Oliver and Henn, Brian and McGibbon, Jeremy and Clark, Spencer K. and Kwa, Anna and Perkins, W. Andre and Wu, Elynn and Harris, Lucas and Bretherton, Christopher S.},
	month = may,
	year = {2025},
	pages = {205},
}

@article{kent2025,
	title = {Skilful global seasonal predictions from a machine learning weather model trained on reanalysis data},
	volume = {8},
	issn = {2397-3722},
	url = {https://www.nature.com/articles/s41612-025-01198-3},
	doi = {10.1038/s41612-025-01198-3},
	language = {en},
	number = {1},
	urldate = {2026-01-15},
	journal = {npj Climate and Atmospheric Science},
	author = {Kent, Chris and Scaife, Adam A. and Dunstone, Nick J. and Smith, Doug and Hardiman, Steven C. and Dunstan, Tom and Watt-Meyer, Oliver},
	month = aug,
	year = {2025},
	pages = {314},
}

@misc{nguyen2023,
	title = {{ClimaX}: {A} foundation model for weather and climate},
	shorttitle = {{ClimaX}},
	url = {http://arxiv.org/abs/2301.10343},
	doi = {10.48550/arXiv.2301.10343},
	language = {en},
	urldate = {2026-01-15},
	publisher = {arXiv},
	author = {Nguyen, Tung and Brandstetter, Johannes and Kapoor, Ashish and Gupta, Jayesh K. and Grover, Aditya},
	month = dec,
	year = {2023},
	note = {arXiv:2301.10343 [cs]},
}

@misc{wang2025,
	title = {{CondensNet}: {Enabling} stable long-term climate simulations via hybrid deep learning models with adaptive physical constraints},
	shorttitle = {{CondensNet}},
	url = {http://arxiv.org/abs/2502.13185},
	doi = {10.48550/arXiv.2502.13185},
	language = {en},
	urldate = {2026-01-15},
	publisher = {arXiv},
	author = {Wang, Xin and Yang, Juntao and Adie, Jeff and See, Simon and Furtado, Kalli and Chen, Chen and Arcomano, Troy and Maulik, Romit and Mengaldo, Gianmarco},
	month = feb,
	year = {2025},
	note = {arXiv:2502.13185 [physics]},
}

@misc{watt-meyer2023,
	title = {{ACE}: {A} fast, skillful learned global atmospheric model for climate prediction},
	shorttitle = {{ACE}},
	url = {http://arxiv.org/abs/2310.02074},
	doi = {10.48550/arXiv.2310.02074},
	language = {en},
	urldate = {2026-01-15},
	publisher = {arXiv},
	author = {Watt-Meyer, Oliver and Dresdner, Gideon and McGibbon, Jeremy and Clark, Spencer K. and Henn, Brian and Duncan, James and Brenowitz, Noah D. and Kashinath, Karthik and Pritchard, Michael S. and Bonev, Boris and Peters, Matthew E. and Bretherton, Christopher S.},
	month = dec,
	year = {2023},
	note = {arXiv:2310.02074 [physics]},
}

@article{weyn2021,
	title = {Sub-seasonal forecasting with a large ensemble of deep-learning weather prediction models},
	volume = {13},
	issn = {1942-2466, 1942-2466},
	url = {http://arxiv.org/abs/2102.05107},
	doi = {10.1029/2021MS002502},
	language = {en},
	number = {7},
	urldate = {2026-01-15},
	journal = {Journal of Advances in Modeling Earth Systems},
	author = {Weyn, Jonathan A. and Durran, Dale R. and Caruana, Rich and Cresswell-Clay, Nathaniel},
	month = jul,
	year = {2021},
	note = {arXiv:2102.05107 [physics]},
	pages = {e2021MS002502},
}

@article{lowe2025,
	title = {Long-{Term} {Predictions} of {Loop} {Current} {Eddy} {Evolutions} {Using} {OceanNet}: {A} {Fourier} {Neural} {Operator}–{Based} {Data}-{Driven} {Ocean} {Emulator}},
	volume = {4},
	copyright = {http://www.ametsoc.org/PUBSReuseLicenses},
	issn = {2769-7525},
	shorttitle = {Long-{Term} {Predictions} of {Loop} {Current} {Eddy} {Evolutions} {Using} {OceanNet}},
	url = {https://journals.ametsoc.org/view/journals/aies/4/3/AIES-D-24-0039.1.xml},
	doi = {10.1175/AIES-D-24-0039.1},
	language = {en},
	number = {3},
	urldate = {2026-01-16},
	journal = {Artificial Intelligence for the Earth Systems},
	author = {Lowe, Anna B. and Gray, Michael and Chattopadhyay, Ashesh and Wu, Tianning and He, Ruoying},
	month = jul,
	year = {2025},
	pages = {e240039},
}

@misc{chattopadhyay2024,
	title = {{OceanNet}: {A} principled neural operator-based digital twin for regional oceans},
	shorttitle = {{OceanNet}},
	url = {http://arxiv.org/abs/2310.00813},
	doi = {10.48550/arXiv.2310.00813},
	language = {en},
	urldate = {2026-01-20},
	publisher = {arXiv},
	author = {Chattopadhyay, Ashesh and Gray, Michael and Wu, Tianning and Lowe, Anna B. and He, Ruoying},
	month = sep,
	year = {2024},
	note = {arXiv:2310.00813 [cs]},
}

@article{dheeshjith2025,
	title = {Samudra: {An} {AI} {Global} {Ocean} {Emulator} for {Climate}},
	volume = {52},
	issn = {0094-8276, 1944-8007},
	shorttitle = {Samudra},
	url = {https://agupubs.onlinelibrary.wiley.com/doi/10.1029/2024GL114318},
	doi = {10.1029/2024GL114318},
	language = {en},
	number = {10},
	urldate = {2026-01-20},
	journal = {Geophysical Research Letters},
	author = {Dheeshjith, Surya and Subel, Adam and Adcroft, Alistair and Busecke, Julius and Fernandez‐Granda, Carlos and Gupta, Shubham and Zanna, Laure},
	month = may,
	year = {2025},
	pages = {e2024GL114318},
}

@article{aouni2025,
	title = {{GLONET}: {Mercator}'s end-to-end neural {Global} {Ocean} forecasting system},
	volume = {2},
	issn = {2993-5210, 2993-5210},
	shorttitle = {{GLONET}},
	url = {http://arxiv.org/abs/2412.05454},
	doi = {10.1029/2025JH000686},
	language = {en},
	number = {3},
	urldate = {2026-01-20},
	journal = {Journal of Geophysical Research: Machine Learning and Computation},
	author = {Aouni, Anass El and Gaudel, Quentin and Regnier, Charly and Gennip, Simon Van and Galloudec, Olivier Le and Drevillon, Marie and Drillet, Yann and Lellouche, Jean-Michel},
	month = sep,
	year = {2025},
	note = {arXiv:2412.05454 [physics]},
	pages = {e2025JH000686},
}

@article{guo2025,
	title = {Data-driven global ocean modeling for seasonal to decadal prediction},
	volume = {11},
	issn = {2375-2548},
	url = {https://www.science.org/doi/10.1126/sciadv.adu2488},
	doi = {10.1126/sciadv.adu2488},
	language = {en},
	number = {33},
	urldate = {2026-01-20},
	journal = {Science Advances},
	author = {Guo, Zijie and Lyu, Pumeng and Ling, Fenghua and Bai, Lei and Luo, Jing-Jia and Boers, Niklas and Yamagata, Toshio and Izumo, Takeshi and Cravatte, Sophie and Capotondi, Antonietta and Ouyang, Wanli},
	month = aug,
	year = {2025},
	pages = {eadu2488},
}

@misc{huang2025,
	title = {{FuXi}-{Ocean}: {A} {Global} {Ocean} {Forecasting} {System} with {Sub}-{Daily} {Resolution}},
	shorttitle = {{FuXi}-{Ocean}},
	url = {http://arxiv.org/abs/2506.03210},
	doi = {10.48550/arXiv.2506.03210},
	language = {en},
	urldate = {2026-01-20},
	publisher = {arXiv},
	author = {Huang, Qiusheng and Niu, Yuan and Zhong, Xiaohui and Guo, Anboyu and Chen, Lei and Zhang, Dianjun and Zhang, Xuefeng and Li, Hao},
	month = oct,
	year = {2025},
	note = {arXiv:2506.03210 [cs]},
}

@misc{niu2025,
	title = {A data-driven global ocean forecasting model with sub-daily and eddy-resolving resolution},
	url = {http://arxiv.org/abs/2509.17015},
	doi = {10.48550/arXiv.2509.17015},
	language = {en},
	urldate = {2026-01-20},
	publisher = {arXiv},
	author = {Niu, Yuan and Huang, Qiusheng and Zhong, Xiaohui and Guo, Anboyu and Chen, Lei and Jia, Xiaoyan and Qi, Jiawei and Zhang, Dianjun and Li, Hao and Zhang, Xuefeng},
	month = sep,
	year = {2025},
	note = {arXiv:2509.17015 [physics]},
}

@misc{subel2024,
	title = {Building {Ocean} {Climate} {Emulators}},
	url = {http://arxiv.org/abs/2402.04342},
	doi = {10.48550/arXiv.2402.04342},
	language = {en},
	urldate = {2026-01-20},
	publisher = {arXiv},
	author = {Subel, Adam and Zanna, Laure},
	month = mar,
	year = {2024},
	note = {arXiv:2402.04342 [physics]},
}

@misc{wang2024,
	title = {{XiHe}: {A} {Data}-{Driven} {Model} for {Global} {Ocean} {Eddy}-{Resolving} {Forecasting}},
	shorttitle = {{XiHe}},
	url = {http://arxiv.org/abs/2402.02995},
	doi = {10.48550/arXiv.2402.02995},
	language = {en},
	urldate = {2026-01-20},
	publisher = {arXiv},
	author = {Wang, Xiang and Wang, Renzhi and Hu, Ningzi and Wang, Pinqiang and Huo, Peng and Wang, Guihua and Wang, Huizan and Wang, Senzhang and Zhu, Junxing and Xu, Jianbo and Yin, Jun and Bao, Senliang and Luo, Ciqiang and Zu, Ziqing and Han, Yi and Zhang, Weimin and Ren, Kaijun and Deng, Kefeng and Song, Junqiang},
	month = oct,
	year = {2024},
	note = {arXiv:2402.02995 [physics]},
}

@misc{epicoco2025,
	title = {{MedFormer}: a data-driven model for forecasting the {Mediterranean} {Sea}},
	shorttitle = {{MedFormer}},
	url = {http://arxiv.org/abs/2509.00015},
	doi = {10.48550/arXiv.2509.00015},
	language = {en},
	urldate = {2026-01-20},
	publisher = {arXiv},
	author = {Epicoco, Italo and Donno, Davide and Accarino, Gabriele and Norberti, Simone and Grandi, Alessandro and Giurato, Michele and McAdam, Ronan and Elia, Donatello and Clementi, Emanuela and Nassisi, Paola and Scoccimarro, Enrico and Coppini, Giovanni and Gualdi, Silvio and Aloisio, Giovanni and Masina, Simona and Boccaletti, Giulio and Navarra, Antonio},
	month = aug,
	year = {2025},
	note = {arXiv:2509.00015 [physics]},
}

@article{holmberg2025,
	title = {Accurate {Mediterranean} {Sea} forecasting via graph-based deep learning},
	volume = {15},
	issn = {2045-2322},
	url = {https://www.nature.com/articles/s41598-025-31177-w},
	doi = {10.1038/s41598-025-31177-w},
	language = {en},
	number = {1},
	urldate = {2026-01-20},
	journal = {Scientific Reports},
	author = {Holmberg, Daniel and Clementi, Emanuela and Epicoco, Italo and Roos, Teemu},
	month = dec,
	year = {2025},
	pages = {45051},
}

@article{pinardi2003,
	title = {The {Mediterranean} ocean forecasting system: first phase of implementation (1998–2001)},
	volume = {21},
	copyright = {https://creativecommons.org/licenses/by/3.0/},
	issn = {1432-0576},
	shorttitle = {The {Mediterranean} ocean forecasting system},
	url = {https://angeo.copernicus.org/articles/21/3/2003/},
	doi = {10.5194/angeo-21-3-2003},
	language = {en},
	number = {1},
	urldate = {2026-01-20},
	journal = {Annales Geophysicae},
	author = {Pinardi, N. and Allen, I. and Demirov, E. and De Mey, P. and Korres, G. and Lascaratos, A. and Le Traon, P.-Y. and Maillard, C. and Manzella, G. and Tziavos, C.},
	month = jan,
	year = {2003},
	pages = {3--20},
}

@article{coppini2023,
	title = {The {Mediterranean} {Forecasting} {System} – {Part} 1: {Evolution} and performance},
	volume = {19},
	copyright = {https://creativecommons.org/licenses/by/4.0/},
	issn = {1812-0792},
	shorttitle = {The {Mediterranean} {Forecasting} {System} – {Part} 1},
	url = {https://os.copernicus.org/articles/19/1483/2023/},
	doi = {10.5194/os-19-1483-2023},
	language = {en},
	number = {5},
	urldate = {2026-01-20},
	journal = {Ocean Science},
	author = {Coppini, Giovanni and Clementi, Emanuela and Cossarini, Gianpiero and Salon, Stefano and Korres, Gerasimos and Ravdas, Michalis and Lecci, Rita and Pistoia, Jenny and Goglio, Anna Chiara and Drudi, Massimiliano and Grandi, Alessandro and Aydogdu, Ali and Escudier, Romain and Cipollone, Andrea and Lyubartsev, Vladyslav and Mariani, Antonio and Cretì, Sergio and Palermo, Francesco and Scuro, Matteo and Masina, Simona and Pinardi, Nadia and Navarra, Antonio and Delrosso, Damiano and Teruzzi, Anna and Di Biagio, Valeria and Bolzon, Giorgio and Feudale, Laura and Coidessa, Gianluca and Amadio, Carolina and Brosich, Alberto and Miró, Arnau and Alvarez, Eva and Lazzari, Paolo and Solidoro, Cosimo and Oikonomou, Charikleia and Zacharioudaki, Anna},
	month = oct,
	year = {2023},
	pages = {1483--1516},
}

@article{pinardi2010,
	title = {{Operational} oceanography in the {Mediterranean} {Sea}: the second stage of development\&quot;},
	volume = {6},
	copyright = {https://creativecommons.org/licenses/by/3.0/},
	issn = {1812-0792},
	shorttitle = {Preface \&quot;{Operational} oceanography in the {Mediterranean} {Sea}},
	url = {https://os.copernicus.org/articles/6/263/2010/},
	doi = {10.5194/os-6-263-2010},
	language = {en},
	number = {1},
	urldate = {2026-01-20},
	journal = {Ocean Science},
	author = {Pinardi, N. and Coppini, G.},
	month = feb,
	year = {2010},
	pages = {263--267},
}

@incollection{wang2017,
	address = {Dordrecht},
	title = {El {Niño} and {Southern} {Oscillation} ({ENSO}): {A} {Review}},
	volume = {8},
	isbn = {978-94-017-7498-7 978-94-017-7499-4},
	shorttitle = {El {Niño} and {Southern} {Oscillation} ({ENSO})},
	url = {http://link.springer.com/10.1007/978-94-017-7499-4_4},
	doi = {10.1007/978-94-017-7499-4_4},
	language = {en},
	urldate = {2026-05-24},
	booktitle = {Coral {Reefs} of the {Eastern} {Tropical} {Pacific}},
	publisher = {Springer Netherlands},
	author = {Wang, Chunzai and Deser, Clara and Yu, Jin-Yi and DiNezio, Pedro and Clement, Amy},
	editor = {Glynn, Peter W. and Manzello, Derek P. and Enochs, Ian C.},
	year = {2017},
	note = {Series Title: Coral Reefs of the World},
	pages = {85--106},
}

@misc{oras52021,
	title = {{ORAS5} global ocean reanalysis monthly data from 1958 to present},
	url = {https://cds.climate.copernicus.eu/doi/10.24381/cds.67e8eeb7},
	doi = {10.24381/CDS.67E8EEB7},
	urldate = {2026-06-10},
	publisher = {ECMWF},
	author = {{Copernicus Climate Change Service}},
	year = {2021},
}

@article{behringer1998,
	title = {An {Improved} {Coupled} {Model} for {ENSO} {Prediction} and {Implications} for {Ocean} {Initialization}. {Part} {I}: {The} {Ocean} {Data} {Assimilation} {System}},
	volume = {126},
	issn = {0027-0644, 1520-0493},
	shorttitle = {An {Improved} {Coupled} {Model} for {ENSO} {Prediction} and {Implications} for {Ocean} {Initialization}. {Part} {I}},
	url = {http://journals.ametsoc.org/doi/10.1175/1520-0493(1998)126<1013:AICMFE>2.0.CO;2},
	doi = {10.1175/1520-0493(1998)126<1013:AICMFE>2.0.CO;2},
	language = {en},
	number = {4},
	urldate = {2026-06-10},
	journal = {Monthly Weather Review},
	author = {Behringer, David W. and Ji, Ming and Leetmaa, Ants},
	month = apr,
	year = {1998},
	pages = {1013--1021},
}

@misc{vaswani2023,
	title = {Attention {Is} {All} {You} {Need}},
	url = {http://arxiv.org/abs/1706.03762},
	doi = {10.48550/arXiv.1706.03762},
	language = {en},
	urldate = {2026-06-10},
	publisher = {arXiv},
	author = {Vaswani, Ashish and Shazeer, Noam and Parmar, Niki and Uszkoreit, Jakob and Jones, Llion and Gomez, Aidan N. and Kaiser, Lukasz and Polosukhin, Illia},
	month = aug,
	year = {2023},
	note = {arXiv:1706.03762 [cs.CL]},
}

@misc{ronneberger2015,
	title = {U-{Net}: {Convolutional} {Networks} for {Biomedical} {Image} {Segmentation}},
	copyright = {arXiv.org perpetual, non-exclusive license},
	shorttitle = {U-{Net}},
	url = {https://arxiv.org/abs/1505.04597},
	doi = {10.48550/ARXIV.1505.04597},
	urldate = {2026-06-10},
	publisher = {arXiv},
	author = {Ronneberger, Olaf and Fischer, Philipp and Brox, Thomas},
	year = {2015},
	note = {Version Number: 1},
}

@article{stouffer2006,
	title = {{GFDL}'s {CM2} {Global} {Coupled} {Climate} {Models}. {Part} {IV}: {Idealized} {Climate} {Response}},
	volume = {19},
	issn = {1520-0442, 0894-8755},
	shorttitle = {{GFDL}'s {CM2} {Global} {Coupled} {Climate} {Models}. {Part} {IV}},
	url = {http://journals.ametsoc.org/doi/10.1175/JCLI3632.1},
	doi = {10.1175/JCLI3632.1},
	language = {en},
	number = {5},
	urldate = {2026-06-10},
	journal = {Journal of Climate},
	author = {Stouffer, R. J. and Broccoli, A. J. and Delworth, T. L. and Dixon, K. W. and Gudgel, R. and Held, I. and Hemler, R. and Knutson, T. and Lee, Hyun-Chul and Schwarzkopf, M. D. and Soden, B. and Spelman, M. J. and Winton, M. and Zeng, Fanrong},
	month = mar,
	year = {2006},
	pages = {723--740},
}

@misc{liu2022,
	title = {A {ConvNet} for the 2020s},
	copyright = {Creative Commons Attribution 4.0 International},
	url = {https://arxiv.org/abs/2201.03545},
	doi = {10.48550/ARXIV.2201.03545},
	urldate = {2026-06-10},
	publisher = {arXiv},
	author = {Liu, Zhuang and Mao, Hanzi and Wu, Chao-Yuan and Feichtenhofer, Christoph and Darrell, Trevor and Xie, Saining},
	year = {2022},
	note = {Version Number: 2},
}

@article{adcroft2019,
	title = {The {GFDL} {Global} {Ocean} and {Sea} {Ice} {Model} {OM4}.0: {Model} {Description} and {Simulation} {Features}},
	volume = {11},
	issn = {1942-2466, 1942-2466},
	shorttitle = {The {GFDL} {Global} {Ocean} and {Sea} {Ice} {Model} {OM4}.0},
	url = {https://agupubs.onlinelibrary.wiley.com/doi/10.1029/2019MS001726},
	doi = {10.1029/2019MS001726},
	language = {en},
	number = {10},
	urldate = {2026-06-10},
	journal = {Journal of Advances in Modeling Earth Systems},
	author = {Adcroft, Alistair and Anderson, Whit and Balaji, V. and Blanton, Chris and Bushuk, Mitchell and Dufour, Carolina O. and Dunne, John P. and Griffies, Stephen M. and Hallberg, Robert and Harrison, Matthew J. and Held, Isaac M. and Jansen, Malte F. and John, Jasmin G. and Krasting, John P. and Langenhorst, Amy R. and Legg, Sonya and Liang, Zhi and McHugh, Colleen and Radhakrishnan, Aparna and Reichl, Brandon G. and Rosati, Tony and Samuels, Bonita L. and Shao, Andrew and Stouffer, Ronald and Winton, Michael and Wittenberg, Andrew T. and Xiang, Baoqiang and Zadeh, Niki and Zhang, Rong},
	month = oct,
	year = {2019},
	pages = {3167--3211},
}

@misc{bonev2023,
	title = {Spherical {Fourier} {Neural} {Operators}: {Learning} {Stable} {Dynamics} on the {Sphere}},
	copyright = {arXiv.org perpetual, non-exclusive license},
	shorttitle = {Spherical {Fourier} {Neural} {Operators}},
	url = {https://arxiv.org/abs/2306.03838},
	doi = {10.48550/ARXIV.2306.03838},
	urldate = {2026-06-10},
	publisher = {arXiv},
	author = {Bonev, Boris and Kurth, Thorsten and Hundt, Christian and Pathak, Jaideep and Baust, Maximilian and Kashinath, Karthik and Anandkumar, Anima},
	year = {2023},
	note = {Version Number: 1},
}

@misc{glorys2018,
	title = {Global {Ocean} {Physics} {Reanalysis}},
	url = {https://resources.marine.copernicus.eu/product-detail/GLOBAL_MULTIYEAR_PHY_001_030/INFORMATION},
	doi = {10.48670/MOI-00021},
	language = {en},
	urldate = {2026-06-10},
	publisher = {Mercator Ocean International},
	author = {{European Union-Copernicus Marine Service}},
	year = {2018},
}

@misc{cipollone2021,
	title = {The {Euro}-{Mediterranean} {Center} on {Climate} {Change} ({CMCC}) {Eddy}-permitting {Global} {Ocean} {Physical} {Reanalysis} ({C}-{GLORS} v7, 1993-2019)},
	copyright = {Creative Commons Attribution 4.0 International},
	url = {https://doi.pangaea.de/10.1594/PANGAEA.931485},
	doi = {10.1594/PANGAEA.931485},
	language = {en},
	urldate = {2026-06-10},
	publisher = {PANGAEA},
	author = {Cipollone, Andrea and Masina, Simona and Storto, Andrea},
	year = {2021},
	note = {Artwork Size: 2 data points Pages: 2 data points},
}

@article{dee2011,
	title = {The {ERA}‐{Interim} reanalysis: configuration and performance of the data assimilation system},
	volume = {137},
	issn = {0035-9009, 1477-870X},
	shorttitle = {The {ERA}‐{Interim} reanalysis},
	url = {https://rmets.onlinelibrary.wiley.com/doi/10.1002/qj.828},
	doi = {10.1002/qj.828},
	language = {en},
	number = {656},
	urldate = {2026-06-10},
	journal = {Quarterly Journal of the Royal Meteorological Society},
	author = {Dee, D. P. and Uppala, S. M. and Simmons, A. J. and Berrisford, P. and Poli, P. and Kobayashi, S. and Andrae, U. and Balmaseda, M. A. and Balsamo, G. and Bauer, P. and Bechtold, P. and Beljaars, A. C. M. and Van De Berg, L. and Bidlot, J. and Bormann, N. and Delsol, C. and Dragani, R. and Fuentes, M. and Geer, A. J. and Haimberger, L. and Healy, S. B. and Hersbach, H. and Hólm, E. V. and Isaksen, L. and Kållberg, P. and Köhler, M. and Matricardi, M. and McNally, A. P. and Monge‐Sanz, B. M. and Morcrette, J.‐J. and Park, B.‐K. and Peubey, C. and De Rosnay, P. and Tavolato, C. and Thépaut, J.‐N. and Vitart, F.},
	month = apr,
	year = {2011},
	pages = {553--597},
}

@Article{Vitart2018,
    author={Vitart, Fr{\'e}d{\'e}ricand Robertson, Andrew W.},
    title={The sub-seasonal to seasonal prediction project (S2S) and the prediction of extreme events},
    journal={npj Climate and Atmospheric Science},
    year={2018},
    month={Mar},
    day={12},
    volume={1},
    number={1},
    pages={3},
    issn={2397-3722},
    doi={10.1038/s41612-018-0013-0},
    url={https://doi.org/10.1038/s41612-018-0013-0}
}

@article{LIANG2026,
    title = {Integrating Subseasonal-to-Seasonal Forecasts into Agricultural Decision Support Systems: A Critical Review and Research Agenda},
    journal = {Engineering},
    year = {2026},
    issn = {2095-8099},
    doi = {https://doi.org/10.1016/j.eng.2026.05.015},
    url = {https://www.sciencedirect.com/science/article/pii/S2095809926003334},
    author = {Xin-Zhong Liang},
}

@incollection{BALMASEDA2026271,
    title = {Chapter 8 - The role of the ocean in subseasonal-to-seasonal predictability and prediction},
    editor = {Andrew W. Robertson and Frédéric Vitart},
    booktitle = {Sub-seasonal to Seasonal Prediction (Second Edition)},
    publisher = {Elsevier},
    edition = {Second Edition},
    pages = {271-320},
    year = {2026},
    isbn = {978-0-443-31538-1},
    doi = {https://doi.org/10.1016/B978-0-443-31538-1.00010-5},
    url = {https://www.sciencedirect.com/science/article/pii/B9780443315381000105},
    author = {Magdalena A. Balmaseda and Charlotte DeMott and Carolyn A. Reynolds and Christopher D. Roberts and Aneesh Subramanian},
}

@Article{Bracco2025,
    author={Bracco, Annalisa and Brajard, Julien and Dijkstra, Henk A. and Hassanzadeh, Pedram and Lessig, Christian and Monteleoni, Claire},
    title={Machine learning for the physics of climate},
    journal={Nature Reviews Physics},
    year={2025},
    month={Jan},
    day={01},
    volume={7},
    number={1},
    pages={6-20},
    issn={2522-5820},
    doi={10.1038/s42254-024-00776-3},
    url={https://doi.org/10.1038/s42254-024-00776-3}
}

@Article{Bauer2015,
    author={Bauer, Peter and Thorpe, Alan and Brunet, Gilbert},
    title={The quiet revolution of numerical weather prediction},
    journal={Nature},
    year={2015},
    month={Sep},
    day={01},
    volume={525},
    number={7567},
    pages={47-55},
    issn={1476-4687},
    doi={10.1038/nature14956},
    url={https://doi.org/10.1038/nature14956}
}

@INPROCEEDINGS{efanov2021,
  author={Efanov, Andrey A. and Ivliev, Sergey A. and Shagraev, Alexey G.},
  booktitle={2021 3rd International Youth Conference on Radio Electronics, Electrical and Power Engineering (REEPE)}, 
  title={Welford’s algorithm for weighted statistics}, 
  year={2021},
  volume={},
  number={},
  pages={1-5},
  doi={10.1109/REEPE51337.2021.9387973}
}

@misc{chen2021,
	title = {{AdaSpeech}: {Adaptive} {Text} to {Speech} for {Custom} {Voice}},
	shorttitle = {{AdaSpeech}},
	url = {http://arxiv.org/abs/2103.00993},
	doi = {10.48550/arXiv.2103.00993},
	language = {en},
	urldate = {2026-06-10},
	publisher = {arXiv},
	author = {Chen, Mingjian and Tan, Xu and Li, Bohan and Liu, Yanqing and Qin, Tao and Zhao, Sheng and Liu, Tie-Yan},
	month = mar,
	year = {2021},
	note = {arXiv:2103.00993 [eess.AS]},
}

@misc{kovachki2024,
	title = {Neural {Operator}: {Learning} {Maps} {Between} {Function} {Spaces}},
	shorttitle = {Neural {Operator}},
	url = {http://arxiv.org/abs/2108.08481},
	doi = {10.5555/3648699.3648788},
	urldate = {2026-06-10},
	author = {Kovachki, Nikola and Li, Zongyi and Liu, Burigede and Azizzadenesheli, Kamyar and Bhattacharya, Kaushik and Stuart, Andrew and Anandkumar, Anima},
	month = may,
	year = {2024},
	note = {arXiv:2108.08481 [cs.LG]},
}

@misc{cmcc_hpcc,
  author       = {{CMCC}},
  title        = {High Performance Computing Center – HPCC},
  howpublished = {\url{https://www.cmcc.it/what-we-do/high-performance-computing-center-hpcc}},
  year         = {2026},
  note         = {Accessed: 2026-06-10},
}

@article{kessler1990,
	title = {Observations of long {Rossby} waves in the northern tropical {Pacific}},
	volume = {95},
	copyright = {http://onlinelibrary.wiley.com/termsAndConditions\#vor},
	issn = {0148-0227},
	url = {https://agupubs.onlinelibrary.wiley.com/doi/10.1029/JC095iC04p05183},
	doi = {10.1029/JC095iC04p05183},
	language = {en},
	number = {C4},
	urldate = {2026-06-11},
	journal = {Journal of Geophysical Research: Oceans},
	author = {Kessler, William S.},
	month = apr,
	year = {1990},
	pages = {5183--5217},
}

@article{saji1999,
	title = {A dipole mode in the tropical {Indian} {Ocean}},
	volume = {401},
	copyright = {http://www.springer.com/tdm},
	issn = {0028-0836, 1476-4687},
	url = {https://www.nature.com/articles/43854},
	doi = {10.1038/43854},
	language = {en},
	number = {6751},
	urldate = {2026-06-11},
	journal = {Nature},
	author = {Saji, N. H. and Goswami, B. N. and Vinayachandran, P. N. and Yamagata, T.},
	month = sep,
	year = {1999},
	pages = {360--363},
}

@article{webster1999,
	title = {Coupled ocean–atmosphere dynamics in the {Indian} {Ocean} during 1997–98},
	volume = {401},
	copyright = {http://www.springer.com/tdm},
	issn = {0028-0836, 1476-4687},
	url = {https://www.nature.com/articles/43848},
	doi = {10.1038/43848},
	language = {en},
	number = {6751},
	urldate = {2026-06-11},
	journal = {Nature},
	author = {Webster, Peter J. and Moore, Andrew M. and Loschnigg, Johannes P. and Leben, Robert R.},
	month = sep,
	year = {1999},
	pages = {356--360},
}

@incollection{yamagata2013,
	address = {Washington, D. C.},
	title = {Coupled {Ocean}-{Atmosphere} {Variability} in the {Tropical} {Indian} {Ocean}},
	isbn = {978-1-118-66594-7 978-0-87590-412-2},
	url = {https://onlinelibrary.wiley.com/doi/10.1029/147GM12},
	doi = {10.1029/147GM12},
	urldate = {2026-06-11},
	booktitle = {Geophysical {Monograph} {Series}},
	publisher = {American Geophysical Union},
	author = {Yamagata, Toshio and Behera, Swadhin K. and Luo, Jing-Jia and Masson, Sebastien and Jury, Mark R. and Rao, Suryachandra A.},
	editor = {Wang, C. and Xie, S.P. and Carton, J.A.},
	month = mar,
	year = {2013},
	pages = {189--211},
}

@article{liu2024,
	title = {Indian {Ocean} {Dipole} {Changes} {During} the {Last} {Interglacial} {Modulated} by the {Mean} {Oceanic} {Climatology}},
	volume = {51},
	issn = {0094-8276, 1944-8007},
	url = {https://agupubs.onlinelibrary.wiley.com/doi/10.1029/2023GL106153},
	doi = {10.1029/2023GL106153},
	language = {en},
	number = {1},
	urldate = {2026-06-11},
	journal = {Geophysical Research Letters},
	author = {Liu, Shanshan and Yuan, Chaoxia and Behera, Swadhin and Luo, Jing‐Jia and Yamagata, Toshio},
	month = jan,
	year = {2024},
	pages = {e2023GL106153},
}

@misc{subich2025,
	title = {Fixing the {Double} {Penalty} in {Data}-{Driven} {Weather} {Forecasting} {Through} a {Modified} {Spherical} {Harmonic} {Loss} {Function}},
	url = {http://arxiv.org/abs/2501.19374},
	doi = {10.48550/arXiv.2501.19374},
	language = {en},
	urldate = {2026-06-11},
	publisher = {arXiv},
	author = {Subich, Christopher and Husain, Syed Zahid and Separovic, Leo and Yang, Jing},
	month = may,
	year = {2025},
	note = {arXiv:2501.19374 [cs.LG]},
}

@misc{brolly2026,
	title = {Stochasticity and probabilistic trajectory scoring are essential for data-driven closures of chaotic systems},
	url = {http://arxiv.org/abs/2603.28671},
	doi = {10.48550/arXiv.2603.28671},
	language = {en},
	urldate = {2026-06-11},
	publisher = {arXiv},
	author = {Brolly, Martin Thomas},
	month = mar,
	year = {2026},
	note = {arXiv:2603.28671 [math.DS]},
}

@article{brier1950,
	title = {{VERIFICATION} {OF} {FORECASTS} {EXPRESSED} {IN} {TERMS} {OF} {PROBABILITY}},
	volume = {78},
	issn = {0027-0644, 1520-0493},
	url = {http://journals.ametsoc.org/doi/10.1175/1520-0493(1950)078<0001:VOFEIT>2.0.CO;2},
	doi = {10.1175/1520-0493(1950)078<0001:VOFEIT>2.0.CO;2},
	language = {en},
	number = {1},
	urldate = {2026-06-11},
	journal = {Monthly Weather Review},
	author = {Brier, Glenn W.},
	month = jan,
	year = {1950},
	pages = {1--3},
}

@article{li2024,
	title = {Generative emulation of weather forecast ensembles with diffusion models},
	volume = {10},
	issn = {2375-2548},
	url = {https://www.science.org/doi/10.1126/sciadv.adk4489},
	doi = {10.1126/sciadv.adk4489},
	language = {en},
	number = {13},
	urldate = {2026-07-29},
	journal = {Science Advances},
	author = {Li, Lizao and Carver, Robert and Lopez-Gomez, Ignacio and Sha, Fei and Anderson, John},
	month = mar,
	year = {2024},
	pages = {eadk4489},
}

@article{lorenz1969,
	title = {Atmospheric {Predictability} as {Revealed} by {Naturally} {Occurring} {Analogues}},
	volume = {26},
	issn = {0022-4928, 1520-0469},
	url = {http://journals.ametsoc.org/doi/10.1175/1520-0469(1969)26<636:APARBN>2.0.CO;2},
	doi = {10.1175/1520-0469(1969)26<636:APARBN>2.0.CO;2},
	language = {en},
	number = {4},
	urldate = {2026-07-29},
	journal = {Journal of the Atmospheric Sciences},
	author = {Lorenz, Edward N.},
	month = jul,
	year = {1969},
	pages = {636--646},
}

@article{larson2024,
	title = {Signature of the western boundary currents in local climate variability},
	volume = {634},
	issn = {0028-0836, 1476-4687},
	url = {https://www.nature.com/articles/s41586-024-08019-2},
	doi = {10.1038/s41586-024-08019-2},
	language = {en},
	number = {8035},
	urldate = {2026-08-05},
	journal = {Nature},
	author = {Larson, James G. and Thompson, David W. J. and Hurrell, James W.},
	month = oct,
	year = {2024},
	pages = {862--867},
}

@article{sohail2025,
	title = {Decline of {Antarctic} {Circumpolar} {Current} due to polar ocean freshening},
	volume = {20},
	issn = {1748-9326},
	url = {https://iopscience.iop.org/article/10.1088/1748-9326/adb31c},
	doi = {10.1088/1748-9326/adb31c},
	number = {3},
	urldate = {2026-08-05},
	journal = {Environmental Research Letters},
	author = {Sohail, Taimoor and Gayen, Bishakhdatta and Klocker, Andreas},
	month = mar,
	year = {2025},
	pages = {034046},
}

@article{germe2022,
	title = {Chaotic {Variability} of the {Atlantic} {Meridional} {Overturning} {Circulation} at {Subannual} {Time} {Scales}},
	volume = {52},
	copyright = {http://www.ametsoc.org/PUBSReuseLicenses},
	issn = {0022-3670, 1520-0485},
	url = {https://journals.ametsoc.org/view/journals/phoc/52/5/JPO-D-21-0100.1.xml},
	doi = {10.1175/JPO-D-21-0100.1},
	number = {5},
	urldate = {2026-08-05},
	journal = {Journal of Physical Oceanography},
	author = {Germe, Agathe and Hirschi, Joël J.-M. and Blaker, Adam T. and Sinha, Bablu},
	month = may,
	year = {2022},
	pages = {929--949},
}

\appendix
\section{Appendix}
\label{sect:appendix}
\counterwithin{figure}{section}

In the Appendix section, we leave additional results for the Neptune global ocean model.

\subsection{Statistical Evaluation}
\label{sect:appendix_statistical_evaluation}


\begin{figure}
    \centering
    \includegraphics[width=\textwidth]{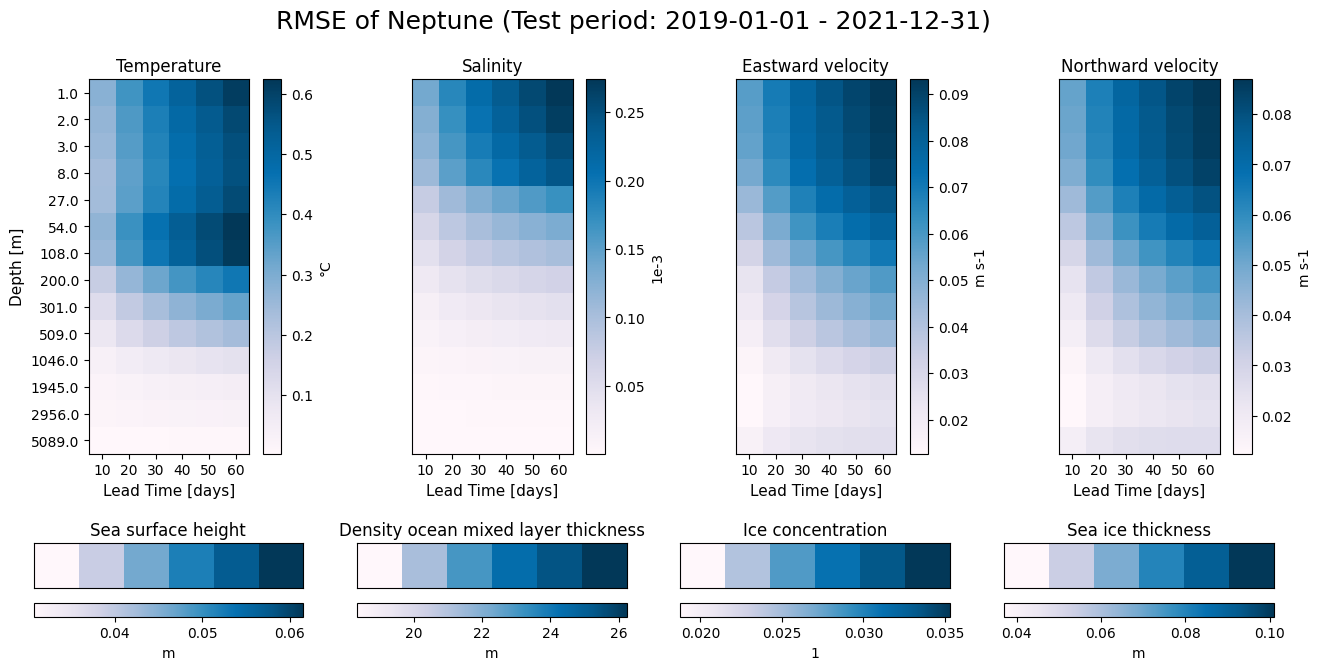}
    \caption{
    Neptune RMSE scorecard for lead times spanning from t+10 to t+60 days of forecast. Lower RMSE values are represented in white, while higher values are reported in blue.
    }
    \label{fig:app_rmse_1}
\end{figure}

\begin{figure}
    \centering
    \includegraphics[width=\textwidth]{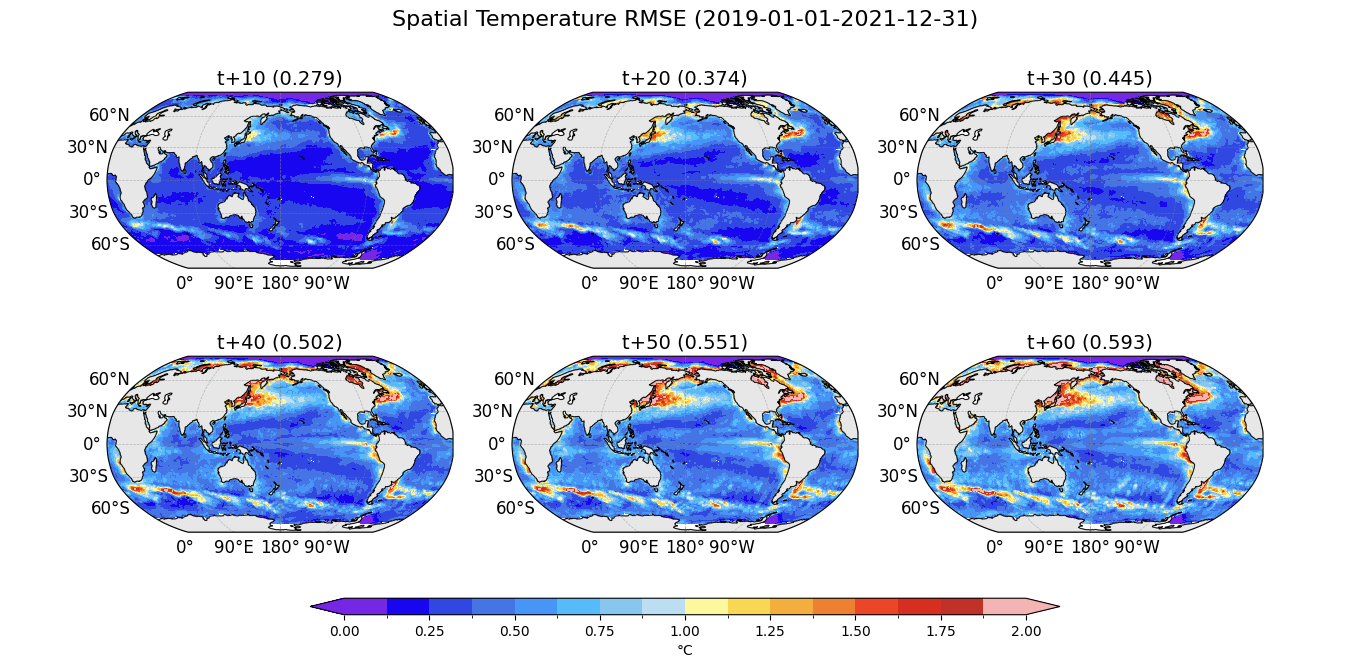}
    \caption{
    Neptune Temperature spatial RMSE error for lead times spanning from t+10 to t+60 days of forecast.
    }
    \label{fig:app_rmse_2}
\end{figure}

\begin{figure}
    \centering
    \includegraphics[width=\textwidth]{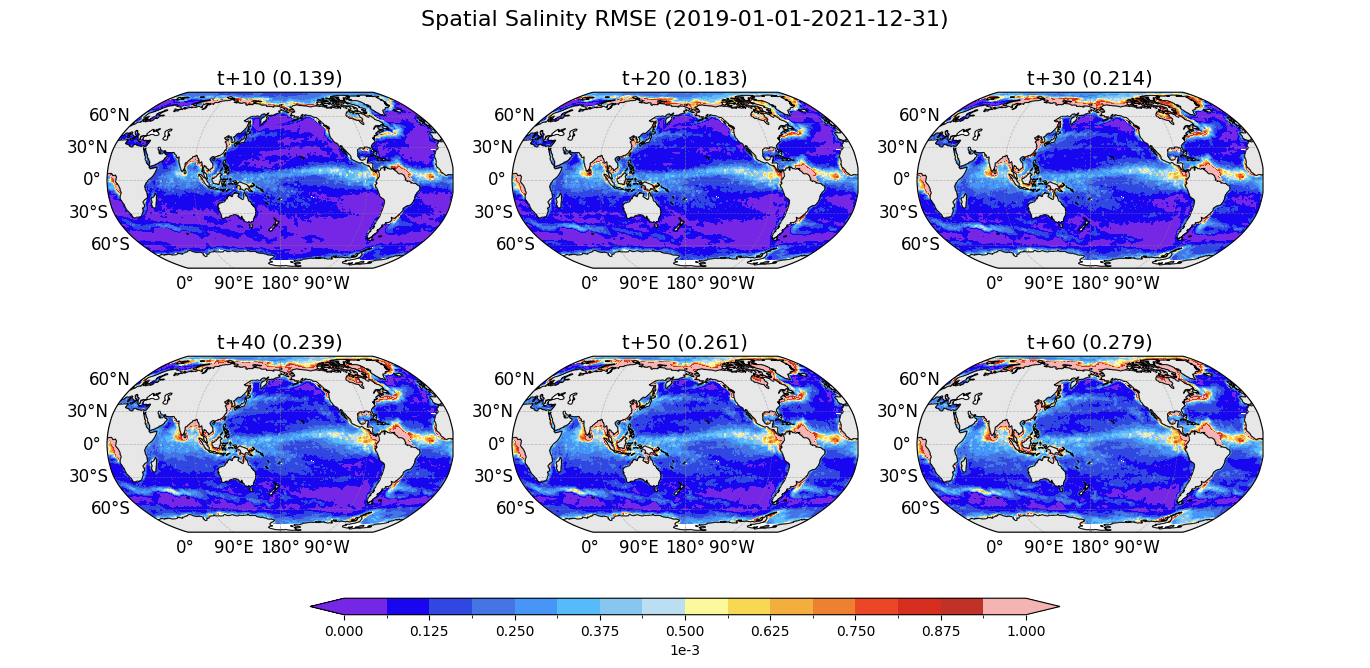}
    \caption{
    Neptune Salinity spatial RMSE error for lead times spanning from t+10 to t+60 days of forecast.
    }
    \label{fig:app_rmse_3}
\end{figure}

\begin{figure}
    \centering
    \includegraphics[width=\textwidth]{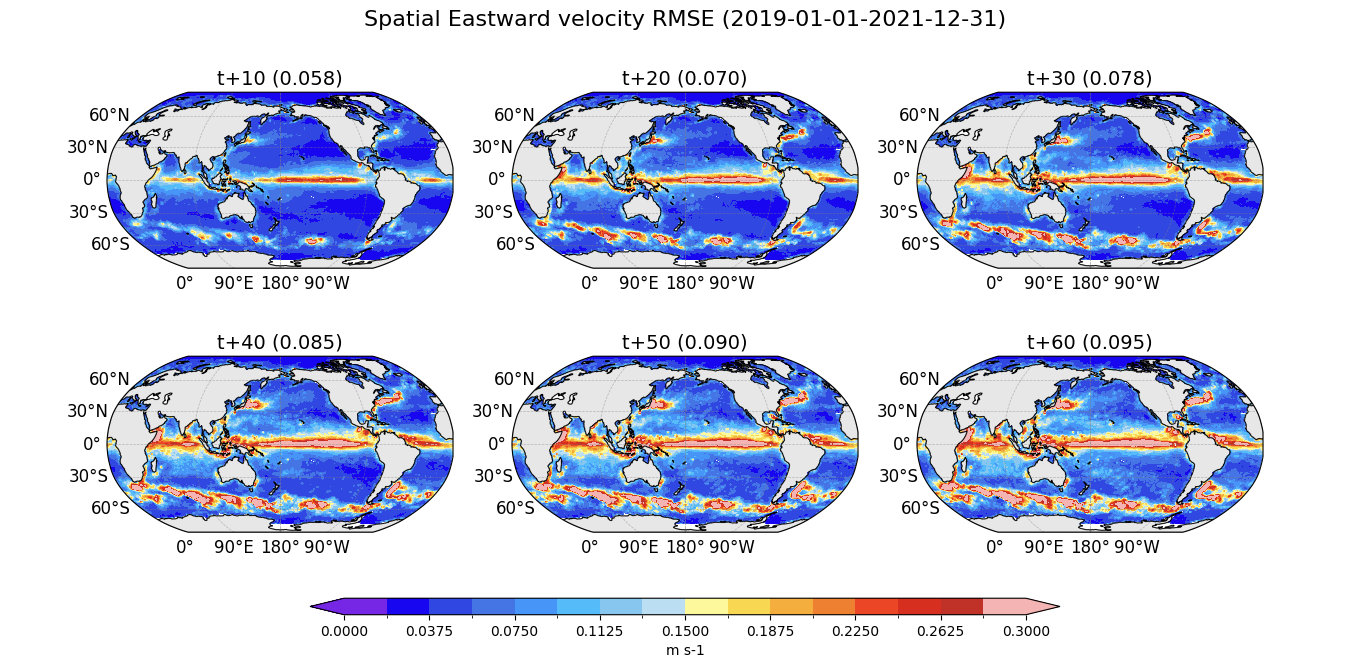}
    \caption{
    Neptune Eastward velocity spatial RMSE error for lead times spanning from t+10 to t+60 days of forecast.
    }
    \label{fig:app_rmse_4}
\end{figure}

\begin{figure}
    \centering
    \includegraphics[width=\textwidth]{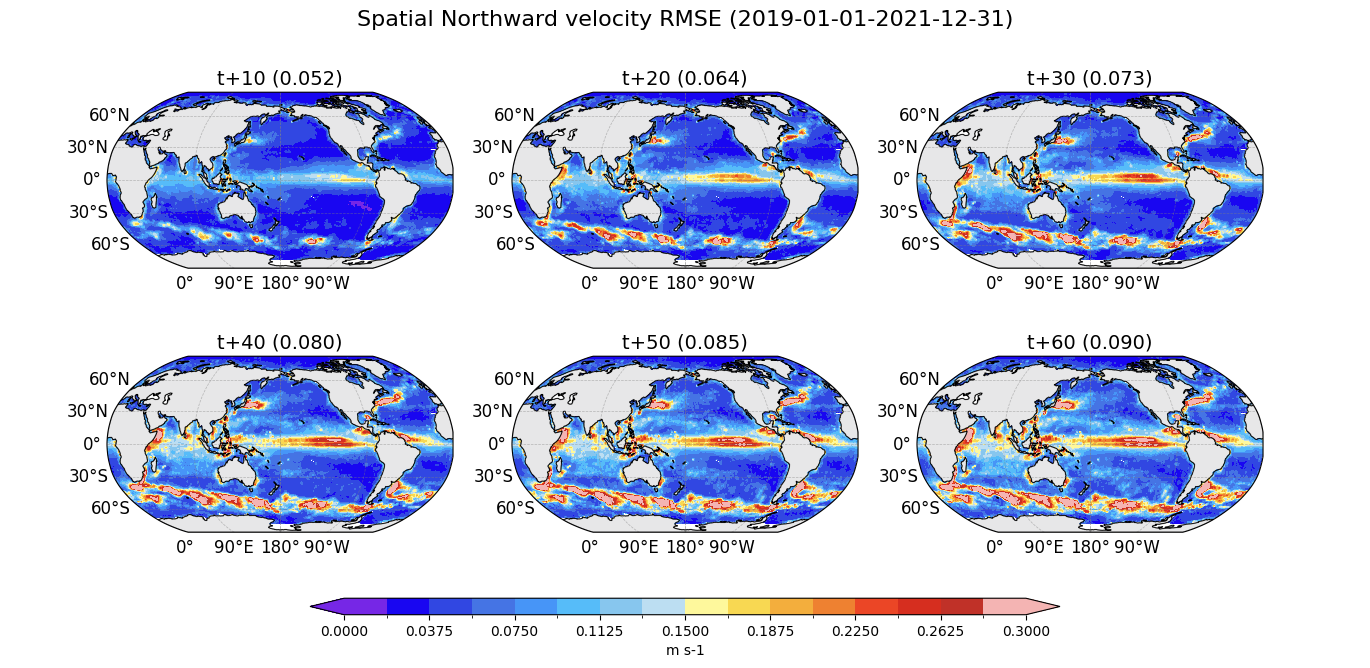}
    \caption{
    Neptune Northward velocity spatial RMSE error for lead times spanning from t+10 to t+60 days of forecast.
    }
    \label{fig:app_rmse_5}
\end{figure}

\begin{figure}
    \centering
    \includegraphics[width=\textwidth]{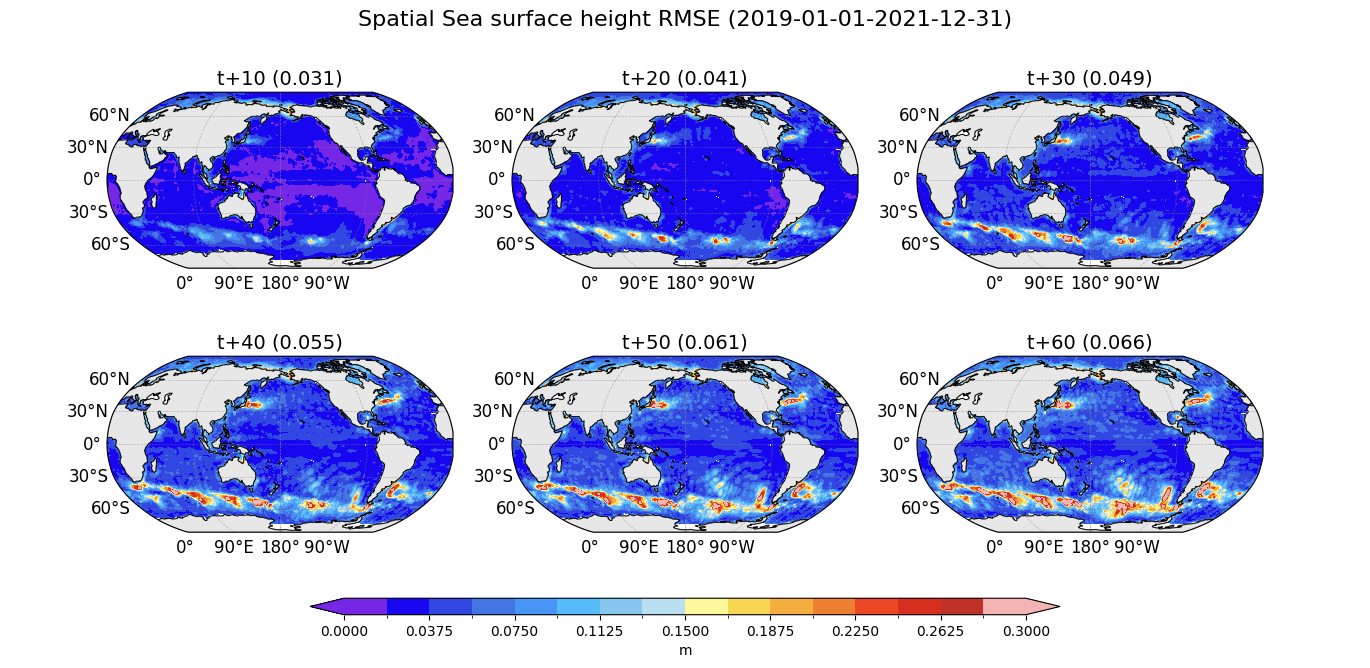}
    \caption{
    Neptune Sea Surface Height spatial RMSE error for lead times spanning from t+10 to t+60 days of forecast.
    }
    \label{fig:app_rmse_6}
\end{figure}

\begin{figure}
    \centering
    \includegraphics[width=\textwidth]{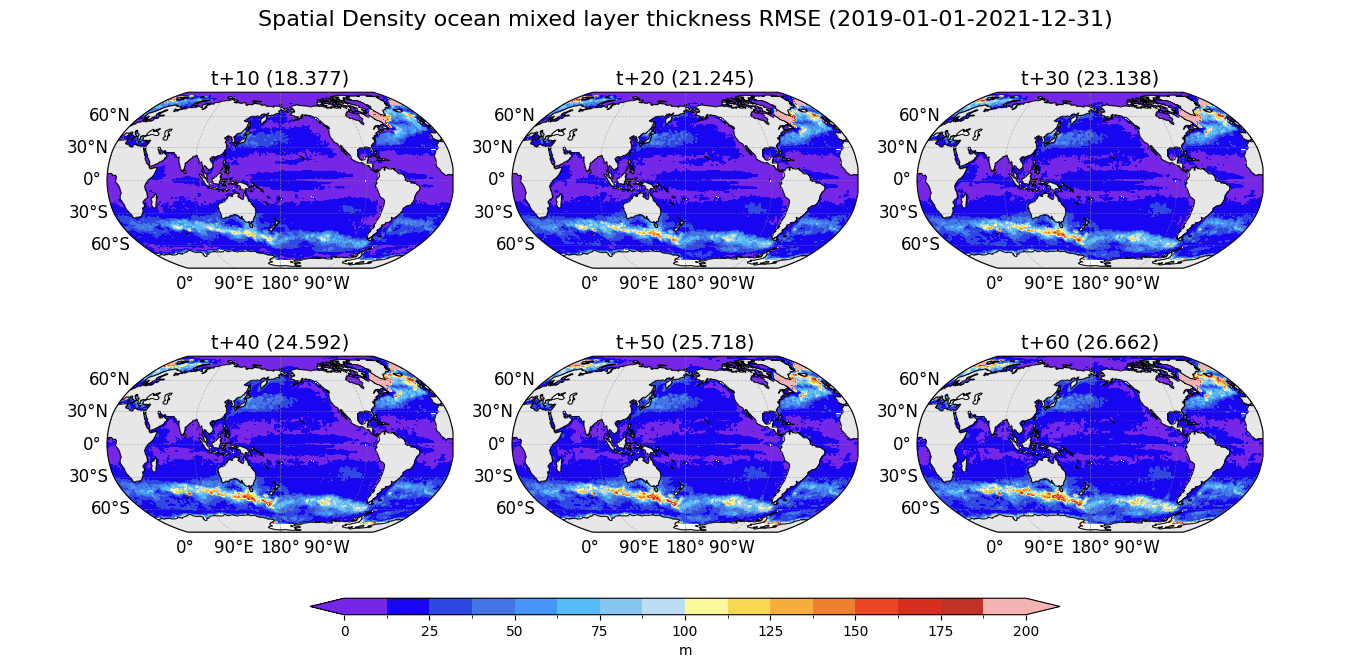}
    \caption{
    Neptune Mixed layer depth spatial RMSE error for lead times spanning from t+10 to t+60 days of forecast.
    }
    \label{fig:app_rmse_7}
\end{figure}

\begin{figure}
    \centering
    \includegraphics[width=\textwidth]{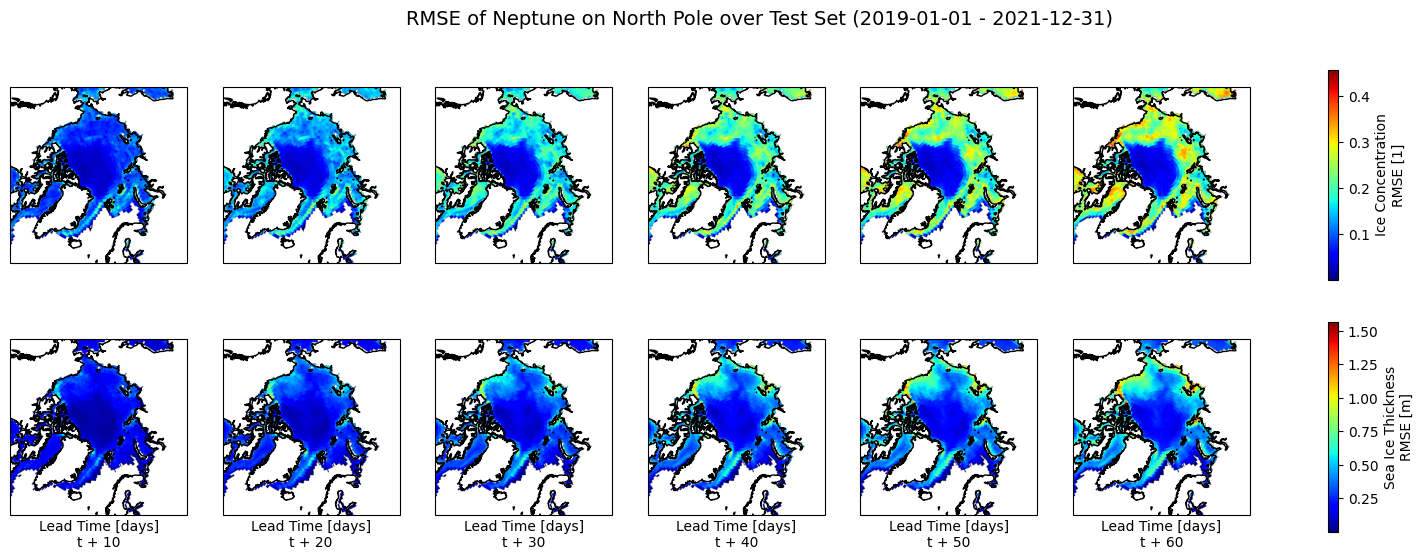}
    \caption{
    Neptune Sea Ice Concentration and Thickness (top and bottom row, respectively) spatial RMSE error over North Pole for lead times spanning from t+10 to t+60 days of forecast.
    }
    \label{fig:app_rmse_8}
\end{figure}

\begin{figure}
    \centering
    \includegraphics[width=\textwidth]{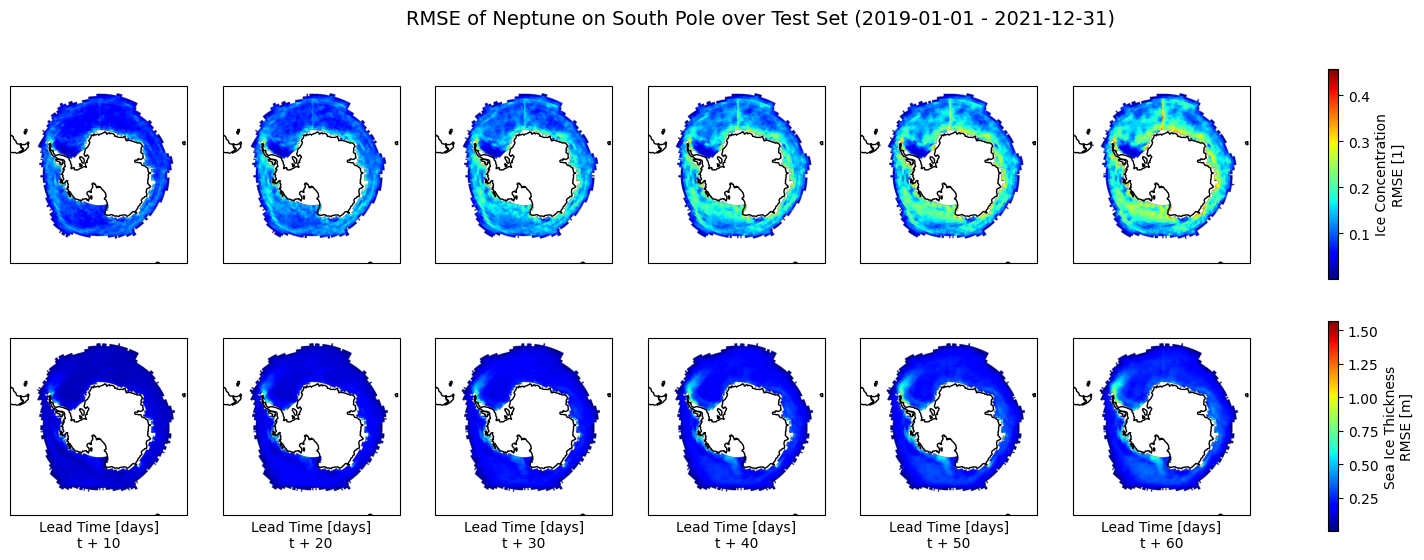}
    \caption{
    Neptune Sea Ice Concentration and Thickness (top and bottom row, respectively) spatial RMSE error over South Pole for lead times spanning from t+10 to t+60 days of forecast.
    }
    \label{fig:app_rmse_9}
\end{figure}


\begin{figure}
    \centering
    \includegraphics[width=\textwidth]{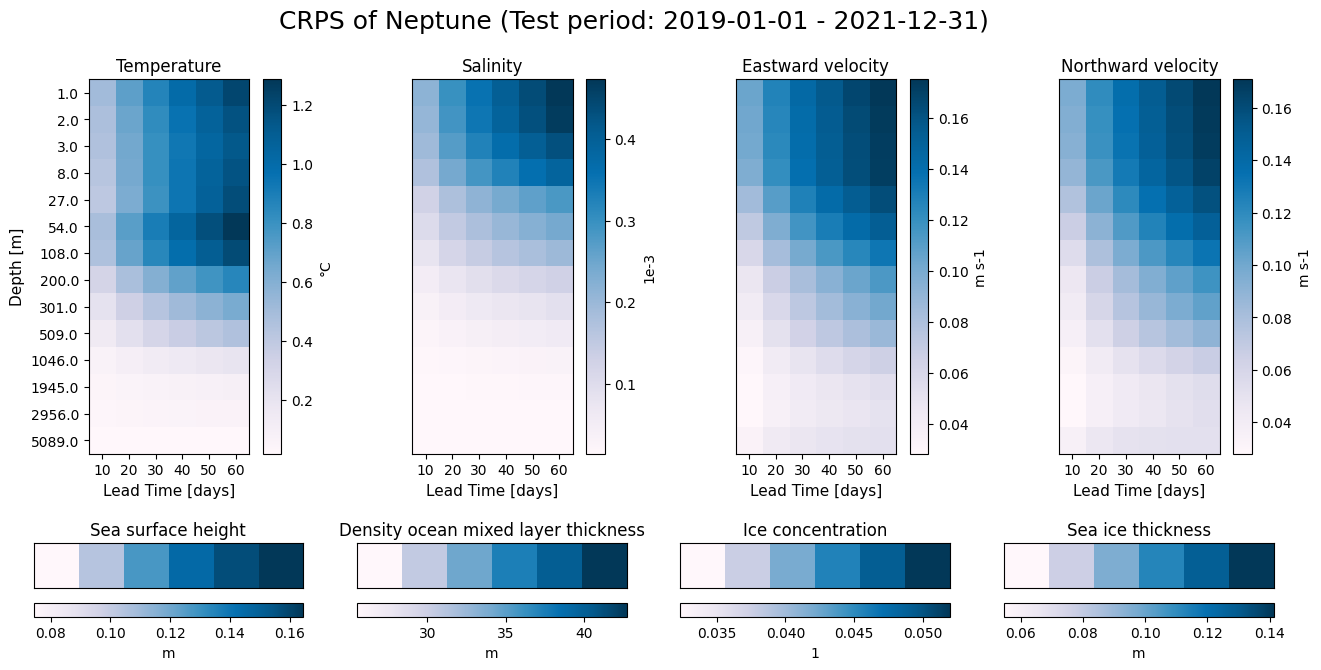}
    \caption{
    Neptune CRPS scorecard for lead times spanning from t+10 to t+60 days of forecast. Lower CRPS values are represented in white, while higher values are reported in blue.
    }
    \label{fig:app_crps_1}
\end{figure}

\begin{figure}
    \centering
    \includegraphics[width=\textwidth]{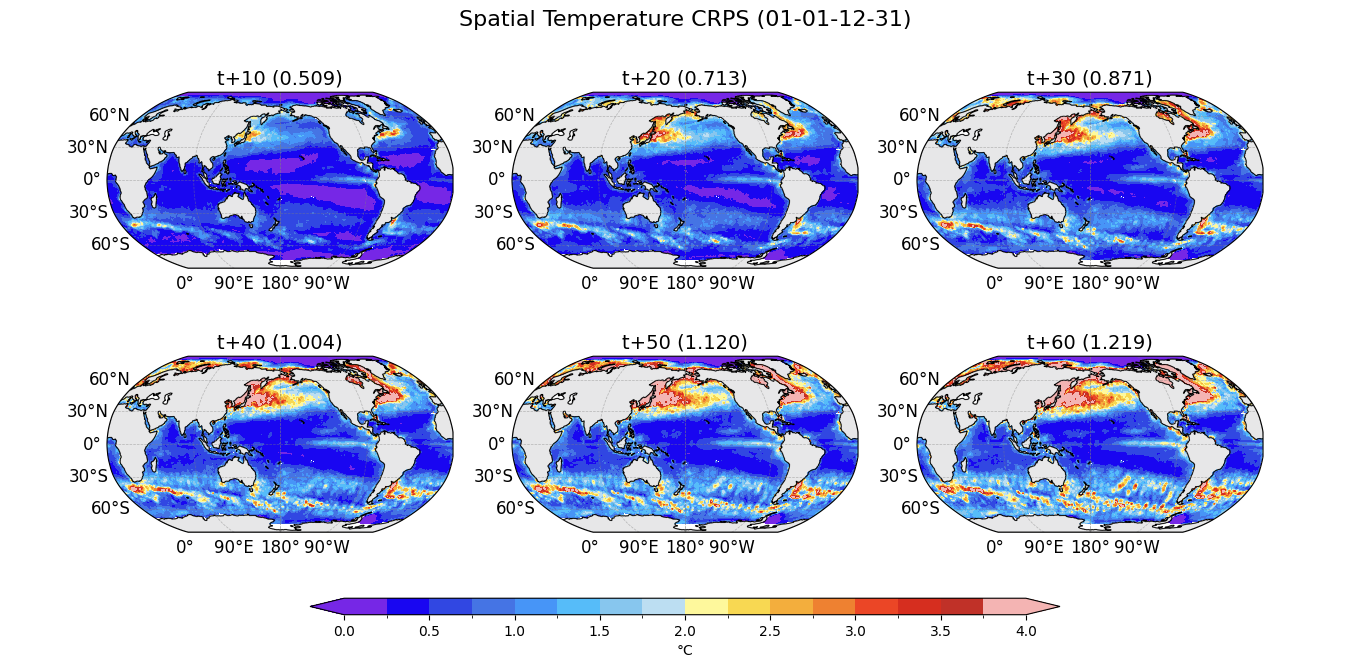}
    \caption{
    Neptune Temperature spatial CRPS error for lead times spanning from t+10 to t+60 days of forecast.
    }
    \label{fig:app_crps_2}
\end{figure}

\begin{figure}
    \centering
    \includegraphics[width=\textwidth]{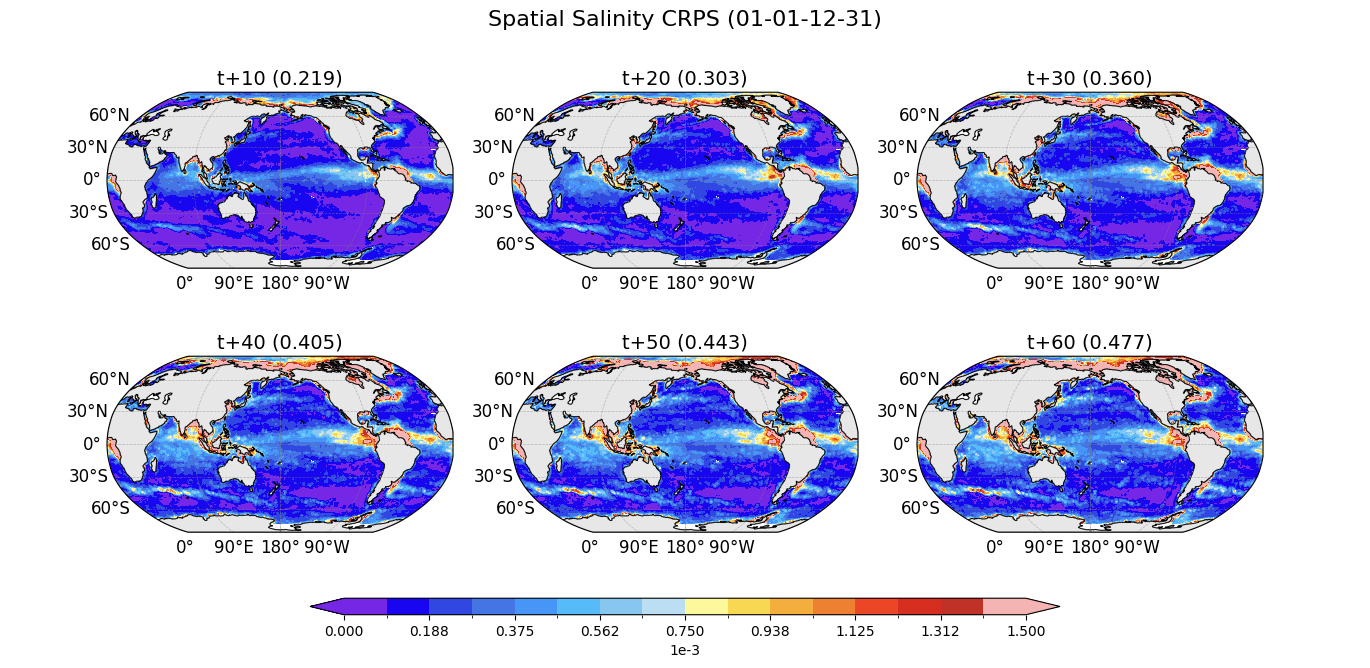}
    \caption{
    Neptune Salinity spatial CRPS error for lead times spanning from t+10 to t+60 days of forecast.
    }
    \label{fig:app_crps_3}
\end{figure}

\begin{figure}
    \centering
    \includegraphics[width=\textwidth]{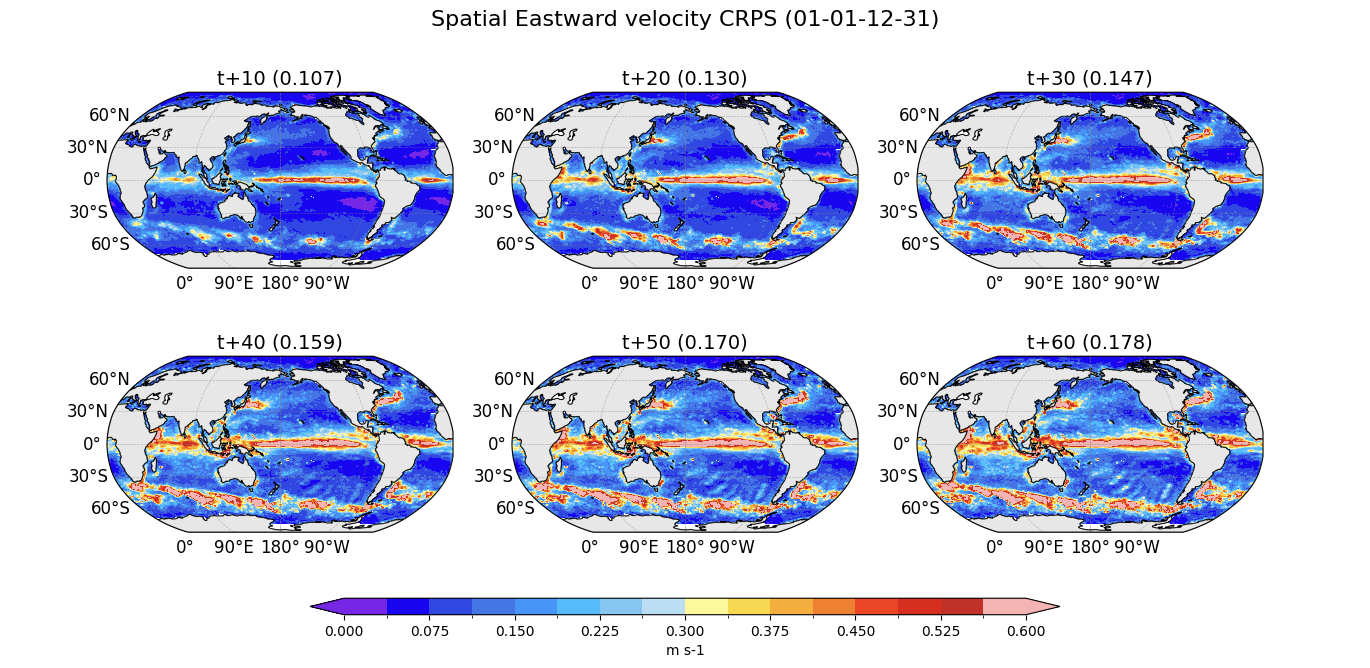}
    \caption{
    Neptune Meridional velocity spatial CRPS error for lead times spanning from t+10 to t+60 days of forecast.
    }
    \label{fig:app_crps_4}
\end{figure}

\begin{figure}
    \centering
    \includegraphics[width=\textwidth]{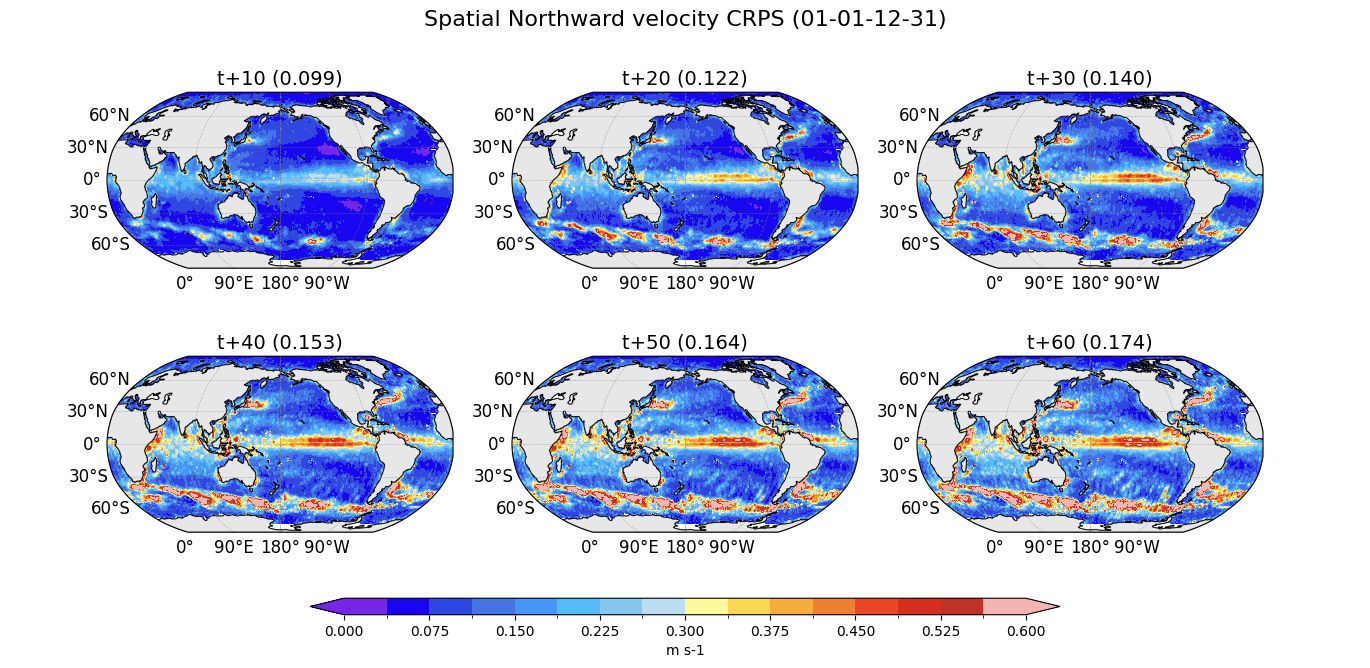}
    \caption{
    Neptune Northward velocity spatial CRPS error for lead times spanning from t+10 to t+60 days of forecast.
    }
    \label{fig:app_crps_5}
\end{figure}

\begin{figure}
    \centering
    \includegraphics[width=\textwidth]{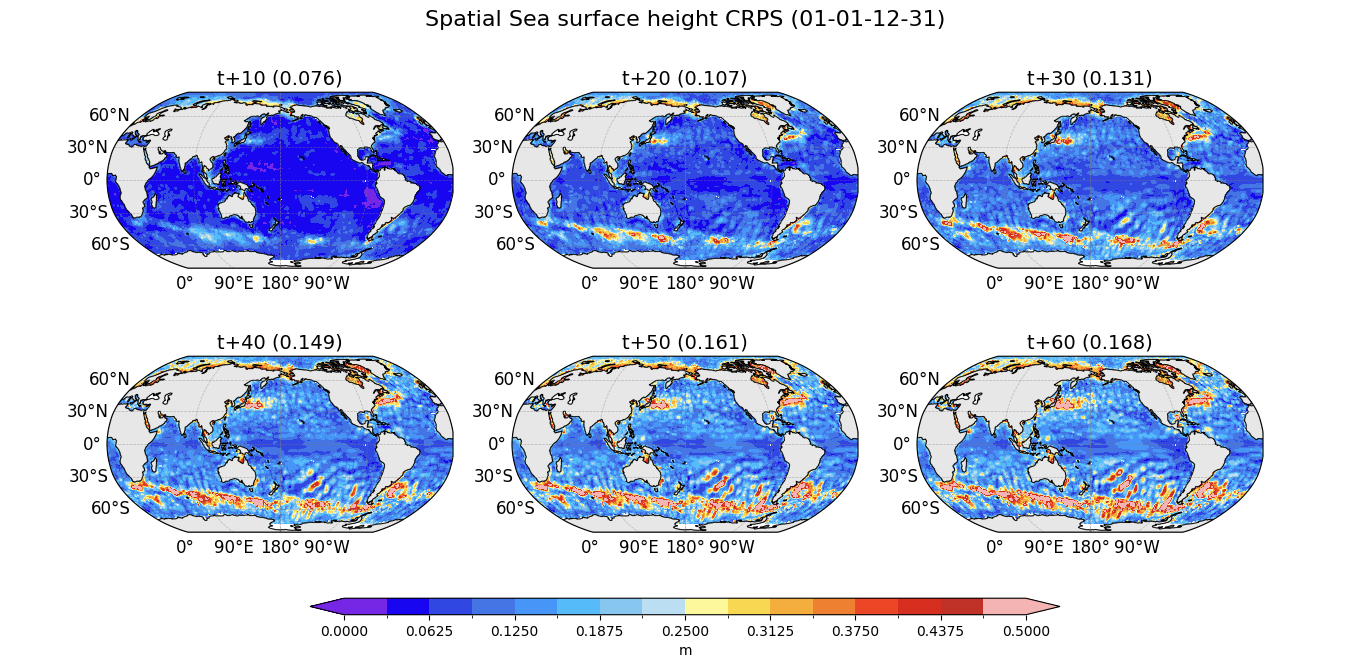}
    \caption{
    Neptune Sea Surface Height spatial CRPS error for lead times spanning from t+10 to t+60 days of forecast.
    }
    \label{fig:app_crps_6}
\end{figure}

\begin{figure}
    \centering
    \includegraphics[width=\textwidth]{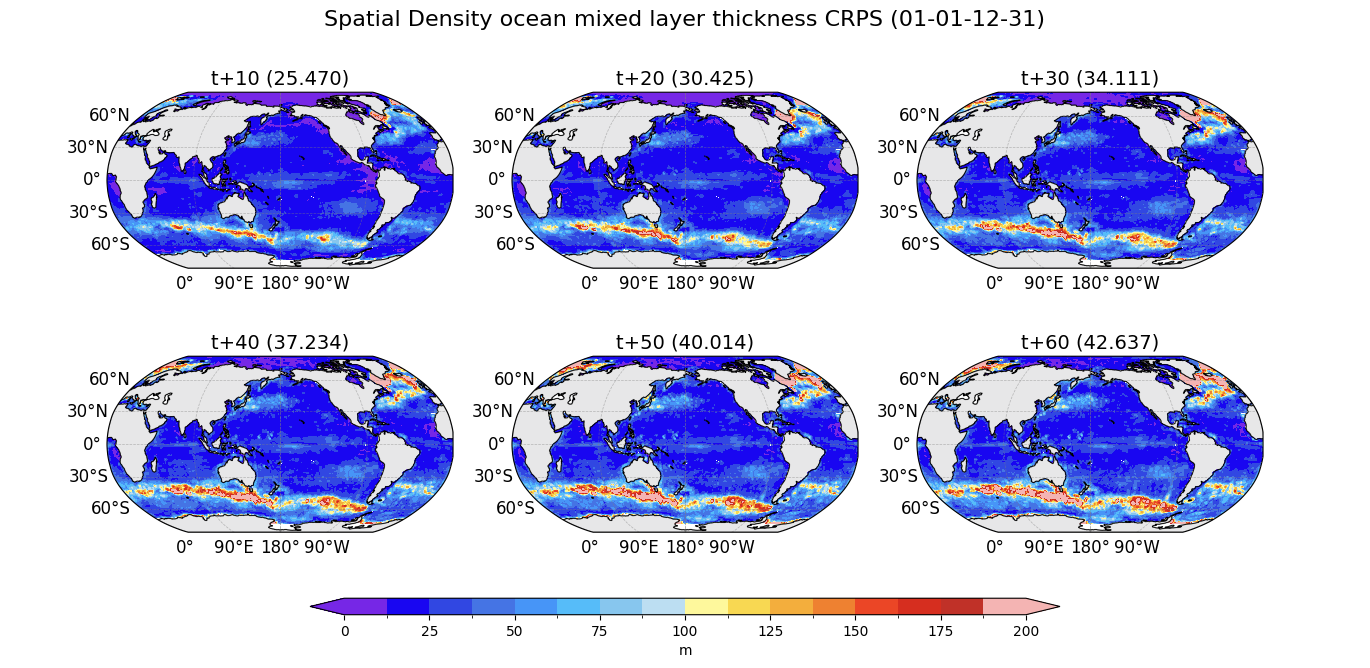}
    \caption{
    Neptune Mixed layer depth spatial CRPS error for lead times spanning from t+10 to t+60 days of forecast.
    }
    \label{fig:app_crps_7}
\end{figure}

\begin{figure}
    \centering
    \includegraphics[width=\textwidth]{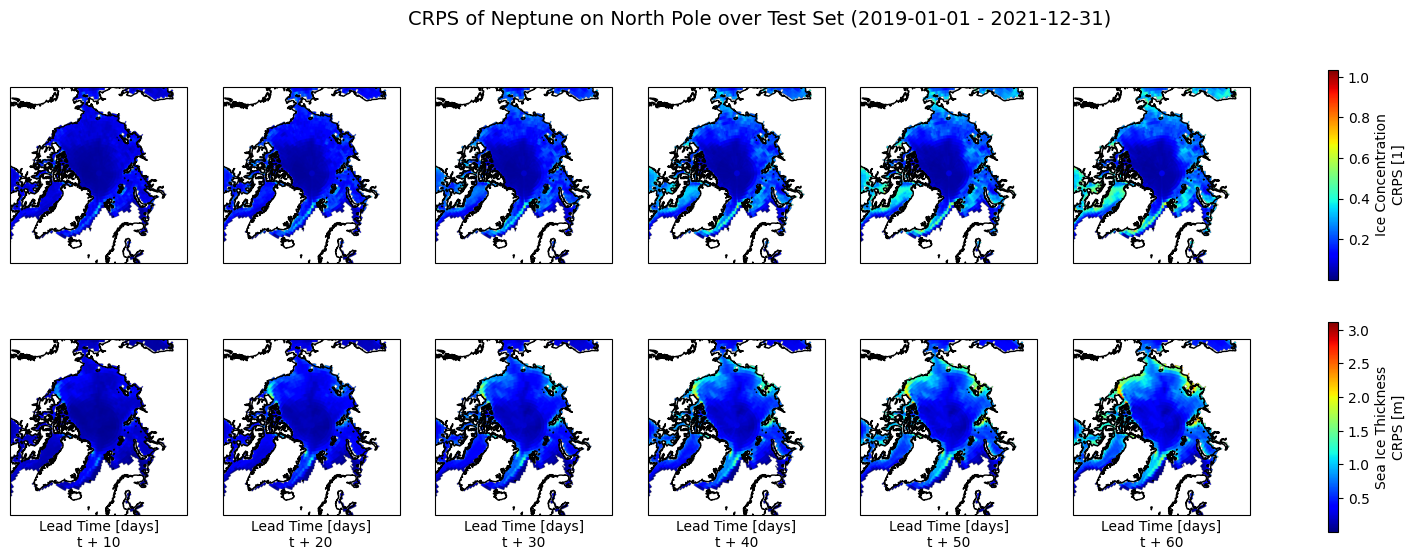}
    \caption{
    Neptune Sea Ice Concentration and Thickness (top and bottom row, respectively) spatial CRPS error over North Pole for lead times spanning from t+10 to t+60 days of forecast.
    }
    \label{fig:app_crps_8}
\end{figure}

\begin{figure}
    \centering
    \includegraphics[width=\textwidth]{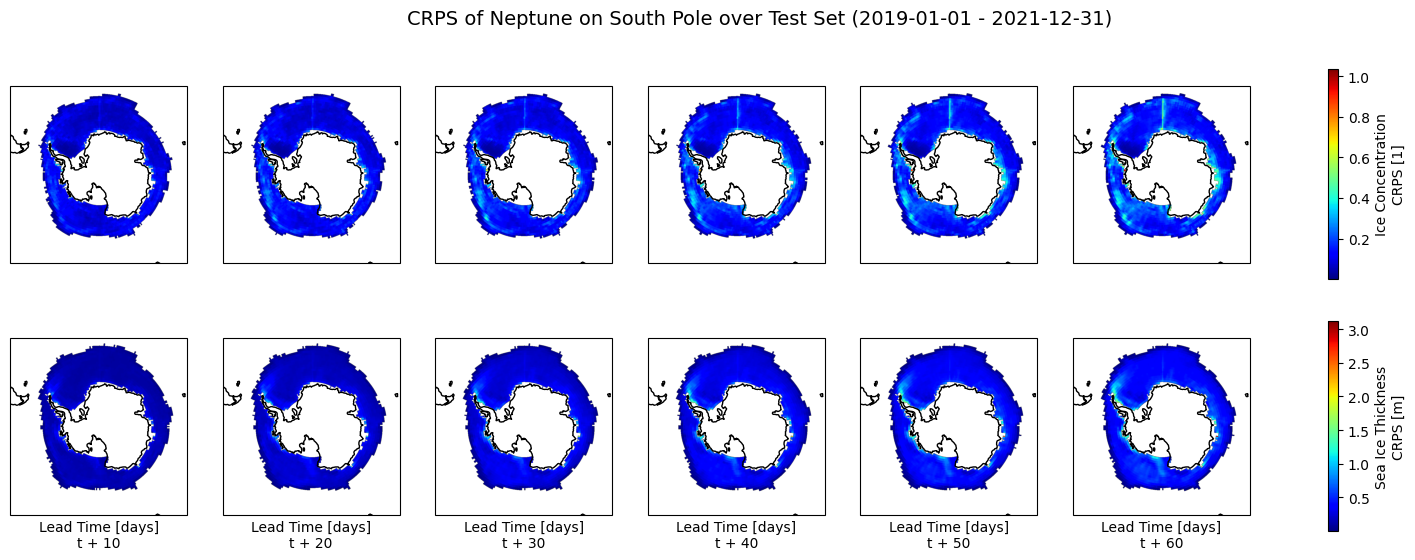}
    \caption{
    Neptune Sea Ice Concentration and Thickness (top and bottom row, respectively) spatial CRPS error over South Pole for lead times spanning from t+10 to t+60 days of forecast.
    }
    \label{fig:app_crps_9}
\end{figure}


\begin{figure}
    \centering
    \includegraphics[width=\textwidth]{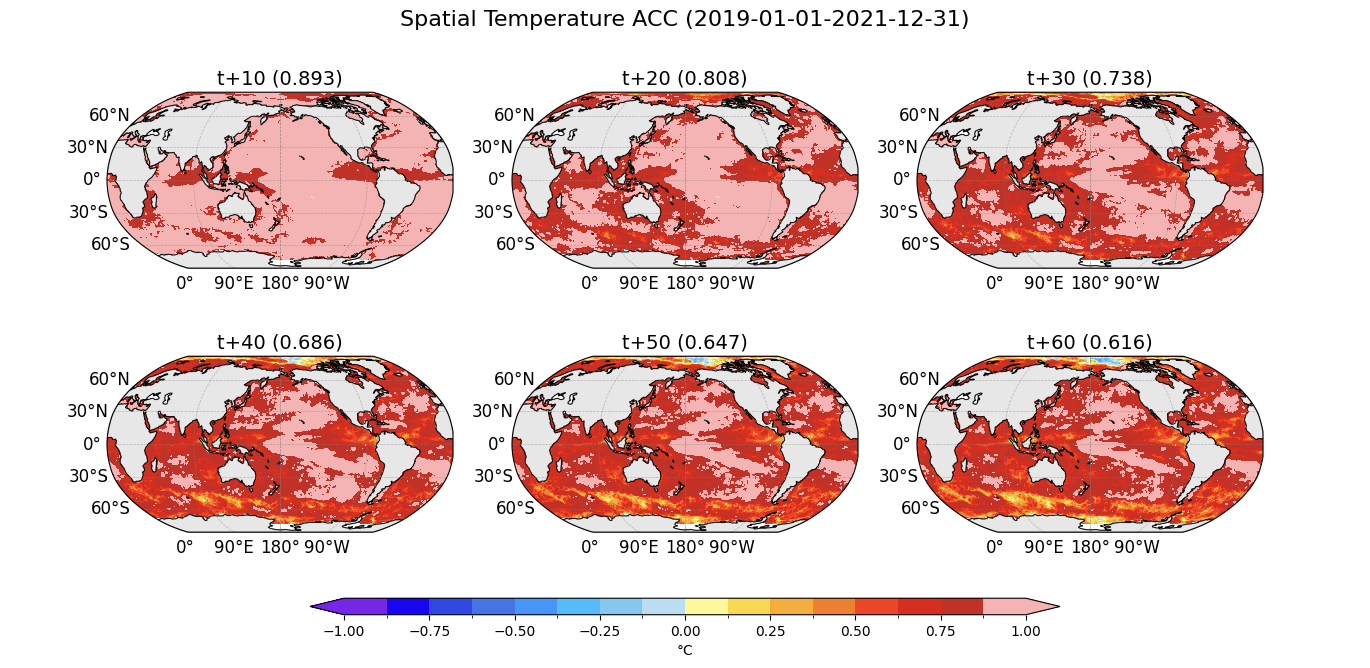}
    \caption{
    Neptune Temperature spatial ACC error for lead times spanning from t+10 to t+60 days of forecast.
    }
    \label{fig:app_acc_2}
\end{figure}

\begin{figure}
    \centering
    \includegraphics[width=\textwidth]{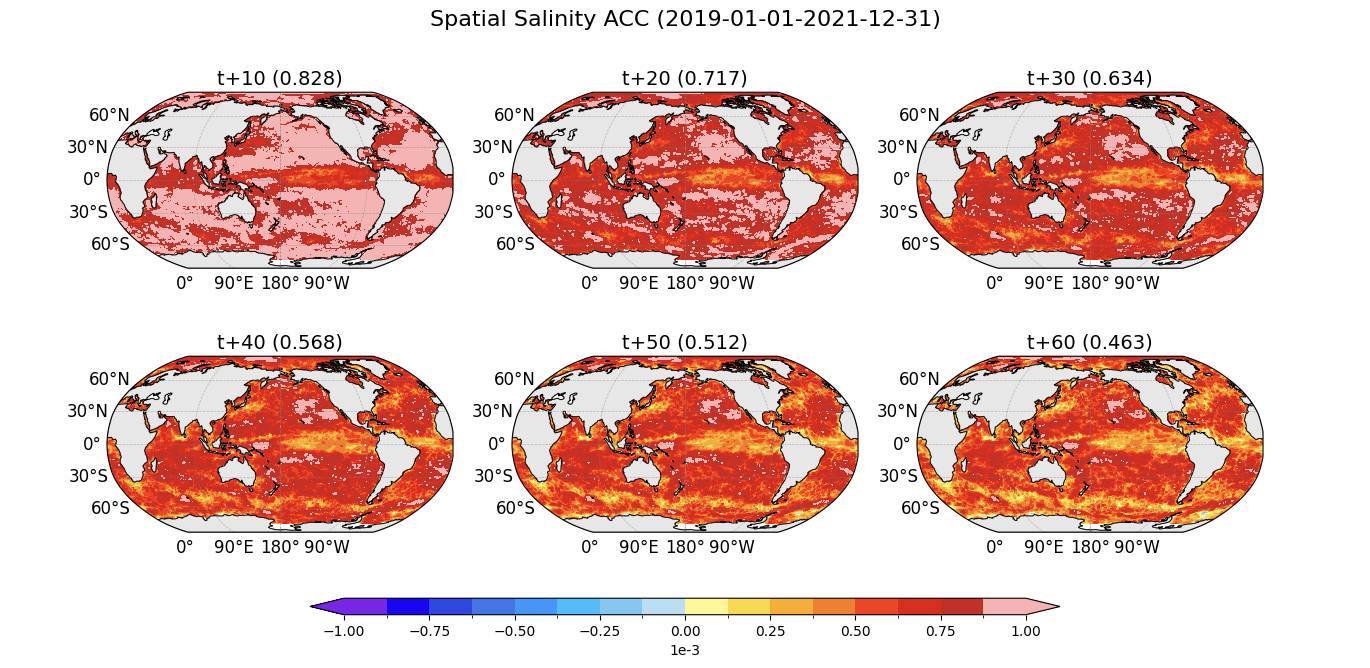}
    \caption{
    Neptune Salinity spatial ACC error for lead times spanning from t+10 to t+60 days of forecast.
    }
    \label{fig:app_acc_3}
\end{figure}

\begin{figure}
    \centering
    \includegraphics[width=\textwidth]{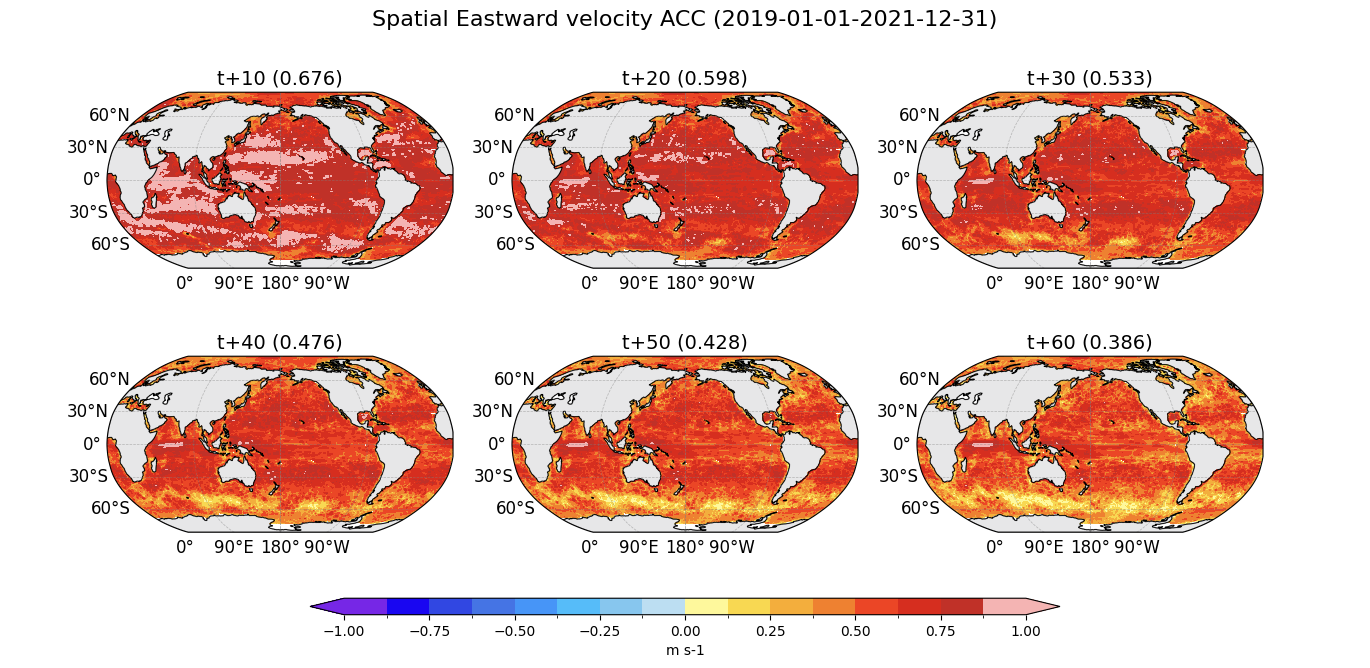}
    \caption{
    Neptune Eastward velocity spatial ACC error for lead times spanning from t+10 to t+60 days of forecast.
    }
    \label{fig:app_acc_4}
\end{figure}

\begin{figure}
    \centering
    \includegraphics[width=\textwidth]{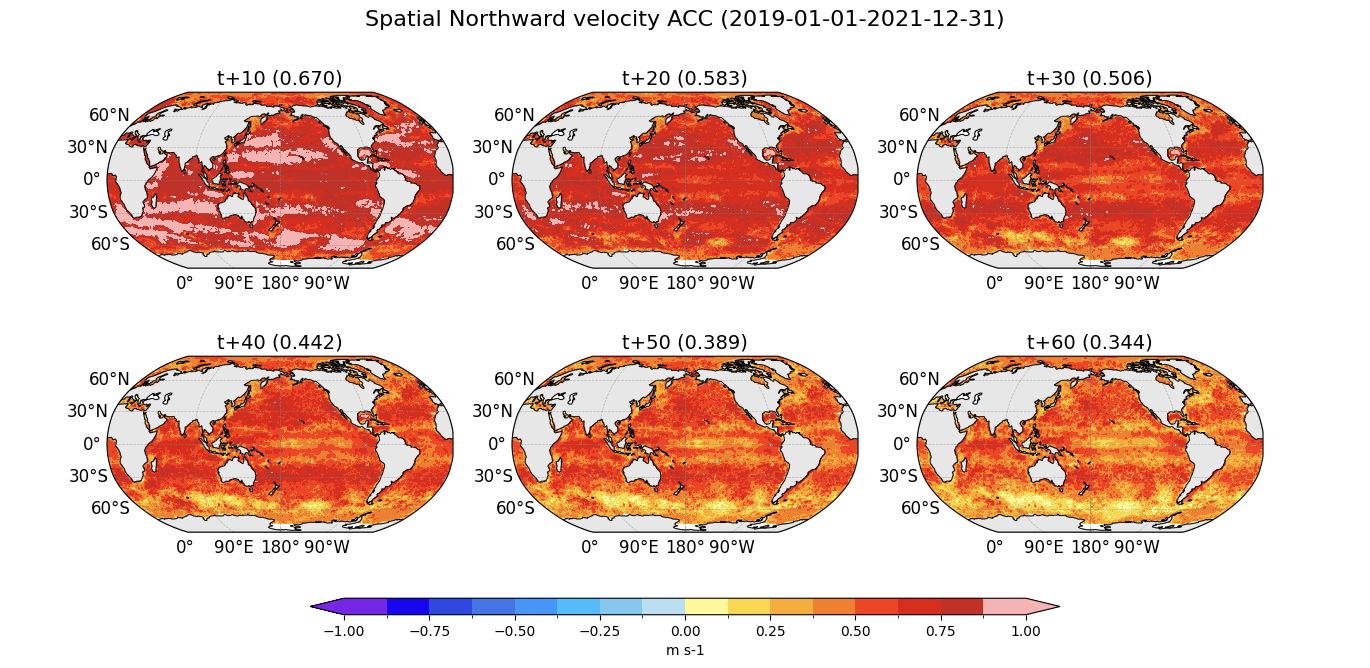}
    \caption{
    Neptune Northward velocity spatial ACC error for lead times spanning from t+10 to t+60 days of forecast.
    }
    \label{fig:app_acc_5}
\end{figure}

\begin{figure}
    \centering
    \includegraphics[width=\textwidth]{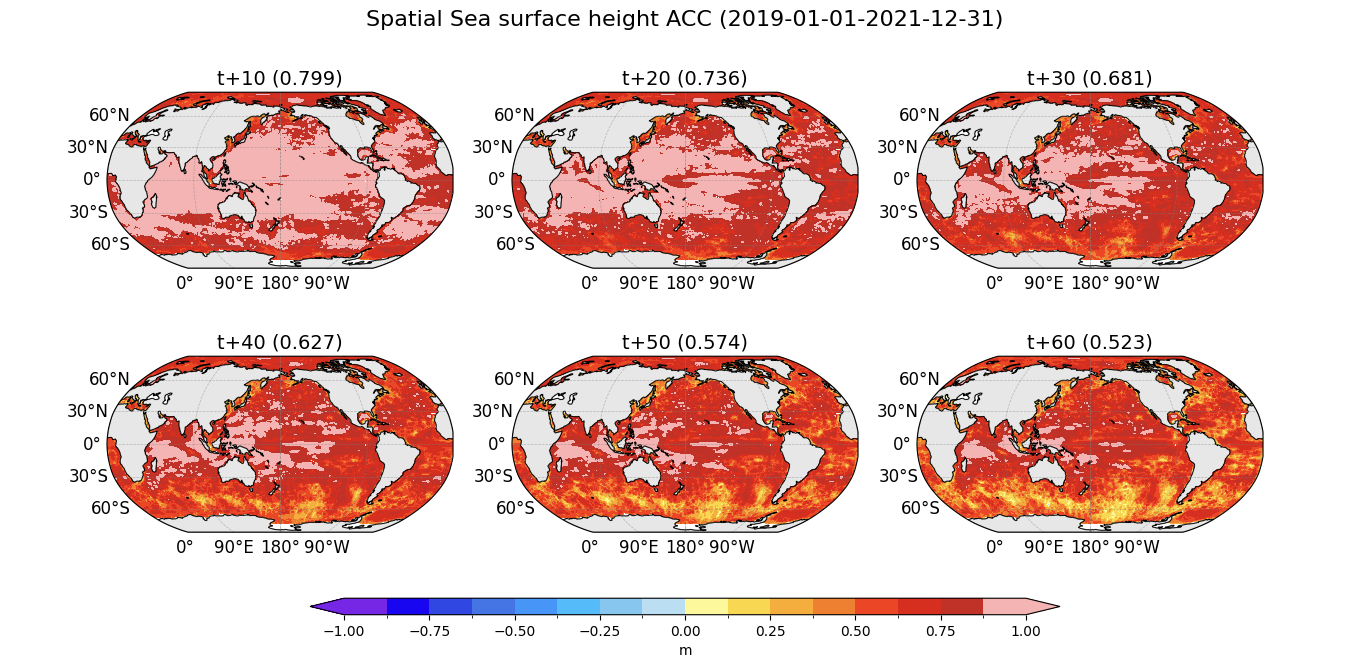}
    \caption{
    Neptune Sea Surface Height spatial ACC error for lead times spanning from t+10 to t+60 days of forecast.
    }
    \label{fig:app_acc_6}
\end{figure}

\begin{figure}
    \centering
    \includegraphics[width=\textwidth]{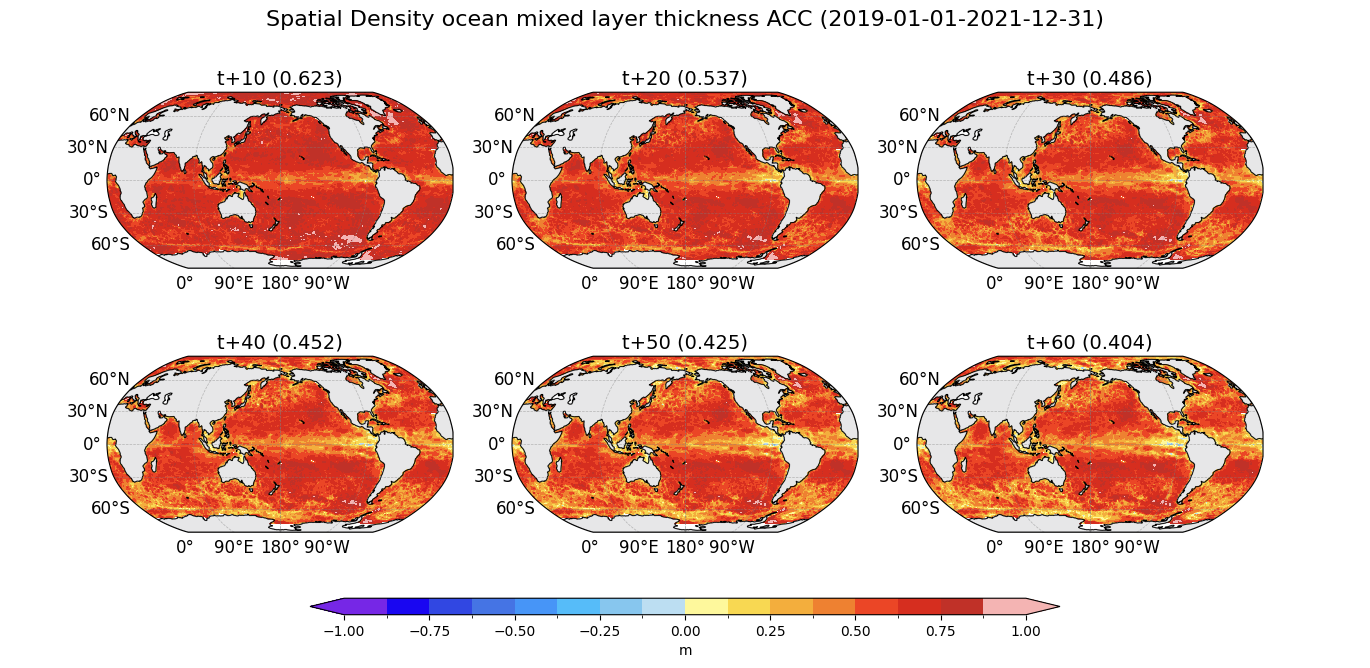}
    \caption{
    Neptune Mixed layer depth spatial ACC error for lead times spanning from t+10 to t+60 days of forecast.
    }
    \label{fig:app_acc_7}
\end{figure}

\begin{figure}
    \centering
    \includegraphics[width=\textwidth]{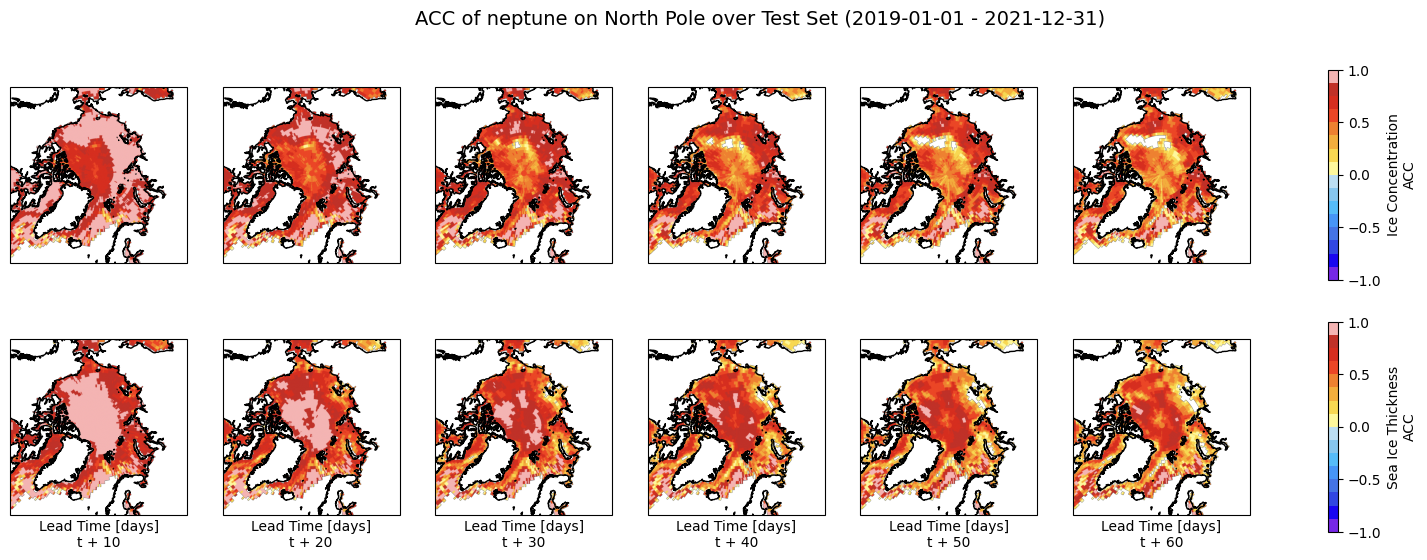}
    \caption{
    Neptune Sea Ice Concentration and Thickness (top and bottom row, respectively) spatial ACC error over North Pole for lead times spanning from t+10 to t+60 days of forecast.
    }
    \label{fig:app_acc_8}
\end{figure}

\begin{figure}
    \centering
    \includegraphics[width=\textwidth]{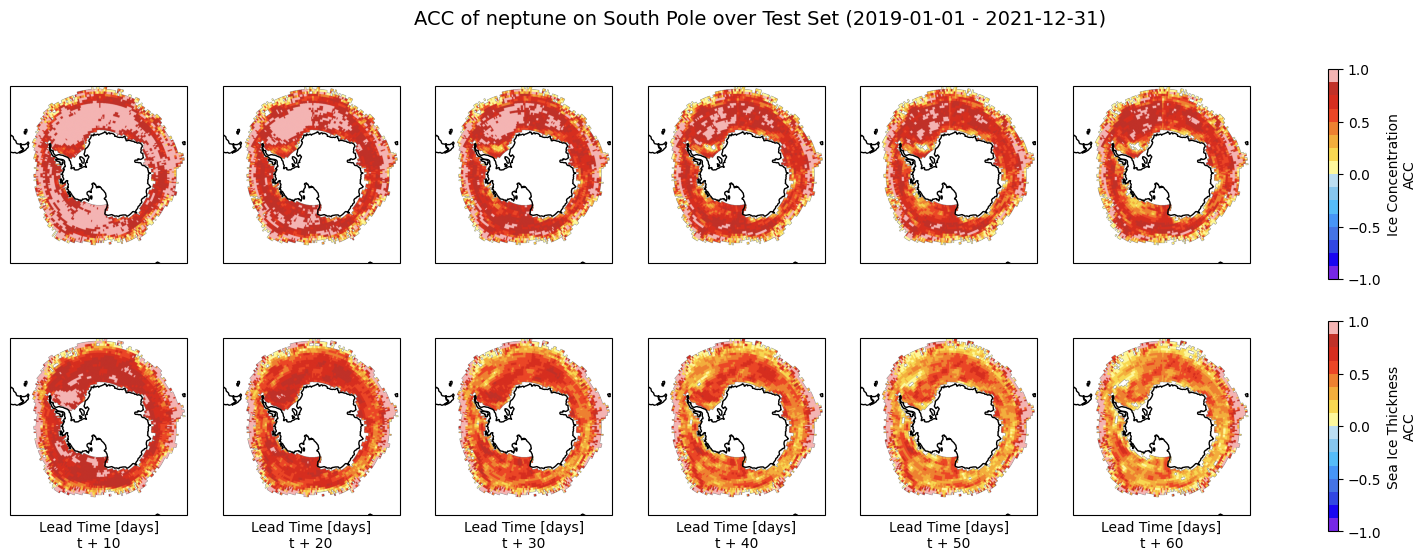}
    \caption{
    Neptune Sea Ice Concentration and Thickness (top and bottom row, respectively) spatial ACC error over South Pole for lead times spanning from t+10 to t+60 days of forecast.
    }
    \label{fig:app_acc_9}
\end{figure}

\subsection{Physical Coherency}
\label{sect:appendix_physical_coherency}

\begin{figure}
    \centering
    \begin{subfigure}{0.3\textwidth}
        \centering
        \caption{Kuroshio current}
        \includegraphics[width=\textwidth]{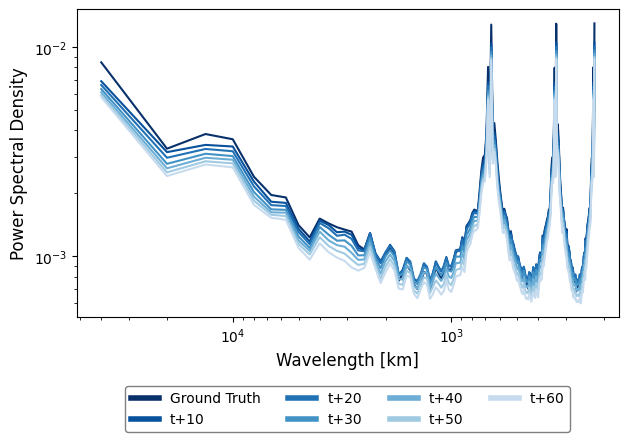}
        \label{fig:eke_kur}
    \end{subfigure}
    \centering
    \begin{subfigure}{0.3\textwidth}
        \centering
        \caption{Gulf Stream current}
        \includegraphics[width=\textwidth]{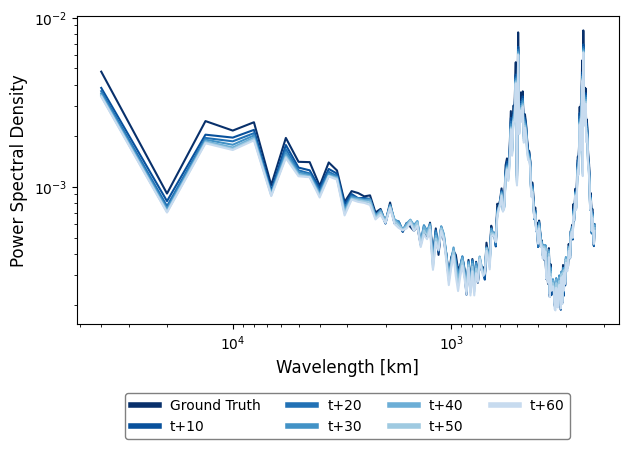}
        \label{fig:eke_gulf}
    \end{subfigure}
    \centering
    \begin{subfigure}{0.3\textwidth}
        \centering
        \caption{Brazil-Malvinas Confluence}
        \includegraphics[width=\textwidth]{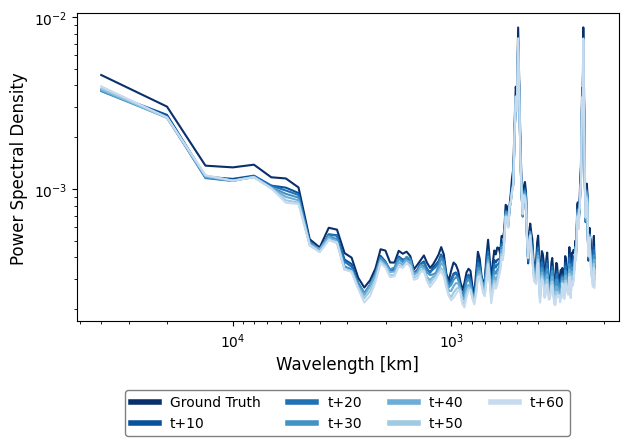}
        \label{fig:eke_brazmav}
    \end{subfigure}
    \centering
    \begin{subfigure}{0.3\textwidth}
        \centering
        \caption{Antarctic Circumpolar Current}
        \includegraphics[width=\textwidth]{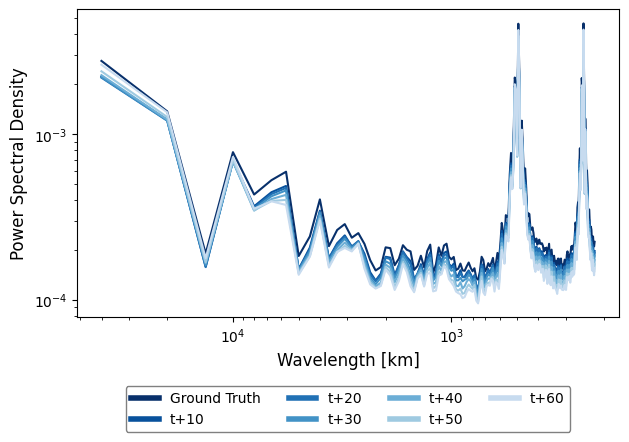}
        \label{fig:eke_acc}
    \end{subfigure}
    \centering
    \begin{subfigure}{0.3\textwidth}
        \centering
        \caption{Agulhas}
        \includegraphics[width=\textwidth]{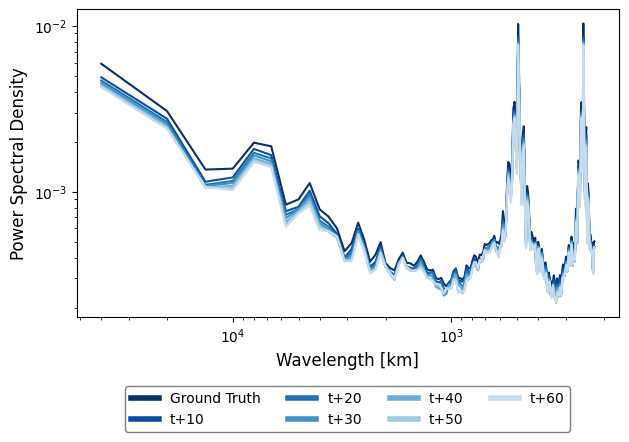}
        \label{fig:eke_agu}
    \end{subfigure}
    \caption{
    Power Spectral Density (PSD) plots computed for 5 highly-energetic ocean regions: a) Kuroshio, b) Gulf Stream, c) Brazil-Malvinas Confluence, d) Antarctic Circumpolar Current, e) Agulhas.
    In each sub-panel, x-axis is reported the wavelength (in km), representing the spatial scale, while y-axis reports the PSD of the signal.
    Ground truth is reported in dark-blue while forecasts-spanning from t+10 to t+60-are reported in progressively lighter-blue colors.
    }
    \label{fig:psd_high_energetic_regions}
\end{figure}

\subsection{Oceanic Indices}
\label{sect:appendix_oceanic_indices}


\end{document}